\documentclass[fleqn,usenatbib]{mnras}

\usepackage{newtxtext,newtxmath}

\usepackage[T1]{fontenc}
\usepackage{ae,aecompl}

\usepackage{graphicx}	
\usepackage{amsmath}	

\usepackage{longtable}
\usepackage{pdflscape}
\usepackage[version=4]{mhchem}
\usepackage{xr-hyper}

\title[Planets or asteroids? A geochemical method to constrain the masses of White Dwarf pollutants]{Planets or asteroids? A geochemical method to constrain the masses of White Dwarf pollutants}

\author[A. M. Buchan et al.]{
Andrew M. Buchan$^{1}$\thanks{E-mail: amb237@cam.ac.uk (AMB)}, Amy Bonsor$^{1}$,
Oliver Shorttle$^{1,2}$, Jon Wade$^{3}$, John Harrison$^{1}$,\newauthor
Lena Noack$^{4}$, Detlev Koester$^{5}$
\\
$^{1}$Institute of Astronomy, University of Cambridge, Madingley Road, Cambridge, CB3 0HA, UK\\
$^2$Department of Earth Sciences, University of Cambridge, Downing Street, Cambridge, CB2 3EQ, UK\\
$^3$Department of Earth Sciences, South Parks Road, Oxford, OX1 3AN, UK\\
$^4$Department of Earth Sciences, Freie Universit\"{a}t Berlin, Malteserstr. 74-100, 12249 Berlin, Germany\\
$^5$Institut f\"{u}r Theoretische Physik und Astrophysik, University of Kiel, 24098 Kiel, Germany
}

\date{Accepted XXX. Received YYY; in original form ZZZ}

\pubyear{2021}

\begin{document}
\label{firstpage}
\pagerange{\pageref{firstpage}--\pageref{lastpage}}
\maketitle

\begin{abstract}
Polluted white dwarfs that have accreted planetary material provide a unique opportunity to probe the geology of exoplanetary systems. However, the nature of the bodies which pollute white dwarfs is not well understood: are they small asteroids, minor planets, or even terrestrial planets? We present a novel method to infer pollutant masses from detections of Ni, Cr and Si. During core--mantle differentiation, these elements exhibit variable preference for metal and silicate at different pressures (i.e., object masses), affecting their abundances in the core and mantle. We model core--mantle differentiation self-consistently using data from metal--silicate partitioning experiments. We place statistical constraints on the differentiation pressures, and hence masses, of bodies which pollute white dwarfs by incorporating this calculation into a Bayesian framework. We show that Ni observations are best suited to constraining pressure when pollution is mantle-like, while Cr and Si are better for core-like pollution. We find 3 systems (WD0449-259, WD1350-162 and WD2105-820) whose abundances are best explained by the accretion of fragments of small parent bodies ($<0.2M_\oplus$). For 2 systems (GD61 and WD0446-255), the best model suggests the accretion of fragments of Earth-sized bodies, although the observed abundances remain consistent ($<3\sigma$) with the accretion of undifferentiated material. This suggests that polluted white dwarfs potentially accrete planetary bodies of a range of masses. However, our results are subject to inevitable degeneracies and limitations given current data. To constrain pressure more confidently, we require serendipitous observation of (nearly) pure core and/or mantle material.
\end{abstract}




\section{Introduction}

Polluted white dwarfs provide a unique opportunity to probe the interiors of rocky bodies by revealing their composition. Due to the high surface gravity of white dwarfs, elements heavier than H or He are expected to sink through the observable part of their atmospheres on time-scales which are short compared to their cooling time-scales \citep{Fontaine1979,Paquette1986a,Paquette1986b}. \citet{Koester2014} found that between 27\% and 50\% of young white dwarfs are `polluted’ with heavy elements. This suggests recent or ongoing accretion of external material.

The pollutants are thought to be remnants of planetary objects which were able to survive into their hosts' post-main sequence lifetime. Post-main sequence stellar mass loss perturbs the orbits of any companions. Asteroids and planetesimals can be scattered onto eccentric orbits which bring them close to the white dwarf \citep{Debes2002}, especially if inner planets are present \citep{Bonsor2011,Mustill2017,Maldonado2020}. Bodies which pass within the white dwarf's Roche radius can be tidally disrupted \citep{Jura2003, Veras2014b}, with the resulting debris ultimately accreting onto the white dwarf via a variety of possible mechanisms \citep{Brouwers2021}, causing pollution.

Elements which have been detected in white dwarf atmospheres include the key rock-forming elements Mg, Si and O, as well as Fe. These elements trace the composition of accreted bodies. Fe, as well as other siderophilic (\textit{lit.} `iron loving') elements such as Ni and Cr, traces the formation of planetary cores. Mg, Si and O are lithophilic (\textit{lit.} `rock loving'), and trace the formation of mantles and crusts.

Measurements of the relative abundance of siderophiles and lithophiles in white dwarf atmospheres provide evidence that core--mantle differentiation, the segregation of metallic core from silicate mantle, is ubiquitous in the formation of rocky planetary bodies \citep{JuraYoung2014}. This process is witnessed in the Solar system by meteorites which record planetesimal differentiation occurring very early in the proto-planetary nebula, when bodies reached a few tens to hundreds of kilometres in diameter \citep{Righter1996,Kleine2009,DeSanctis2012}. For example, \citet{Melis2011} reported observations of Fe-rich pollution in the white dwarf GALEX J1931+0117, which they suggested could be due to a differentiated body with an Fe-rich core that lost its outer layers. Similar examples are given by \citet{Gaensicke2012}, \citet{Wilson2015}, \citet{Hollands2018} and \citet{Hollands2021}. \citet{Klein2010} and \citet{Zuckerman2011} inferred crust--mantle differentiation in 2 systems via a similar analysis. If these white dwarfs have accreted fragments from collisions between larger bodies, \citet{Bonsor2020} concluded that at least 60\% of all polluted white dwarfs have accreted fragments of differentiated bodies (even if we cannot tell that they are differentiated), based on the tendency of collisional evolution to yield fragments whose core mass fractions are similar to that of their parents (and which are therefore indistinguishable from undifferentiated material). \citet{Harrison2018} modelled white dwarf pollution by considering the formation histories of pollutants, focussing on incomplete condensation and core--mantle differentiation. This model was incorporated into a Bayesian framework by \citet{Harrison2021} in order to estimate the most likely origin of a given composition observed in a white dwarf atmosphere. In several cases, even after accounting for other relevant processes, their model shows that differentiation is still statistically required.

The source of white dwarf pollutants, and the mechanism of their delivery to the white dwarf, remains an open question. The ubiquitous nature of pollution necessitates a process common to many systems. \citet{Jura2003} proposed that pollution could be caused by the accretion of many asteroids which pass within the white dwarf's Roche radius and are tidally disrupted. Planets, whose orbits were perturbed by mass loss from the white dwarf's progenitor, can in turn perturb a belt of asteroids or comets \citep{Debes2002,Bonsor2011,Debes2012,Mustill2017}. \citet{Wyatt2014} suggest that the accretion of many small bodies can explain the discrepancy in average accretion rates between white dwarfs of spectral type DA and DB. However, pollution by much larger objects has also been proposed. Exomoons can be liberated from their companions and scatter towards the white dwarf, polluting it \citep{Payne2016a,Payne2016b}. The discovery of Be in white dwarfs \citep{Klein2021} has been attributed to accretion of an icy exomoon \citep{Doyle2021}. Abundances in a number of white dwarfs point to the presence of crustal material (e.g., \citealt{Klein2010,Zuckerman2011,Melis2011}), implying that their pollutants might be terrestrial-like minor planets. Simulations by \citet{Veras2013} show that it is possible for inner planets to directly impact the white dwarf following post-main-sequence dynamical instability. Transit photometry has revealed planets orbiting close to white dwarfs, confirming that planets are able to reach close-in orbits (e.g., \citealt{Vanderburg2015,Vanderburg2020}). Moreover, the abundances of pollutants in the atmosphere of WD J0914+1914 (and its gaseous circumstellar disc) suggest that it is accreting material from an icy giant planet \citep{Gaensicke2019}.

Constraining the masses of white dwarf pollutants therefore allows for these scenarios to be distinguished from each other. Additionally, mass constraints would provide valuable context to other chemically derived information. For example, if a pollutant is derived from a parent which is both differentiated and low mass, an additional heat source (besides release of gravitational potential energy) must be present to facilitate differentiation.

In this paper, we aim to constrain the masses of the objects which give rise to white dwarf pollution by modelling how their differing geochemistry affects the composition of pollutants. Our model builds on the work of \citet{Harrison2021}, who used a Bayesian framework to model the compositions observed in white dwarf atmospheres. We add the ability to trace planetary cores and mantles of varying composition, as this allows us to constrain the conditions under which the iron core formed.

Our working hypothesis is that prior to accretion onto the white dwarf, pollutants may experience collisions. We refer to a collisionally unprocessed body as a parent body, and the resulting post-collision bodies as fragments. If a parent body has differentiated into a core and mantle, its fragments will be composed of the same core and mantle material. However, the core:mantle ratio in any given fragment will not necessarily be the same as in its parent. This could account for observations of core--rich or mantle--rich material. More sophisticated treatments of collisions track fragment composition as a function of collision parameters (e.g., \citealt{Marcus2010,Bonsor2015,Carter2015}), but such an analysis is beyond the scope of this paper. Other non-collisional explanations of core- and mantle-like compositions exist, such as wind-stripping of outer layers (as in \citealt{Melis2011}).

\begin{figure}
    \centering
    \includegraphics[width=8cm]{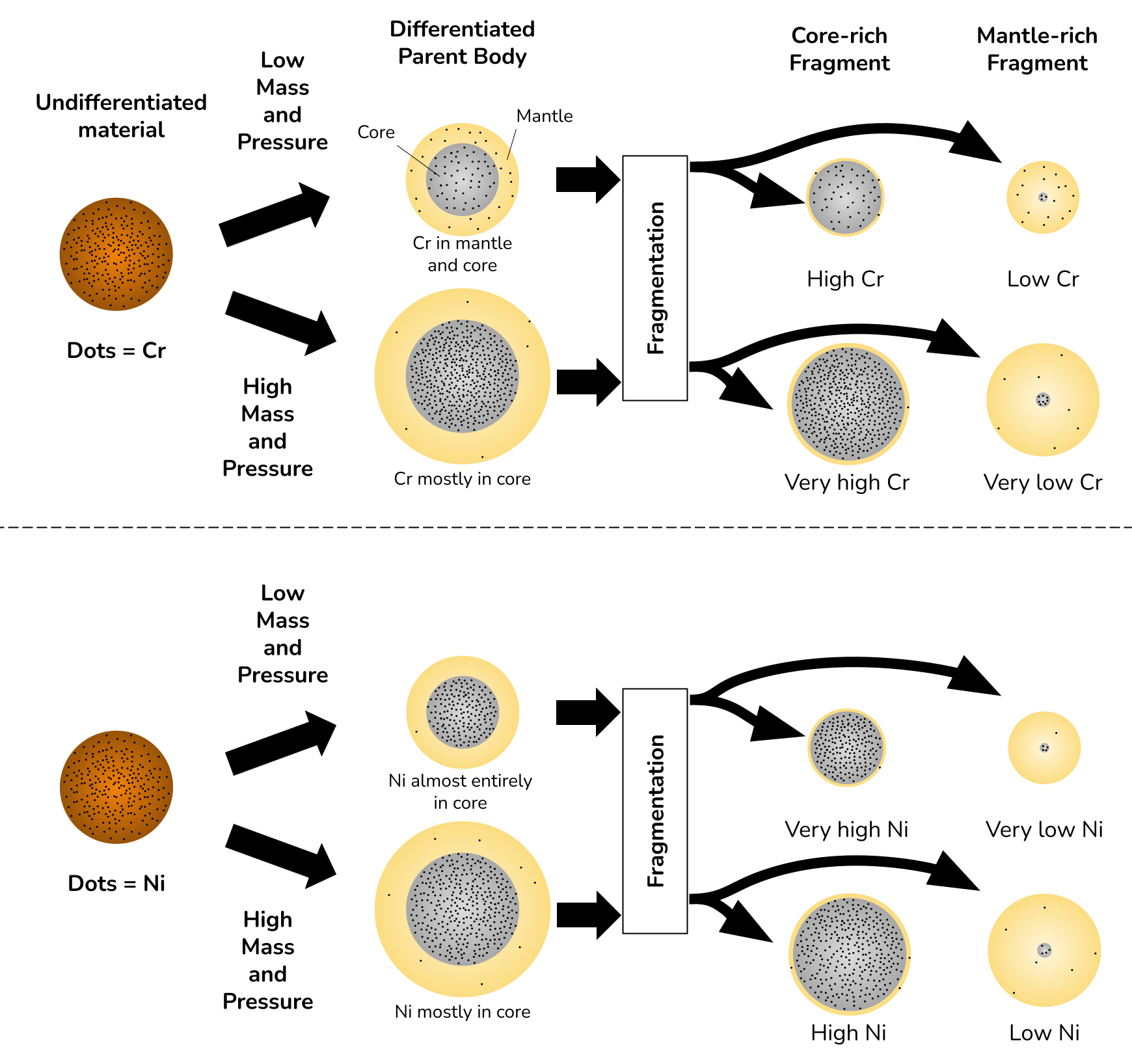}
    \caption{Top panel: A schematic illustrating how the Cr content of planetary cores and mantles changes in planetary bodies of different sizes. The aim is to detect these differences in planetary bodies accreted by white dwarfs. Assuming planetary differentiation occurs with little change in the attendant redox conditions, higher pressures of core--mantle segregation result in Cr exhibiting increasingly siderophile behaviour and becoming more concentrated in the core. This altered compositional signature remains present in any fragments derived from the parent, enabling constraints on the size of the parent body from the composition of fragments observed in polluted white dwarfs. In our working hypothesis, fragmentation occurs via collisional processing, but our model does not include any treatment of the fragmentation mechanism so this choice is inconsequential. The increased bulk concentration of Cr in the high pressure parent body (compared to the low pressure parent body) is to aid visual clarity. Si exhibits similar behaviour to Cr, becoming increasingly siderophile with the increasing temperature concomitant with rising core formation pressures. Bottom panel: Similar to top panel, but illustrating Ni instead of Cr. Ni exhibits the opposite behaviour, becoming more lithophile with increasing pressure. Silicate mantle Ni content therefore increases as pressure, and hence planetary body size, increases. Because Ni is always highly siderophilic, the core Ni content is less strongly affected.}
    \label{fig:schematic}
\end{figure}

The premise of our work is illustrated in Figure \ref{fig:schematic}. The distribution of certain elements during core--mantle differentiation is significantly affected by the mass of the differentiating body. One such element is Cr. For larger, more massive parent bodies with higher internal pressure, Cr becomes increasingly concentrated in the core and consequently the mantle exhibits lower Cr concentrations. This parent body may subsequently be disrupted into core-rich and/or mantle-rich fragments. If a white dwarf becomes polluted with core-like or mantle-like fragments, the relative abundance of Cr acts as a proxy for the internal pressure of metal--silicate segregation of the fragment's parent. We expect this pressure to increase with the mass of the parent body: core--mantle segregation is commonly treated as occurring at a fixed fraction of mantle depth (e.g., \citealt{Wade2005}). In this work, we focus on Cr, Ni and Si. These elements have been detected in several white dwarf atmospheres and show the greatest sensitivity to pressure of all the elements we model (see Section \ref{sec:d_behaviour}).

To determine the behaviour of Cr, Ni and Si, we make use of the results of liquid metal-- liquid silicate partitioning experiments performed at elevated pressures and temperatures. In this context, the term ‘metal’ refers to those elements which comprise the chemically metallic core of a differentiated planet, and the term ‘silicate’ to the rocky portion, dominated by oxides of the rock forming elements, principally Ca, Mg, Si, and Al. During metal--silicate partitioning experiments, a sample containing both an Fe-rich metallic phase and a silicate phase is subjected to high pressure and temperature (e.g., \citealt{Wade2005,Corgne2008,Fischer2015}). The partitioning behaviour of element(s) of interest (i.e., its preference for the metal or silicate phase) can then be parametrized as a function of pressure, temperature and oxygen fugacity. Metal--silicate partitioning experiments are used to constrain the conditions under which the Earth differentiated (e.g., \citealt{Wade2005,Siebert2013,Badro2015,Fischer2015}). Such modelling efforts can find good agreement with estimates of Earth's core and mantle composition, arriving at a general consensus that Earth underwent metal--silicate differentiation at peak pressures in the range of 40-60\;GPa. We present a novel application of this methodology to extrasolar systems.

\section{Methods}
\label{sec:Methods}

To explain observed pollution abundances in a sample of white dwarfs, and to identify those which require pollution by bodies which have undergone core-mantle differentiation at elevated pressures, we explored 42 white dwarf systems. We selected a sample of systems with a confirmed Fe detection, as well as a detection of (or upper bound on) at least one of Cr, Ni and Si. The properties of these systems are summarised in Tables \ref{tab:sample} and \ref{tab:abundances}. We model the abundances of 12 elements which have been observed in white dwarfs, although typically only 6-7 of these elements were present in a given system, to constrain the models. The 12 elements are the lithophile (\textit{lit.} `rock loving') Al, Ti, Ca, and Mg, the moderately siderophile (\textit{lit.} `iron loving') Ni, Fe, Cr and Si, and the atmophile Na, O, C and N. These are among the most commonly observed elements in white dwarfs, all of which can affect the interpretation of pollutant material. A notable absence on the list of elements is S, whose stellar abundance is difficult to constrain (see note in Section \ref{sec:ModelCaveats}). Abundances are quoted relative to H or He, depending on which of these elements dominates the white dwarf atmosphere. Observational errors vary considerably, but are typically on the order of ~0.1 dex. We fit the data using an adapted version of the Bayesian model presented in \citet{Harrison2021} with up to 9 parameters. The parameters are:

\begin{itemize}
    \item Stellar metallicity, $[\textrm{Fe}/\textrm{H}]_{\textrm{index}}$
    \item Time since accretion started onto the white dwarf, $t$
    \item White dwarf atmosphere pollution fraction, $f_{\textrm{pol}}$
    \item White dwarf accretion event lifetime, $t_{\textrm{event}}$
    \item Formation distance, $d_{\textrm{formation}}$ (optional)
    \item Feeding zone size for planetesimal formation, $z_{\textrm{formation}}$ (optional)
    \item Fragment core fraction, $f_c$ (optional)
    \item Core--mantle equilibration pressure, $P$ (optional)
    \item Core--mantle oxygen fugacity, $f_{\textrm{O}_{2}}$ (optional)
\end{itemize}

The key adaptation of the model detailed in \citet{Harrison2021} is the estimation of elemental core (metal) and mantle (silicate) abundances which result from differentiation under non-Earth-like conditions. In this section we mainly focus on this new treatment of differentiation and we refer the reader to \citet{Harrison2021} for a comprehensive description of the remainder of the model.

\subsection{Outline of White Dwarf Pollution Model}

We assume that the pollution present in the atmosphere of a white dwarf represents the remains of a single body which formed from the same material as the white dwarf progenitor. The model calculates the initial composition of such a body, which forms in a disc at a fixed distance from its host star, and how this composition is modified by various processes.  To achieve this, the initial step sets the composition of the material in a disc which is available for planet-formation. We would ideally use a distribution of initial abundances for the planet-forming material, but in the absence of such a distribution we instead use local stars as a proxy. We take a sample of 958 stars from \citet{Brewer2016} ordered by metallicity (calculated as [Fe/H]). The disc composition can take 958 different values, each of which is simply taken from one of the local stars.

The composition of a body as it forms within the disc is then estimated. \citet{Harrison2021} use a slightly modified version of the irradiated, viscous alpha-disc model of \citet{Chambers2009} to calculate a temperature at a given distance from the star, assuming formation occurs after 1.5\;Myr. We determine which elements are solid, and hence available for the formation of a rocky body, by minimising the Gibbs free energy at the relevant temperature and pressure. The body may also have a feeding zone, which is an annulus within the disc from which it accretes material. In this case, the body's composition is a weighted average over the feeding zone.

A rocky body may undergo collisional evolution, in which an impact causes it to fragment. If the parent (i.e., pre-impact) body has differentiated into a core and a mantle, then the bulk composition of any given fragment will inevitably differ from the bulk composition of the parent body. The fragment composition will depend on the relative abundance of core-like and mantle-like material in the fragment, which is described in terms of the fragment core fraction. The method of calculating the parent's core and mantle composition is new to this work, so is described in more detail in Section \ref{sec:nonearthlike_differentiation}.

The observed pollutant abundances differ from those of the polluting body itself, because different elements sink through the observable part of the white dwarf’s atmosphere at different rates \citep{Koester2009}. The model calculates the observed pollution as a function of time and also allows the duration of the accretion event to be varied. Before the accretion event ends, atmospheric pollutants build up (and may reach a steady state), but after accretion ends all pollutant abundances decay exponentially. The fraction of the white dwarf’s atmosphere which is composed of the observed pollution is the final variable we adopt from \citet{Harrison2021}.

The parameters which we carry over from \citet{Harrison2021} are therefore the stellar metallicity $[\textrm{Fe}/\textrm{H}]_{\textrm{index}}$, the formation distance $d_{\textrm{formation}}$ (optional), the feeding zone size $z_{\textrm{formation}}$ (optional), the fragment core fraction $f_c$ (optional), the time since accretion started $t$, the accretion event lifetime $t_{\textrm{event}}$ and the pollution fraction $f_{\textrm{pol}}$.

White dwarf pollution may be explicable without invoking all the above phenomena. For example, the pollution may not display any signatures of differentiation such as increased Fe content. To establish which phenomena are present, the model is embedded within a Bayesian framework. We use the nested sampling algorithm MultiNest \citep{Feroz2008,Feroz2009,Feroz2019}, as included in PyMultiNest \citep{Buchner2014} to calculate the Bayesian evidence of the model. For a more detailed description, we refer the reader to \citet{Harrison2021}.

We group parameters into 4 sets, as follows:
\begin{itemize}
    \item Set 1 (always present): $[\textrm{Fe}/\textrm{H}]_{\textrm{index}}$, $t$, $f_{\textrm{pol}}$, $t_{\textrm{event}}$
    \item Set 2: $d_{\textrm{formation}}$
    \item Set 3: $z_{\textrm{formation}}$
    \item Set 4: $f_c$, $P$, $f_{\textrm{O}_{2}}$
\end{itemize}
Set 1 is the most basic description of white dwarf pollution in our framework (i.e., primitive material modified only by the white dwarf's atmosphere). The other sets parametrize additional phenomena which may or may not be present. We create 8 models corresponding to every possible combination of sets of parameters (there are 8 in total, since set 1 must always be present). This is similar to the setup shown in Table 2 of \citet{Harrison2021}, but with $f_c$ always accompanied by $P$ and $f_{\textrm{O}_{2}}$, crustal differentiation omitted and additional parameter combinations allowed. We identify the model with the greatest Bayesian evidence to determine which phenomena are present. This process favours models with fewer parameters. Additionally, we calculate the chi-squared value of a given model to ensure it fits the data well. In cases where the favoured model includes core--mantle differentiation (i.e., set 4), we calculate the sigma significance of differentiation by comparing it to the best model (i.e., highest Bayesian evidence) of those which did not invoke differentiation.

\subsection{Modelling Non-Earth-like differentiation}
\label{sec:nonearthlike_differentiation}

\citet{Harrison2021} modelled the composition of a differentiated body by combining core- and mantle- like material in arbitrary proportions. However, the compositions of core- and mantle- like material they considered was restricted to Earth-like compositions, with potential for some modification by varying the core:mantle:crust ratio within the parent body. We relax the assumption of Earth-like core and mantle compositions by calculating them using a self-consistent partitioning model.

This work is motivated by the potential for information about the parent body, such as its size, to be encoded in its core and mantle compositions. Elements such as Cr and Ni can partition into a metallic phase (analogous to a planetary core) more or less efficiently under varying pressures and temperatures \citep{Bouhifd2011,Siebert2012,Fischer2015}. Therefore, a large rocky body (which differentiates at high pressure) may have a different core and mantle composition from an otherwise identical smaller body, as shown in Figure \ref{fig:schematic}.

We removed the consideration of crustal components from the model altogether, since the objective was to investigate systems which appear to be core- or mantle-rich, but with modified abundances of pressure sensitive elements.

We make use of empirical data from partitioning experiments, which measure the partitioning behaviour of elements as a function of pressure, temperature, oxygen fugacity and interaction with other elements (given by interaction parameters) \citep{Wade2005,Corgne2008,Cottrell2009,Siebert2013,Boujibar2014,Fischer2015,Blanchard2019}. These allow us to calculate the composition of the core. After eliminating temperature (see Section \ref{sec:liquidus}), the remaining variables are pressure, oxygen fugacity and elemental interactions. The dependence on interactions is a significant complication because these depend on the composition of the core, which is what we aim to calculate. We therefore require an iterative approach in order to arrive at a self-consistent solution.

\subsubsection{Self-consistent Partitioning}
\label{sec:Self-consistent Partitioning}

We aim to calculate molar partition coefficients for a variety of elements. The molar partition coefficient $D^{*}_{M}$ of an element $M$ is defined as

\begin{equation}
    D^{*}_{M} = \frac{X^{met}_M}{X^{sil}_M},
	\label{eq:d_def_sec2}
\end{equation} where $X^{phase}_M$ is the concentration by number of $M$ in the specified phase, either metal (i.e., core) or silicate (i.e., mantle).

We calculate this quantity in three different ways, depending on the element. The equations we use are 

\begin{equation}
\begin{aligned}
\log{D^*_M} = & \,  a + \frac{b}{T} + \frac{c \cdot P}{T} + d \cdot N - \frac{T_0}{T}\cdot \log{\gamma^{met}_M(T_0)} \\
      & - \frac{v \cdot \textrm{fO}_2}{4} + \frac{v}{2} \cdot \log{\gamma^{sil}_{\textrm{FeO}}},
\label{eq:logdm_rudge}
\end{aligned}
\end{equation}


\begin{equation}
\begin{aligned}
\log{D^*_M} = & \,  a + \frac{b}{T} + \frac{c \cdot P}{T}  - \frac{v \cdot \textrm{fO}_2}{4} + \frac{v}{2} \cdot \log{\gamma^{sil}_{\textrm{FeO}}} \\
      & - \frac{v}{2}\cdot\log{\gamma^{met}_{\textrm{Fe}}}-\frac{T_0}{T}\cdot\log{\gamma^{met}_M(T_0)} \\
      & + \log{\gamma^{met}_{\textrm{Fe}}} + \log{\gamma^0_M},
\label{eq:logdm_fischer2}
\end{aligned}
\end{equation}

\begin{equation}
    \log{D^*_M} = \log{\gamma^{sil}_{\textrm{FeO}}} - \frac{\textrm{fO}_2}{2} - \log{\gamma^{met}_{\textrm{Fe}}},
	\label{eq:logdm_dfe}
\end{equation} where $a$, $b$, $c$ and $d$ are empirically derived coefficients, $T$ is temperature, $P$ is pressure, $N$ is the molar ratio of non-bridging oxygens to tetrahedral cations (NBO/T) in the silicate melt (which we take to be similar to the primitive terrestrial mantle, with an NBO/T of 2.7), $T_0$ is a reference temperature, $\gamma^{phase}_M$ is the activity coefficient of $M$ in the specified phase at temperature $T$, $v$ is the valence of $M$, $\textrm{fO}_2$ is the oxygen fugacity in log units relative to the Iron-W{\"u}stite (Fe - FeO) buffer, $\gamma^0_M$ are element-specific terms which have a temperature dependence and all logarithms are base 10. We take $\gamma^{sil}_{\textrm{FeO}}$ to be equal to 3. Equation \ref{eq:logdm_rudge} is the same as equation G.5 in \citet{Rudge2010}. Equation \ref{eq:logdm_fischer2} is equivalent to Equation \ref{eq:logdm_rudge}, but has been modified for compatibility with alternative parametrizations of empirical data.

The choice of which equation to apply to any given element is determined by how the empirical data for that element was parametrized, which is done in different ways by different sources. The equation used for each element is shown in Table \ref{tab:partition}. Note that equation \ref{eq:logdm_dfe} is used exclusively for Fe, and follows from the definition of $\textrm{fO}_2$:

\begin{equation}
    \textrm{fO}_2 = 2 \log \left( \frac{a_{\textrm{FeO}}^{sil}}{a_{\textrm{Fe}}^{met}}\right),
	\label{eq:fo2_def}
\end{equation} where $a^{phase}_{M}$ is the chemical activity of $M$ in the specified phase, which can in turn be calculated as $a^{phase}_{M} = \gamma^{phase}_{M} \cdot X^{phase}_{M}$.

Some elements (N, Na, Mg, Al, Ti and Ca) are assumed to be solely lithophile and not partition into the core at all, however. Such elements are assigned a partition coefficient of 0.






\subsubsection{Interaction Parameters}
\label{sec:Interaction_Parameters}

An element's activity in a phase is affected by its interaction with the other elements present. \citet{Corgne2008}, following the approach of \citet{Ma2001}, express $\gamma^{met}_M$ as

\begin{equation}
\begin{aligned}
&\ln{\gamma^{met}_M} = \ln{\gamma^{met}_{\textrm{Fe}}} + \ln{\gamma^0_M} - \varepsilon^M_M\ln(1-X^{met}_M) \\
&- \sum_{j=2, j\neq M}^{N}\left[\varepsilon^j_MX^{met}_j\left(1 + \frac{\ln{(1 - X^{met}_j)}}{X^{met}_j} - Y_{M}\right)\right] \\
&+ \sum_{j=2, j\neq M}^{N}\left[\varepsilon^j_M(X^{met}_j)^{2}X^{met}_M\left(Y_{M} + Y_{j} + Z_{M} - 1\right)\right],
\label{eq:gamma_def_eps}
\end{aligned}
\end{equation} where

\begin{equation}
    Y_{j} = \frac{1}{1-X^{met}_j},
	\label{eq:y_def_eps}
\end{equation}

\begin{equation}
    Z_{j} = \frac{X^{met}_j}{2(1-X^{met}_j)^2},
	\label{eq:z_def_eps}
\end{equation}

\begin{equation}
\begin{split}
    \varepsilon^j_M = \frac{T_0}{T}\varepsilon^j_M(T_0),
	\label{eq:epsilon_def}
\end{split}
\end{equation} and the $\varepsilon^j_M(T_0)$ terms are empirically determined pair-specific constants called interaction parameters.

\citet{Fischer2015} provide parametrizations for certain elements alongside an accompanying parametrization for their interactions with certain other elements. Their equation 4, which is in turn taken from the Steelmaking Data Sourcebook \citep{SDS1988}, is

\begin{equation}
    \varepsilon^{i}_k = \frac{e^{i}_{k}M_{i}T_{ref}}{0.242T} - \frac{M_i}{55.85}+1,
	\label{eq:fischer_eps_def}
\end{equation} where $\varepsilon^{i}_k$ gives the effect of interaction with element $k$ on element $i$, $M_i$ is the molar mass of element $i$ in $g/mol$, and $e^{i}_{k}$ are empirical values determined at a reference temperature $T_{ref}$. For element-element pairs which were parametrized in this way, we calculate their interaction parameters using equation \ref{eq:fischer_eps_def} with the accompanying parameter values from \citet{Fischer2015}. Otherwise, we use equation \ref{eq:epsilon_def} with parameter values from the Steelmaking Data Sourcebook \citep{SDS1988}.

\subsubsection{Pressure-Temperature profile}
\label{sec:liquidus}

We eliminate $T$ as a variable by calculating it as a function of pressure, $P$. We use an adaptation of the peridotite liquidus given by \cite{Schaefer2016}:

\begin{equation}
  \begin{gathered}
    T = \begin{cases}
    \beta_1 + \alpha_1(P-P_c) & \text{if } P > P_c \\
    \beta_2 + \alpha_2P & \text{if } P \leq P_c \\
    \end{cases} \\
    P_c = \frac{\beta_1-\beta_2}{\alpha_2},
    \label{eq:liquidus}
  \end{gathered}
\end{equation} with $\alpha_1=26.53$\,K\,GPa$^{-1}$, $\alpha_2=104.42$\,K\,GPa$^{-1}$, $\beta_1=2425$\,K and $\beta_2=2020$\,K. This P-T profile is motivated by our assumption that metal ponds at the base of a magma ocean, which can be no hotter than the peridotite liquidus. It is also motivated by simplicity: at temperatures cooler than the liquidus, a third phase is present (i.e., the solid precipitate).

\subsubsection{Composition Calculations}
\label{sec:comp_calc}

Assuming we can calculate a partition coefficient for any element, we now need to calculate the resulting core and mantle compositions. For element $M$, we calculate 

\begin{equation}
\begin{gathered}
    N^{met}_M = \frac{N^{total}_MD_Mw_{met}}{w_{sil} + D_Mw_{met}} \\
    N^{sil}_M = \frac{N^{total}_Mw_{sil}}{w_{sil} + D_Mw_{met}},
	\label{eq:board_calcs}
\end{gathered}
\end{equation} where $N^{phase}_M$ is the total number abundance of $M$ in the specified phase and $w_{phase}$ is the number fraction of the specified phase, i.e., the core and mantle number fractions. We require $w_{sil} + w_{met} = 1$. We then find the number concentrations, $X^{phase}_M$ as

\begin{equation}
\begin{gathered}
    X^{met}_M = \frac{N^{met}_M}{\sum_{j}N^{met}_j}\\
    X^{sil}_M = \frac{N^{sil}_M}{\sum_{j}N^{sil}_j}.
	\label{eq:xm_calcs_o}
\end{gathered}
\end{equation}



We recalculate the core number fraction as

\begin{equation}
    w_{met} = \frac{\sum_{j}N^{met}_{j}}{\sum_{j}N^{total}_{j}}.
	\label{eq:w_met_calc_o}
\end{equation}

\subsubsection{Partition Calculation Algorithm}

To calculate the mantle and core composition due to differentiation at a specified pressure $P$ and oxygen fugacity $\textrm{fO}_2$, assuming we have an initial guess of core composition $X^{metal}_M$ for all $M$ of interest, and an initial guess of $w_{met}$, we follow these steps:

\begin{enumerate}
  \item Calculate $T$ using equation \ref{eq:liquidus}
  \item Calculate $\gamma^{met}_M$ for all elements $M$ using equation \ref{eq:gamma_def_eps}
  \item Calculate $D^*_{\textrm{Fe}}$ using equation \ref{eq:logdm_dfe}
  \item Calculate $D^{*}_{M}$ for all other elements $M$ using either equation \ref{eq:logdm_rudge} or \ref{eq:logdm_fischer2} according to Table \ref{tab:partition}
  \item Calculate $N_{M}$ for all $M$ using equations \ref{eq:board_calcs}.
  \item Calculate $X_{M}$ for all $M$ using equations \ref{eq:xm_calcs_o}.
  \item Calculate $w_{met}$ using equation \ref{eq:w_met_calc_o}
\end{enumerate}

Repeat until convergence of the $D^*_M$ for elements of interest, or until 1000 iterations have been completed without convergence.

The calculation of the $\gamma^{met}_M$ is a noteworthy step, because it ensures self-consistency between the partition coefficients and the resulting composition of the metallic melt. This self-consistent approach has previously been used by \citet{Badro2015}.

\subsubsection{Model assumptions}
\label{sec:ModelCaveats}

We briefly note two key assumptions of our model. We discuss its caveats in more detail in Section \ref{sec:ModelCaveats2}.

Earth (and, by extension, other bodies) is not thought to differentiate at a single value of pressure, temperature or oxygen fugacity. Instead, these may vary over the course of accretion (e.g., \citealt{Wade2005,Badro2015}). We use a single-stage differentiation model which assumes a single value of pressure (as well as oxygen fugacity). This should be thought of as an `effective value', representing the average pressure of core--mantle differentiation, which can be compared to similar values quoted for Earth, Mars etc. We refer this pressure as $P_{\textrm{diff}}$.

In Section \ref{sec:comp_calc}, we ignore the possibility of partial differentiation - that is, disequilibrium between segregating core and mantle. We assume the whole inventory of a given element equilibrates with the metallic core and silicate mantle at the base of a magma ocean according to its partition coefficient.

\subsection{Calculating the size of rocky bodies}
\label{sec:size_calc}

We wish to find the mass and radius of a rocky body which underwent differentiation at a given pressure. Previous work has made use of interior structure models to derive parametrizations relating properties such as planetary radius $R_p$, mass $M_p$ and core-mantle boundary pressure $P_{\textrm{CMB}}$ (e.g., \citealt{Noack2020}). The core-mantle boundary pressure is not the same as the differentiation pressure our model constrains. For example, the Earth's core--mantle boundary pressure is 136\;GPa \citep{McDonough2003}, while the differentiation pressure - determined principally by the Ni and Co abundance in the mantle - appears significantly lower at $\approx$50\;GPa \citep{Fischer2015}.

We therefore used the same methodology outlined in \citet{Noack2020} to derive parametrizations in terms of the mid-mantle pressure $P_{mm}$, which initial modelling suggested was a significantly better proxy for differentiation pressure than $P_{\textrm{CMB}}$ for both Earth and Mars. We considered masses between 0.001 and 2 $M_{\oplus}$, and Fe weight fractions between 0.15 and 0.75.

We found that

\begin{equation}
    M_{p} = \left(\frac{R_{p}}{7008.42-1829X_{\textrm{Fe}}}\right)^{1/0.313},
    \label{eq:Rp}
\end{equation} where $M_{p}$ is the planet's mass (in units of Earth masses), $R_{p}$ is the planet's radius (in $\textrm{km}$) and $X_{\textrm{Fe}}$ is the planet's Fe mass fraction (on a scale from 0 to 1),

\begin{equation}
    R_{c} = 1067.44(100X_{\textrm{Fe}})^{0.329}M_{p}^{0.31},
\end{equation} where $R_{c}$ is the planet's core radius in $\textrm{km}$,

\begin{equation}
	g_s = \frac{GM_{\oplus}M_{p}}{(1000R_{p})^{2}},
\end{equation} where $g_s$ is the planet's gravitational acceleration at the surface (in $\textrm{m}/\textrm{s}$), G is the gravitational constant expressed in SI units, and $M_{\oplus}$ is the Earth's mass (in $\textrm{kg}$),

\begin{equation}
	g_c = \frac{GM_{\oplus}M_{p}X_{\textrm{Fe}}}{(1000R_{c})^{2}}, 
\end{equation} where $g_c$ is the planet's gravitational acceleration at the edge of its core (in $\textrm{m}/\textrm{s}$),

\begin{equation}
	g_{um} = 0.75g_s + 0.25g_c,
\end{equation} where $g_{um}$ is the planet's average gravitational acceleration across its upper mantle (in $\textrm{m}/\textrm{s}$),

\begin{equation}
	\rho_{m} = \frac{3M_{\oplus}M_{p}(1-X_{\textrm{Fe}})}{4\pi(R_p^3 - R_c^3)},
\end{equation} where $\rho_{m}$ is the average density of the planet's mantle (in $\textrm{kg}/\textrm{m}^3$),

\begin{equation}
	\rho_{um} = 0.5(\rho_{m} + \rho_{s}),
\end{equation} where $\rho_{um}$ is the average density of the planet's upper mantle (in $\textrm{kg}/\textrm{m}^3$) and $\rho_{s}$ is the density of uncompressed surface rocks (which we take to be $3100\textrm{kg}/\textrm{m}^3$),

\begin{equation}
	D_{mm} = 1000\times0.5(R_{p} - R_{c}),
\end{equation} where $D_{mm}$ is the depth of the mid-mantle (in $\textrm{m}$) and

\begin{equation}
	P_{mm} = 10^{-9}g_{um}\rho_{um}D_{mm},
\end{equation} where $P_{mm}$ is the mid-mantle pressure in $\textrm{GPa}$.

\begin{figure}
    \centering
    \includegraphics[width=8cm]{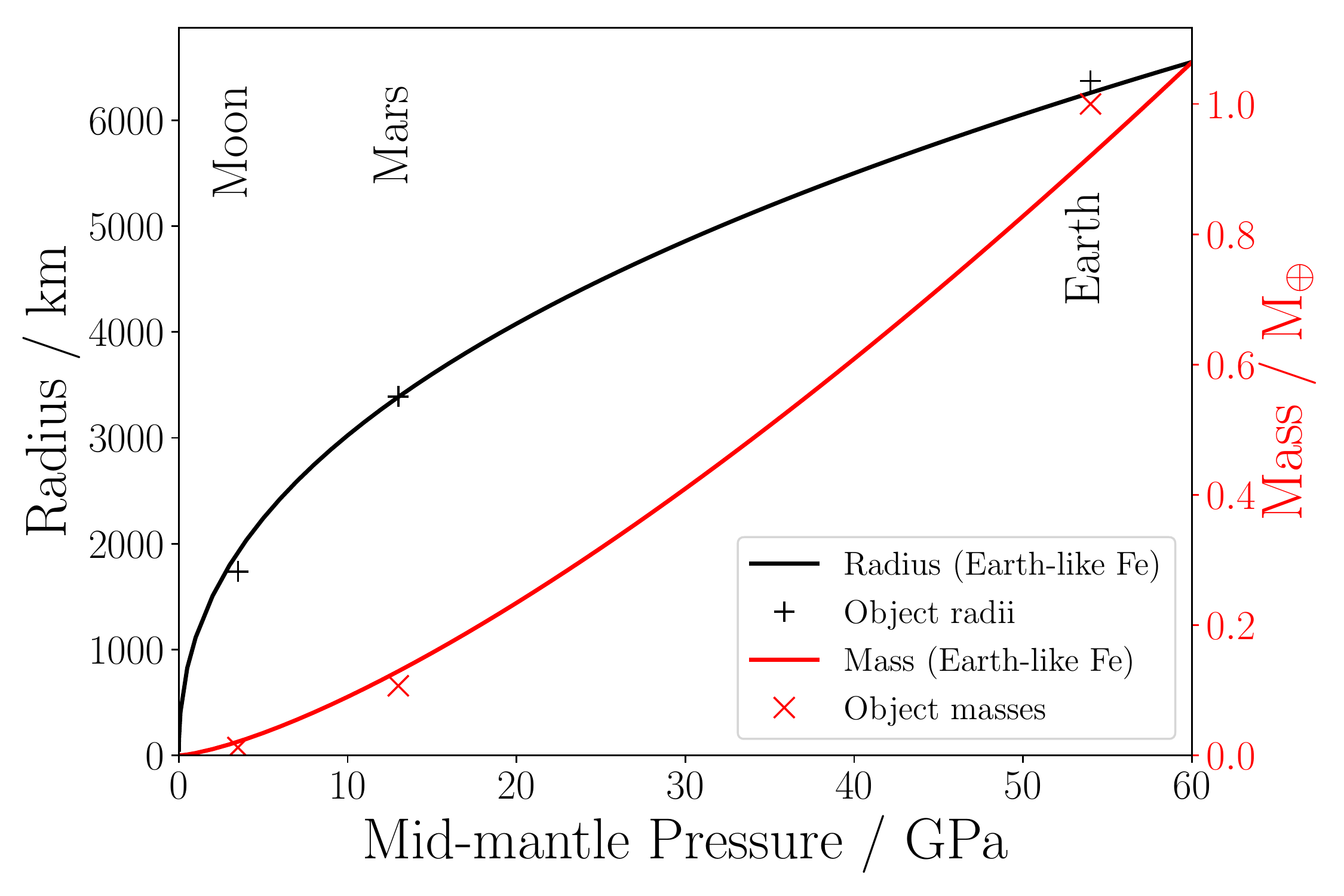}
    \caption{Our parametrized relationship between the mid-mantle pressure within a differentiated body and its radius (black line) and its total mass (red line). The lines shown assume an Earth-like Fe content. Also shown for reference are the mass and radius of Earth, Mars and the Moon, which correspond to pressures of 54\;GPa \citep{Fischer2015}, 13\;GPa \citep{RaivanWestrenen2013} and 3.5\;GPa \citep{Righter1996} respectively.}
    \label{fig:MassRadiusPressure}
\end{figure}

Following these equations, we can calculate a mid-mantle pressure given the planet's radius and Fe content. We invert the calculation, running a binary search to find a value of $R_p$ (and the accompanying value of $M_p$) that recovers our input value of $P_{mm}$ (to within a small tolerance) given the body's Fe content. This is shown in Figure \ref{fig:MassRadiusPressure}.

Bodies with equal mass may have different internal pressures depending on their rotation speed \citep{Lock2019}. We assume no rotation, so the mass/radius values we infer are lower limits. In the corotation limit, the actual mass could be higher by a factor of 2 \citep{Lock2019}, corresponding to a radius increase of roughly 1.25 (using equation \ref{eq:Rp}).

\subsection{Resulting behaviour of Cr, Ni and Si}
\label{sec:d_behaviour}

We illustrate the behaviour of three key differentiation-pressure sensitive elements (Cr, Ni and Si) as predicted by our partitioning model. These elements have all been detected in polluted White Dwarf atmospheres, and are the most promising elements for constraining any pressure signatures.

\begin{figure}
    \centering
    \includegraphics[width=8cm]{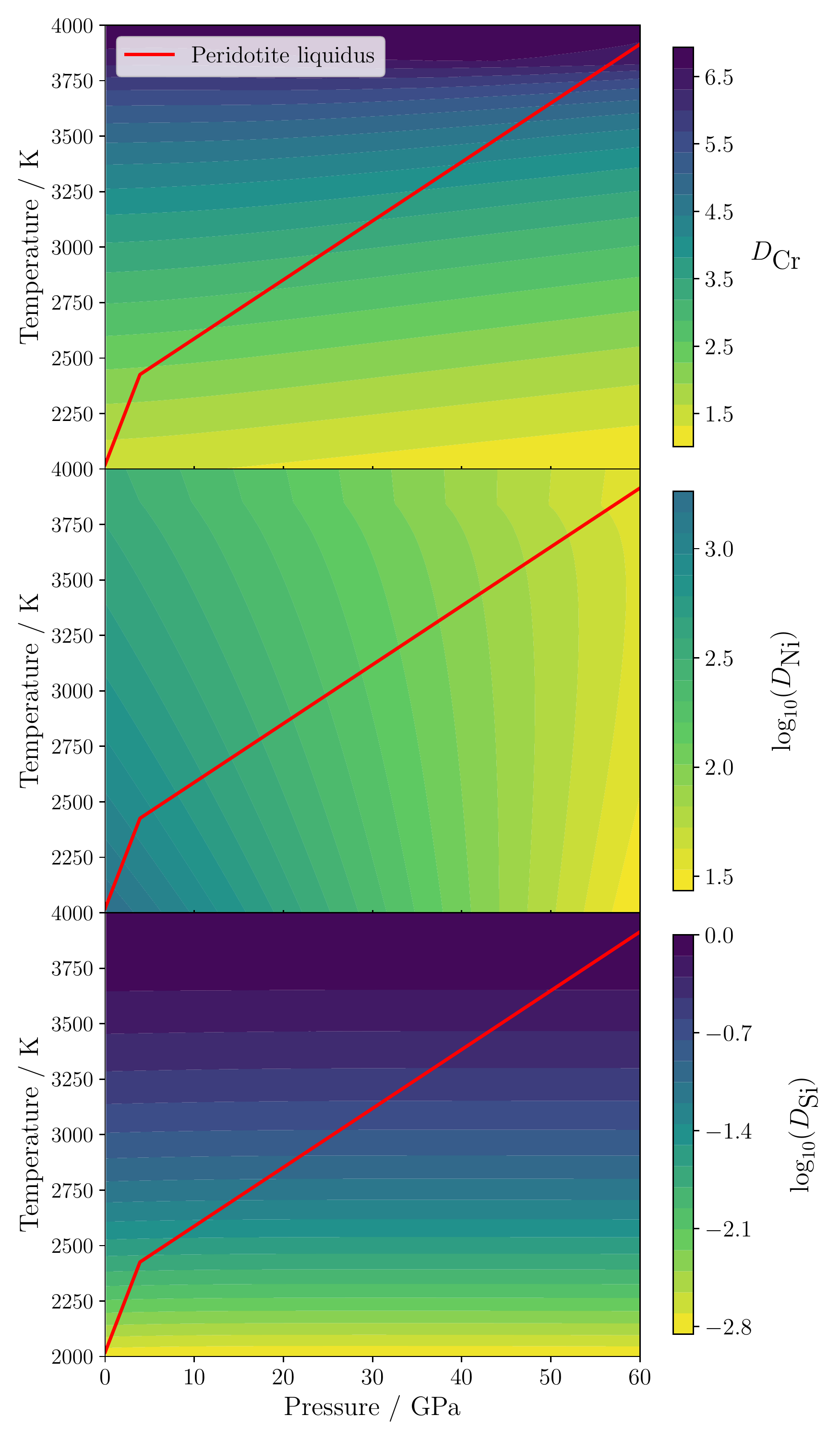}
    \caption{Modelled partitioning behaviour as a function of pressure and temperature for Cr (top panel), Ni (middle) and Si (bottom). Note the use of a log scale (base 10) for Ni and Si. In the full Bayesian model, temperature is treated as a function of pressure, given by the peridotite liquidus (red line, adapted from \citet{Schaefer2016}). Note that $D_{\textrm{Cr}}$ decreases with increasing pressure when temperature is held constant, but this behaviour is reversed when the temperature is determined by the peridotite liquidus. Similarly, Si inherits its pressure dependence from the imposed pressure-temperature constraint. Calculations in this figure were made assuming a bulk-Earth composition and oxygen fugacity of IW - 2.}
    \label{fig:d_el_3d_plot}
\end{figure}

Figure \ref{fig:d_el_3d_plot} shows how the partition coefficient (i.e., the core:mantle concentration ratio) varies for Cr, Ni and Si as a function of pressure and temperature. Cr and Ni are always siderophilic to some extent (i.e., $D_{\textrm{Cr}} > 1$). As described in Section \ref{sec:ModelCaveats}, we cap $D_{\textrm{Si}}$ at a maximum value of 1, forcing it to be lithophilic, but this cap is not reached for the majority of pressure/temperature space. When holding temperature constant, changing pressure causes both Cr and Ni to become less siderophilic, while the behaviour of Si is essentially unaffected.

However, we do not expect pressure and temperature to be independent, and the imposition of a pressure-temperature relationship heavily alters the pressure dependent behaviour. We assume that the pressure and temperature are related via the peridotite liquidus, which we adapt from the relation given in \citet{Schaefer2016}. This relationship is plotted in Figure \ref{fig:d_el_3d_plot} as a red line. When the temperature at which metal--silicate partitioning occurs is constrained in this way, the partitioning behaviour of Cr is reversed: Cr becomes more siderophilic with increasing pressure. The partitioning behaviour of Ni is amplified. Si is very sensitive to temperature, and is therefore also sensitive to pressure when it is related to temperature in this way. We find that that Si increasingly enters the core as the pressures of core-mantle segregation, and hence the temperature, rise.

Figure \ref{fig:d_el_3d_plot} has implications for the circumstances under which we expect each element to be most sensitive to pressure. The partition coefficient of Ni varies over multiple orders of magnitude, but is always much larger than 1 (note that $D_{\textrm{Ni}}$ is shown on a log scale). Recalling that the partition coefficient is the metal:silicate concentration ratio, this implies that the abundance of Ni in the core does not change significantly as a function of differentiation pressure (it contains almost all of the available Ni, irrespective of $P_{\textrm{diff}}$). However, the mantle abundance of Ni changes roughly linearly with $D_{\textrm{Ni}}$, ranging over orders of magnitude.

From the systematics described above, we can identify that Ni is best able to constrain pressure for pollutant material which is mantle-rich. Si is (mostly) highly lithophile, so by a similar argument we infer that it is best suited for constraining the pressure of core-rich material. Cr is an intermediate case, containing $P_{\textrm{diff}}$ information for both core- and mantle-rich fragments.

\subsection{Testing the model against Earth and Mars}
\label{sec:retrieval}

\begin{figure}
    \centering
    \includegraphics[width=\textwidth/2]{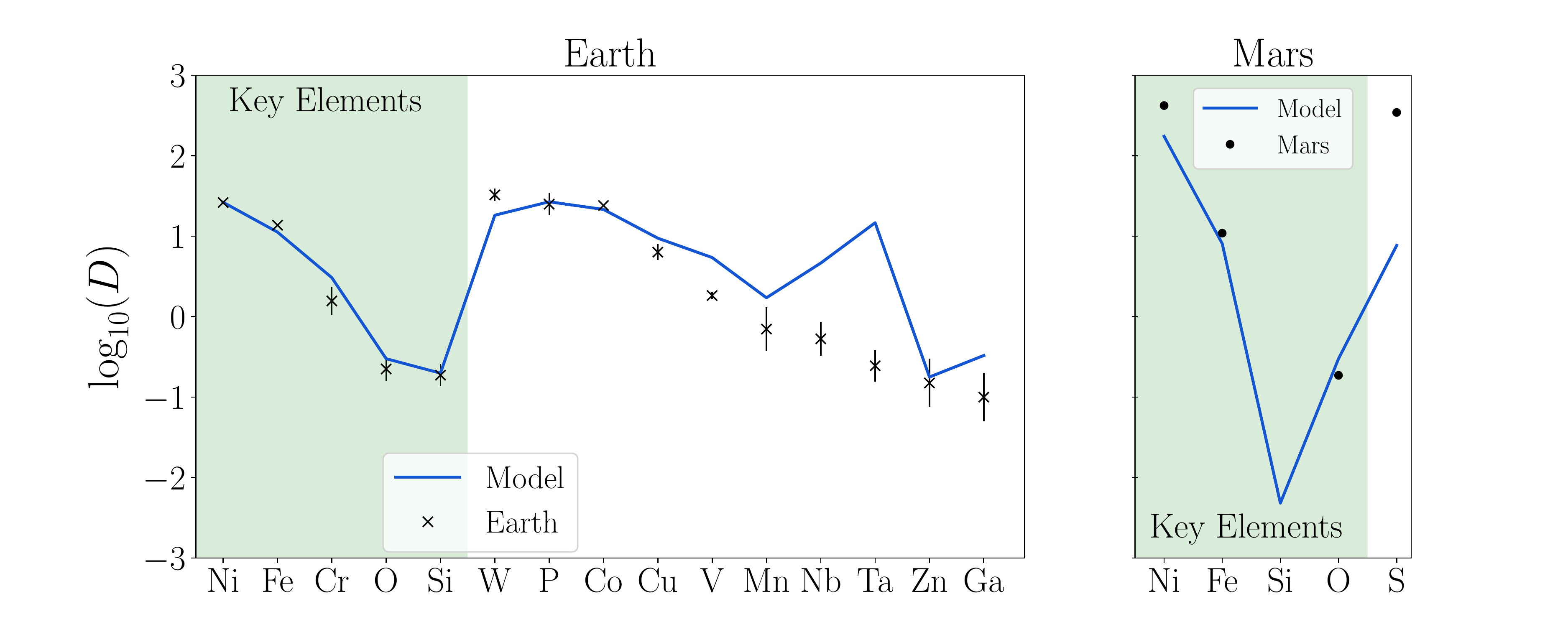}
    \caption{Comparison of published partition coefficients with those given by our model for Earth and Mars. The partition coefficients for Earth were taken from \citet{Rudge2010}. No error was given for Fe because it is well constrained. The partition coefficients for Mars were inferred from the bulk silicate and core compositions given in \citet{Yoshizaki2020}. Comparison with other studies suggests errors are likely to be on the order of a few percent for Fe and Ni, but much larger for O and S. The partition coefficient for Si is inferred to be 0 (since its core abundance is 0) and therefore is not shown on the plot. For the purposes of this plot, we activated S partitioning in our model although this was not used in the rest of our modelling. The model parameters were tuned to match the published coefficients. We used a pressure of 45\;GPa and oxygen fugacity of IW - 1.3 for Earth. The corresponding values for Mars were 5\;GPa and IW - 1.1.}
    \label{fig:partition_coefficients_plot}
\end{figure}

To test the veracity of our approach we evaluate our model's performance on data representative of Earth and Mars. Our model replicates the partition coefficients required to generate the observed mantle abundances of key elements in Earth and Mars (to within 30\% on $\log(D)$, except for Cr). Crucially, it identifies the Earth as having experienced high pressure core--mantle segregation, and is therefore a large body which can be distinguished from planetesimals or asteroids. Figure \ref{fig:partition_coefficients_plot} compares partition coefficients from our model with estimated effective partition coefficients for single stage core--mantle segregation in Earth and Mars. The partition coefficients for Earth are taken from \citet{Rudge2010}, while those for Mars are calculated from the composition found in \citet{Yoshizaki2020}. The model was run assuming the bulk composition given in \citet{McDonough2003} for Earth, and in \citet{Yoshizaki2020} for Mars. Core--mantle segregation pressure and oxygen fugacity were allowed to very in the model in order to best match the reference coefficients. We set the pressure and oxygen fugacity to 45\;GPa and IW\;-\;1.3 respectively for Earth, and to 5\;GPa and IW\;-\;1.1 for Mars.

Of the key elements discussed in this paper, the element which shows the greatest discrepancy with the reference partition coefficients is Cr. Our model consistently overestimates the partition coefficient of Cr when compared to our reference value of 1.6 (our value is 3.1). Using equations \ref{eq:board_calcs}, we find that this implies we overestimate Cr abundances by 0.2 dex for pure core material, and underestimate them by 0.1 dex for pure mantle material. These values are comparable to observational error, so do not dominate over pre-existing sources of uncertainty. We estimate that a change of this magnitude could lead to significant underestimation of $P_{\textrm{diff}}$ (by as much as 20 to 30\;GPa in the cases of high pressure, core-rich material or low pressure mantle-rich material; see Table \ref{tab:observability}). It is difficult to reconcile the terrestrial mantle abundance of Cr with core segregation occurring at a fixed oxygen fugacity, with the Earth’s mantle Cr abundance best fit by core formation progressing under variable oxygen fugacities up to IW - 2. An additional complication is that our model invokes bodies forming at conditions more oxidised than those apparently witnessed by the Earth. At oxygen fugacities above IW - 2 it has been shown that silicate melts become increasingly rich in $\textrm{Cr}^{3+}$ rather that the $\textrm{Cr}^{2+}$ we have modelled here \citep{Berry2004,Wood2008}. As such, we expect the predicted metal silicate partitioning behaviour of Cr to be an increasingly poor predictor of Cr abundance in highly oxidised white dwarf polluters.

\begin{figure}
    \centering
    \includegraphics[width=8cm]{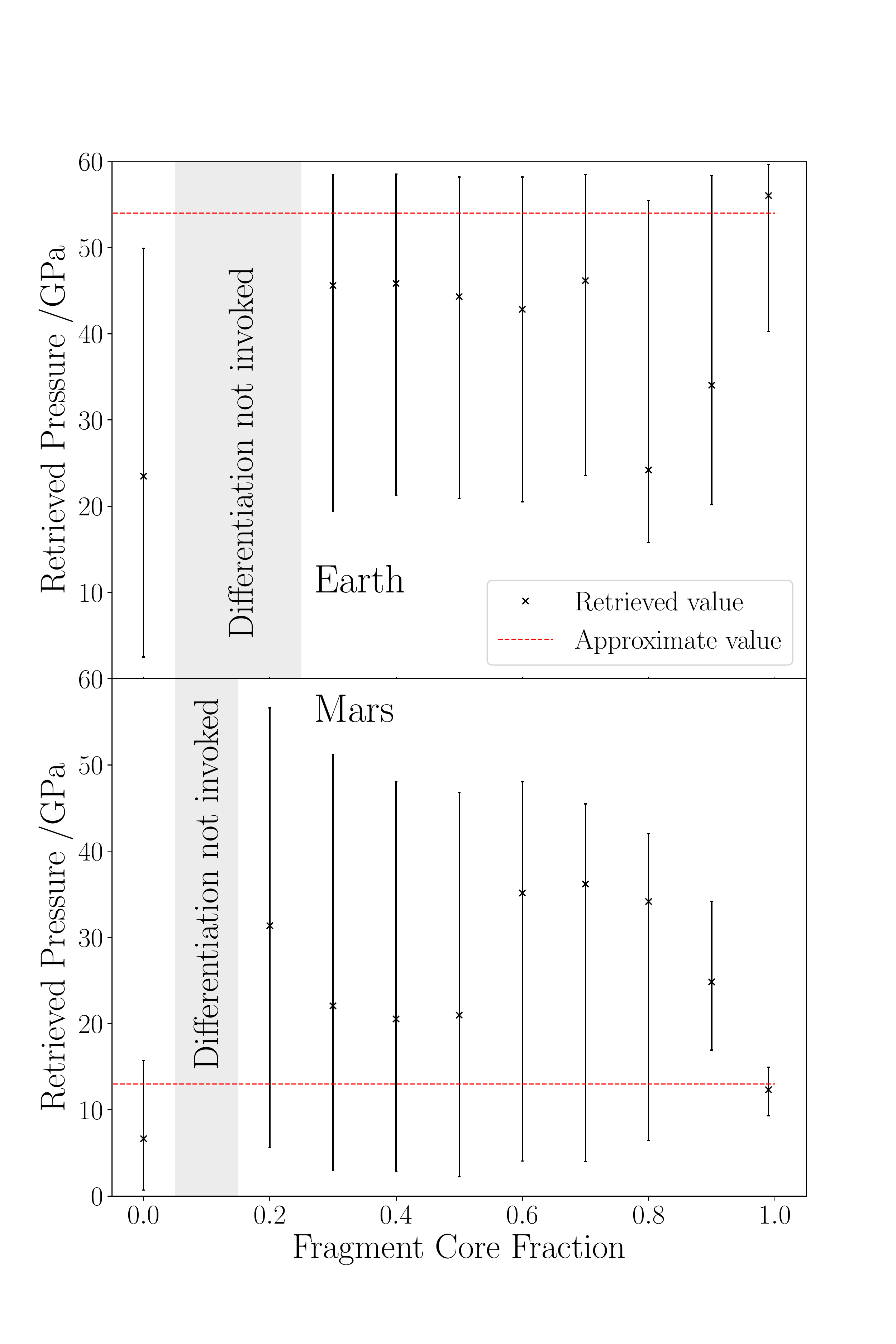}
    \caption{Top panel: Retrieved values of pressure from synthetic input data. The synthetic data was calculated by our partitioning model to correspond to fragments of an Earth-like planet. The fragment compositions are determined by mixing different ratios of core/mantle material. The fragment core fraction sets the proportion of core material in the accreted fragment (the remainder being mantle), which is independent of the pressure at which the parent body formed. We therefore expect to retrieve the same pressure for all fragments. The black crosses indicate synthetic fragments, with the x axis showing the fragment's core fraction and the y axis showing the resulting pressure retrieved by our full Bayesian model (with 90\% confidence intervals). For the fragments with core fraction 0.1 and 0.2, the favoured model did not invoke differentiation, so these fragments have no retrieved pressure and are not shown here. The red dotted line shows the approximate pressure value we would expect to retrieve (54\;GPa, from \citet{Fischer2015}). Bottom panel: Similar to top panel, but using synthetic data corresponding to a Mars-like planet. The target pressure value is 13\;GPa (from \citet{RaivanWestrenen2013}). In this case, the fragment with a core fraction of 0.1 was the only fragment for which differentiation was not invoked.}
    \label{fig:retrieved_pressure}
\end{figure}

We ran the full Bayesian model on synthetic data representing fragments derived from Earth- and Mars-like planets to test whether the model could recover the corresponding core--mantle differentiation pressure. The fragments varied in their core:mantle mixing ratio. Our model requires fragments to be highly core- or mantle- rich in order to be able to constrain pressure well, in which case $P_{\textrm{diff}}$ can be retrieved to within about 10\;GPa or less (with the exception of Earth-like mantle fragments). This is illustrated in Figure \ref{fig:retrieved_pressure}, which shows the pressure (with 90\% confidence intervals) inferred for the synthetic fragments. Importantly, the model identifies all synthetic objects as the product of high pressure differentiation (i.e., not derived from asteroids or planetesimals), with the exception of the Mars-like mantle fragment which is also consistent with low pressure. This is despite the large errors, systematic uncertainties and degeneracies.

For intermediate fragment core fractions (i.e., between 0.1 and 0.9 inclusive), the pressure signature is inherently less clear. Since the model is attempting to constrain $P_{\textrm{diff}}$ by moving material between core and mantle components, if both components are present in significant quantities pressure changes have little effect. This results in less accurate retrieved values, with larger errors (Figure \ref{fig:retrieved_pressure}). In the extreme case of the fragment core fraction closely matching the parent core fraction, the model does not need to invoke differentiation.

Figure \ref{fig:retrieved_pressure} shows that for Earth-like fragments with core fractions of 0.8 and 0.9, the retrieved median pressure was significantly lower than the target value. This is because of a degeneracy in the model. The (high pressure) synthetic data can be matched by low pressure, combined with a slightly increased fragment core fraction and moving to a steady state of accretion. The steady state solution occupies a larger amount of parameter space than the intended build-up phase solution, leading this to become the favoured solution (although the build-up phase solution is still present). A similar degeneracy appears to be present for lower fragment core fractions, although in these cases the intended solution is also the favoured solution. This appears to be because the steady state solution needs some fine tuning to match the higher lithophile abundances in those cases.

The model did not constrain pressure well for the Earth-like mantle fragment. In this case, the pressure was degenerate with oxygen fugacity (and to a lesser extent the fragment core fraction). This implies that there may be high pressure mantle-rich pollutants in our data set which our model is not identifying as such.

The nearly pure core or mantle fragments, which are required for tight pressure constraints, can be identified by the relative abundances of siderophile and lithophile elements. Commonly observed lithophiles include Ca and Mg, while Fe is the most commonly observed siderophile element. In this paper, we use Mg/Fe as a proxy for fragment core fraction, but Ca/Fe would also be suitable. Assuming steady-state accretion, $\log(\textrm{Mg}/\textrm{Fe}) \lesssim-2$ or $\log(\textrm{Ca}/\textrm{Fe}) \lesssim -3$ indicates highly core--rich material, while $\log(\textrm{Mg}/\textrm{Fe}) \gtrsim1$ or $\log(\textrm{Ca}/\textrm{Fe}) \gtrsim0$ indicates highly mantle--rich material (see Table \ref{tab:observability}). However, the relative abundances of Ca, Mg and Fe can be altered by their differing sinking time-scales (typically $t_{\textrm{Fe}} < t_{\textrm{Ca}} < t_{\textrm{Mg}}$), leading to a degeneracy between the fragment core fraction and the phase of accretion. This can be addressed by considering both Ca/Fe and Mg/Fe simultaneously, although there is still the potential for degeneracy due to their differing condensation temperatures (which can be addressed by considering more elements).

\section{Results}
\label{sec:Results}

For all 42 polluted white dwarfs in our sample, we identified the model with the highest Bayesian evidence. This model offers the best explanation for the observed abundances. We divide the sample into 6 categories according to the need for core-mantle differentiation and the constraints which can be placed on the conditions under which it occurred (i.e., the inferred $P_{\textrm{diff}}$). The categories are as follows:
\begin{itemize}
\item Core-rich, low differentiation pressure (i.e., small parent body) (3 systems)
\item Mantle-rich, high differentiation pressure (i.e., large parent body) (3 systems)
\item Differentiation pressure degenerate with oxygen fugacity (4 systems)
\item Differentiation pressure unconstrained (5 systems)
\item No evidence of differentiation (26 systems)
\item Unphysical solution (1 system)
\end{itemize}

We will describe these categories, and salient features of the objects which fall into them, in subsections \ref{sec:LPC} to \ref{sec:PF}. For specific comments on individual white dwarfs, median values of the best model parameters, and our categorisation for each system we refer the reader to Appendix Section \ref{sec:individual_systems}, Table \ref{tab:results1} and Table \ref{tab:results2}.

\begin{figure*}
    \centering
    \includegraphics[width=13cm,keepaspectratio=true]{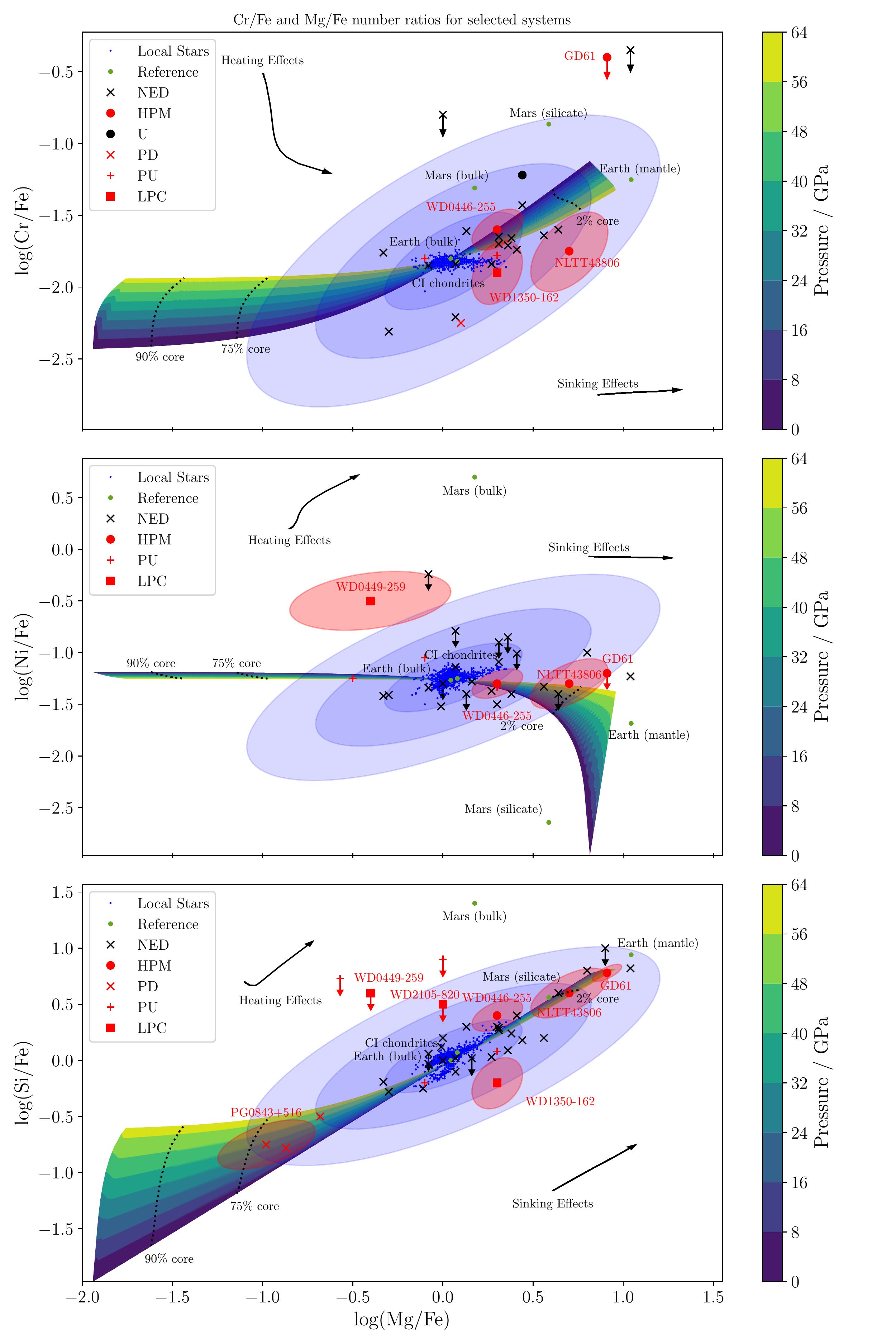}
    \caption{Top panel: Cr/Fe and Mg/Fe number ratios for selected systems. In each panel, we include all systems from our sample of 42 which have the abundance measurements required for placement on the plot. Note that since Cr, Ni and Si are usually not all detected, white dwarf systems are typically absent from at least one panel. Local Main Sequence stars used in our model are shown with blue dots. Selected reference systems are shown with green dots and labelled. White Dwarfs are shown with red or black symbols according to the category we place them in, as indicated by the legend. Abbreviations are as follows: HPM = High Pressure Mantle-rich, LPC = Low Pressure Core-rich, PD = Pressure degenerate with oxygen fugacity, PU = Pressure unconstrained, NED = No Evidence of Differentiation, U = Unphysical. See the main text for discussion of these categories. Arrows indicate upper bounds. Blue ellipses show the 1, 2 and 3 sigma confidence ellipses generated by applying noise to (randomly selected) local stars based on the average error on the White Dwarf observations. White Dwarf errors are shown for selected systems as 1 sigma confidence ellipses. The contours indicate the range of compositions that can be reached by differentiating a bulk Earth composition at a fixed oxygen fugacity of IW - 2 but varying pressure, and then combining the resulting core and mantle material in different proportions. The colour of the contour indicates the pressure, which is a function of the size of the pollutant's parent body, while the core fraction of the resulting body decreases from left to right. Lines for 2\%, 75\% and 90\% core are shown as examples. Our model assumes that the core content of Mg is zero, so pure core material would be arbitrarily far to the left. The contours are cut off at 95\% core. Effects due to sinking and heating are shown with arrows; for details see the main text. Middle and bottom panels: Similar to top panel, but for Ni and Si instead of Cr respectively.}
    \label{fig:bowtie_multipanel}
\end{figure*}

To better understand the significance of the observed elemental abundances across our sample, we plot elemental number ratios for each system in Figure \ref{fig:bowtie_multipanel}. Each panel features a different pressure-sensitive element: Cr in the top panel, Ni in the middle panel and Si in the bottom panel. The x-axis is the same for each panel, showing the Mg/Fe ratio. In our model, Mg is found exclusively in mantle-like material, while Fe is found primarily in core-like material. For systems where the pollutant is a fragment composed of a mix of mantle- and core-like material, the Mg/Fe ratio therefore acts as a proxy for the fragment's mantle:core mixing ratio. Note that this is separate from the mantle:core ratio of the original ('parent') body from which the fragment is derived, as illustrated in Figure \ref{fig:schematic}. On the y-axis we show the abundance of Cr, Ni or Si relative to Fe. This traces the pressure-sensitive partitioning behaviour of these elements.

In Figure \ref{fig:bowtie_multipanel}, we use red/black markers to show data from polluted White Dwarfs. Upper bounds are shown with arrows. Most systems lie close to stellar composition, as represented by local stars (from \citet{Brewer2016}, blue dots). The blue ellipses, centred on the local stars, show representative white dwarf observational errors and the 1, 2 and 3 sigma level. The majority of pollution observations are consistent with stellar material, to within 3 sigma. When considering only the elements shown in a given panel, no additional processes are strictly required to explain these observations. Such an explanation may be inconsistent with constraints provided by other elements, however. Our modelling explores the possibility of finding a more statistically favourable explanation for the data by invoking additional processes.

The key process we invoke is differentiation of a rocky body (the `parent body'), followed by accretion of a fragment of this body. In this scenario, the observed pollution tells us about the composition of the fragment. The contours in Figure \ref{fig:bowtie_multipanel} give a rough indication of the possible compositions of such a fragment. The contours are generated by using our model to calculate the core and mantle composition of a parent body with bulk Earth-like composition which differentiates at a certain pressure (and fixed oxygen fugacity), and then combining the resulting core and mantle material in arbitrary proportions. The colour of the contour indicates the differentiation pressure, $P_{\textrm{diff}}$, (i.e., the size of the parent body), while the core:mantle mixing ratio of the fragment is reflected (approximately) by the Mg/Fe value. Lines for 2\%, 75\% and 90\% core are shown as examples. White Dwarf pollutants which lie within these contours can potentially be explained by invoking differentiation. Such an explanation becomes more statistically favourable the further away the pollutant is from matching stellar material. As an example, PG0843+516 lies at the outer edge of the 3 sigma ellipse in the bottom panel, but lies within the pressure contours. We find that the model with highest evidence invokes differentiation, and is favoured to more than 5 sigma over any model that doesn't invoke differentiation.

Many objects do not lie within the pressure contours of Figure \ref{fig:bowtie_multipanel}. This can be partially attributed to observational errors, but there are also important physical processes which can alter the locations of White Dwarf pollutants.
\begin{itemize}
    \item The intrinsic composition of a pollutant differs from that observed due to sinking effects. Different elements sink through the photosphere of a White Dwarf at different rates. Therefore, as accretion proceeds (and eventually ends), the relative observed elemental abundances are altered \citep{Koester2009}. We show the resulting movement through `abundance space' over time with the arrows marked `Sinking Effects' in Figure \ref{fig:bowtie_multipanel}. A hypothetical pollutant whose composition lies within the pressure contours can therefore move off the contours if observed at a sufficiently late time. The `Sinking Effects' arrows are generated using our model, assuming representative (relative) photospheric sinking timescales. We calculate the change in an arbitrary initial composition as accretion proceeds through build-up, steady state and declining phases. We end our calculation roughly 2 Mg sinking timescales after accretion reaches the declining phase, but in principle the arrow can extend arbitrarily far at arbitrarily late time.
    \item Parent bodies which form nearer to or further from their host stars will be enhanced in refractory or volatile elements, respectively. To a lesser extent, the relative abundances of elements shown in Figure \ref{fig:bowtie_multipanel} are similarly affected. The movement though abundance space due to temperature driven incomplete condensation is roughly indicated with the arrows labelled `Heating Effects'. These arrows are generated using our model. We calculate the change in an arbitrary initial composition resulting from condensation at varying formation distance (and hence temperature).
    \item The initial (stellar) composition of the primitive material can vary, which translates the position of the contours. The spread of local stars in abundance space indicates the strength of this effect.
\end{itemize}

The contours in Figure \ref{fig:bowtie_multipanel} are obtained assuming an oxygen fugacity in the parent body of IW - 2, but their shape would be altered if the oxygen fugacity was different. Similarly, the composition of the parent body slightly alters the contour shape (as well as its position) because the interactions between elements during differentiation is a function of the abundance of those elements in the pre-differentiated body.

Our model is more likely to find fits invoking core-rich fragments than mantle-rich fragments. This is because, as a rough approximation, the heating/sinking arrows in Figure \ref{fig:bowtie_multipanel} tend to point from the core-rich pressure contours towards the mantle-rich contours. This means that systems near mantle-rich contours may be explained as a core-rich fragment modified by heating and/or sinking, but the reverse is not true.

For Cr and Si, core-rich fragments are more sensitive to pressure (i.e., parent body size) than mantle-rich fragments, while for Ni the opposite is true. This is shown in Figure \ref{fig:bowtie_multipanel} by the increased spacing between pressure contours for core-rich fragments (with low Mg/Fe) in the Cr and Si panels, and for mantle-rich fragments (with high Mg/Fe) in the Ni panel. There is a relative lack of observations which lie in the regions of maximum pressure sensitivity. This implies that future observations of white dwarf pollutants found in this region of abundance space may enable our model to draw stronger pressure constraints than the present data set allows (see Section \ref{sec:retrieval} for further discussion).

\subsection{Core-rich, low differentiation pressure}
\label{sec:LPC}

\begin{figure}
    \centering
    \includegraphics[width=8cm]{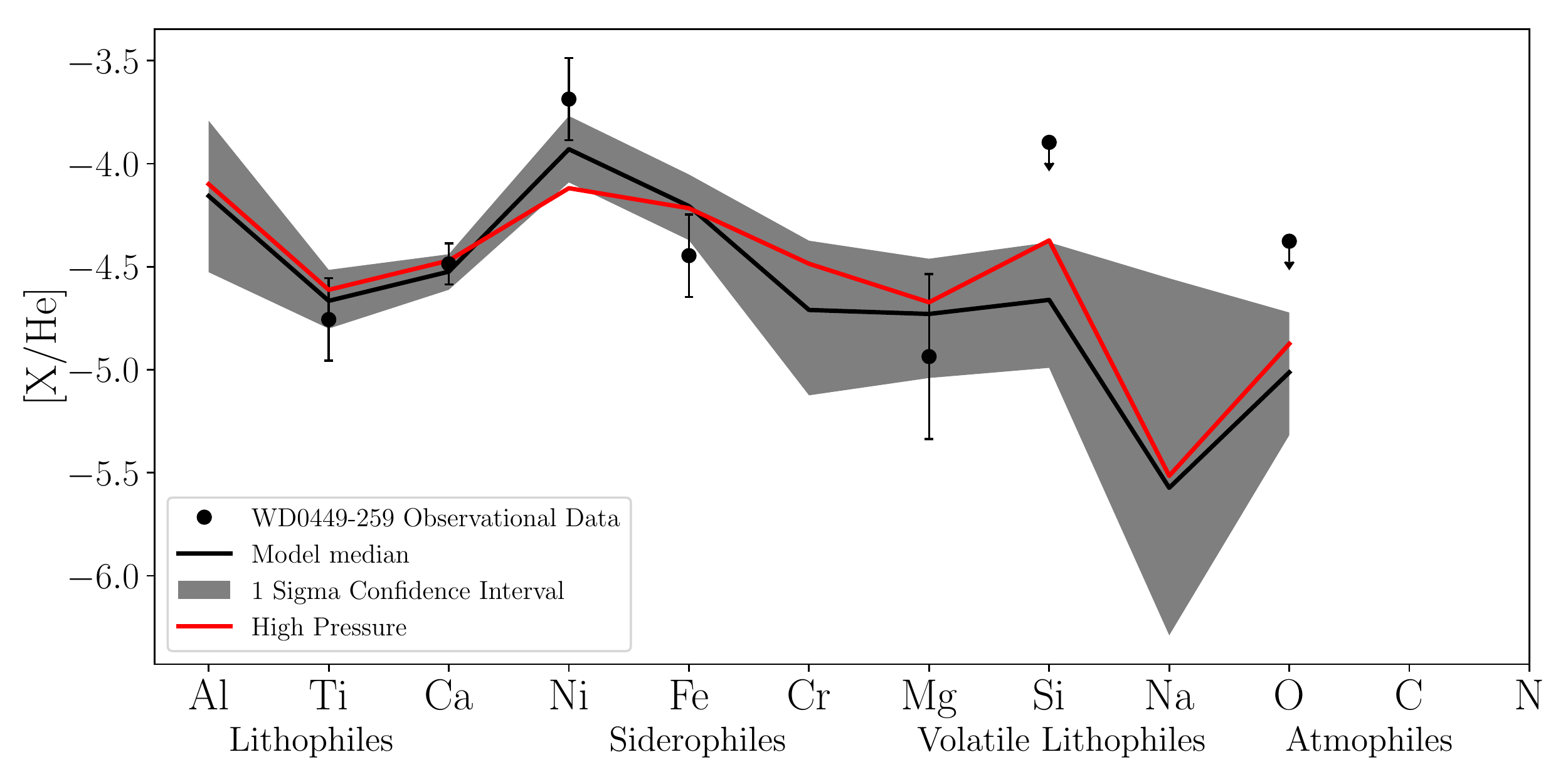}
    \caption{Median fit (with 1 sigma errors) to the observations of \mbox{WD0449-259} pollution for the model with highest Bayesian evidence. The fit was generated by sampling 10,000 sets of posterior values for each parameter and calculating the median of the resulting abundances. Upper limits are shown with arrows. The median pressure is $19.9_{-13.2}^{+19.2}$\;GPa, whilst the red line is the median fit when setting pressure to 60\;GPa.}
    \label{fig:WD0449_comp}
\end{figure}

\begin{figure}
    \centering
    \includegraphics[width=8cm]{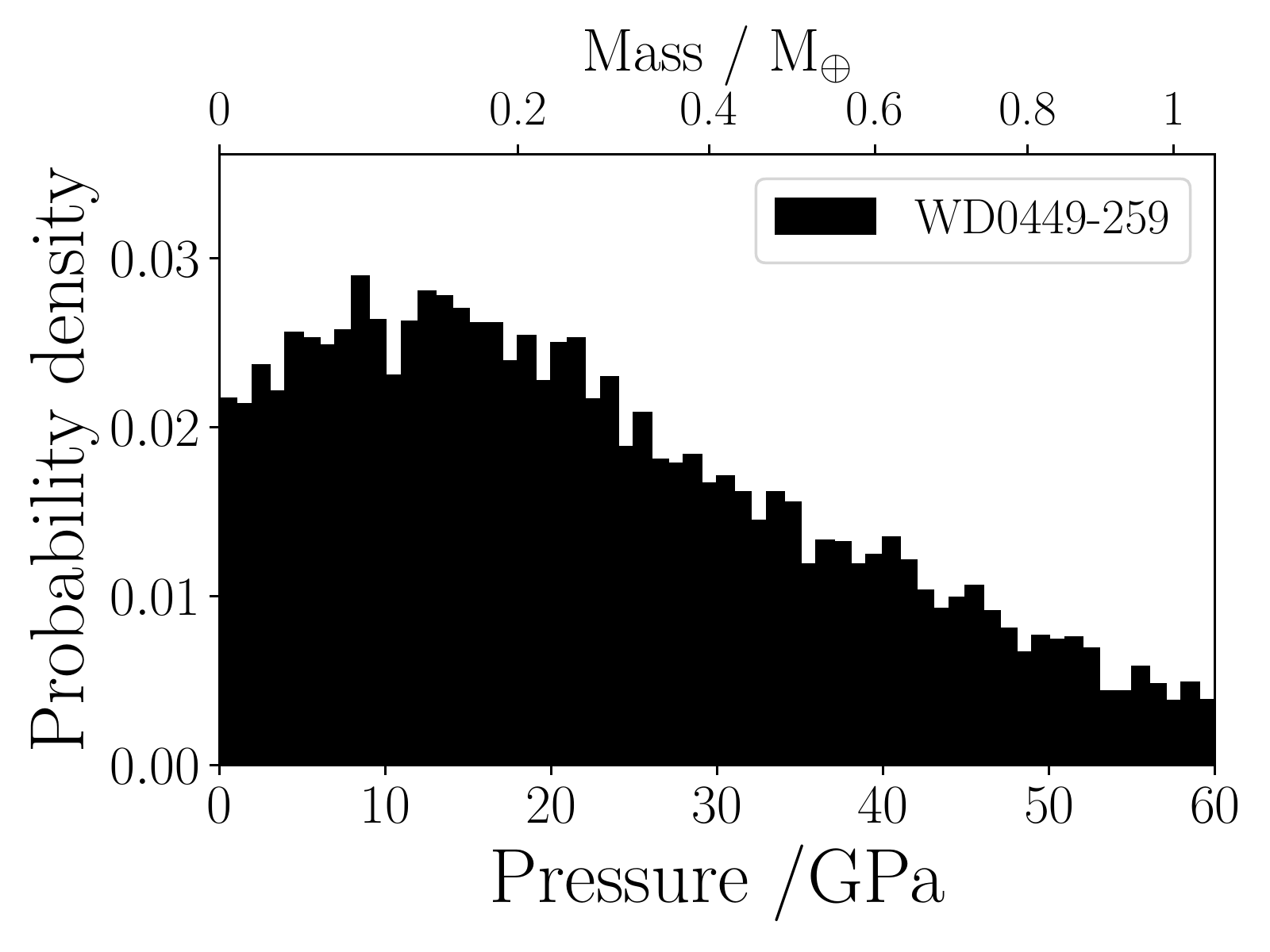}
    \caption{Posterior pressure distribution for WD0449-259, shown in Figure \ref{fig:WD0449_comp}. The model favours low pressure. The pressure shown here is assumed to correspond to the pressure at which the pollutant's parent body underwent core--mantle segregation. The upper x axis shows how low pressure implies a low parent body mass, according to our parametrization (see Section \ref{sec:size_calc} and Figure \ref{fig:MassRadiusPressure}).}
    \label{fig:WD0449_pressuredist}
\end{figure}

We find 3 systems (WD0449-259, WD1350-162 and WD2105-820) for which the model with highest Bayesian evidence invoked both core--mantle differentiation and accretion of a core-rich fragment, as well as favouring low pressure (i.e., a small parent body) over high pressure. The posterior distributions on pressure visibly peak at low pressure (4\;GPa, 8\;GPa and 4\;GPa respectively), which is illustrated for WD0449-259 in Figure \ref{fig:WD0449_pressuredist}. In this case, the key element is Ni. Figure \ref{fig:WD0449_comp} shows how low pressure improves the Ni fit by roughly $1\sigma$ compared to high pressure. We cannot rule out a high pressure solution for any system, because in each case the posterior probability that the pressure is above 45\;GPa is at least 10\%. Moreover, our model contains a degeneracy which can cause high pressure, core--rich objects to appear as low pressure objects. Our results must therefore be treated with caution, especially in the case of WD1350-162. We discuss this important degeneracy in Section \ref{sec:highpressurecorerich}. For further details on individual systems, see Appendix Section \ref{sec:individual_systems}.


These systems are shown with red squares in Figure \ref{fig:bowtie_multipanel}, which also shows the movement in abundance space that can be caused by heating and sinking effects. Given the direction of this movement, it is possible for core-rich objects to appear close to primitive in this plot.

\subsection{Mantle-rich, high differentiation pressure}
\label{sec:HPM}

\begin{figure}
    \centering
    \includegraphics[width=8cm]{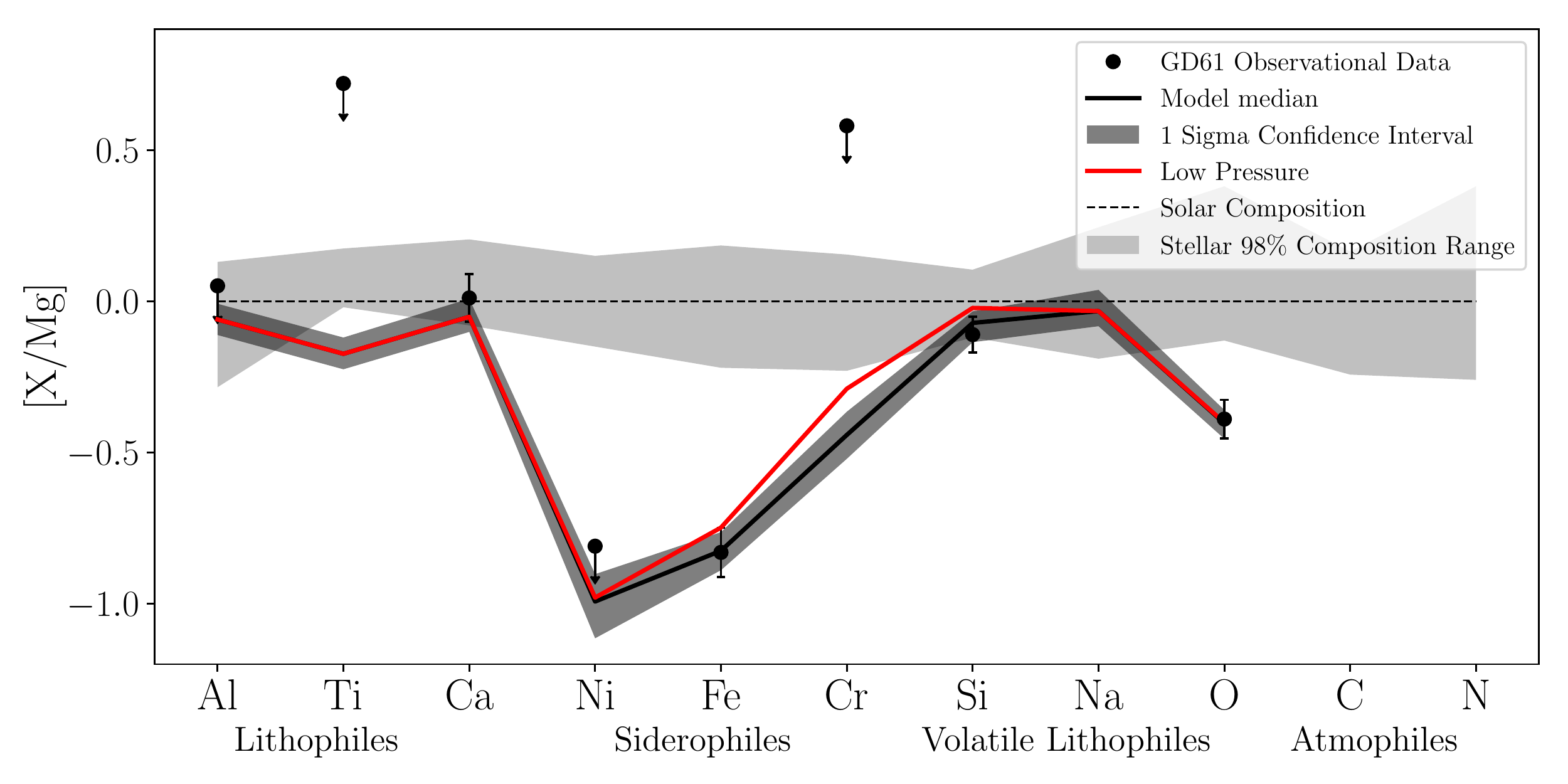}
    \caption{Median fit (with 1 sigma errors) to the observations of GD61 pollution for the model with highest Bayesian evidence. The fit was generated by sampling 10,000 sets of posterior values for each parameter and calculating the median of the resulting abundances. Arrows indicate upper bounds. The median pressure is $40.5_{-18.5}^{+11.7}$, while the red line is the median fit when setting pressure to 0\;GPa. An upper bound on Na which lies above the top of the y-axis was included during modelling, but is not shown here to aid visual clarity.}
    \label{fig:GD61_comp}
\end{figure}

\begin{figure}
    \centering
    \includegraphics[width=8cm]{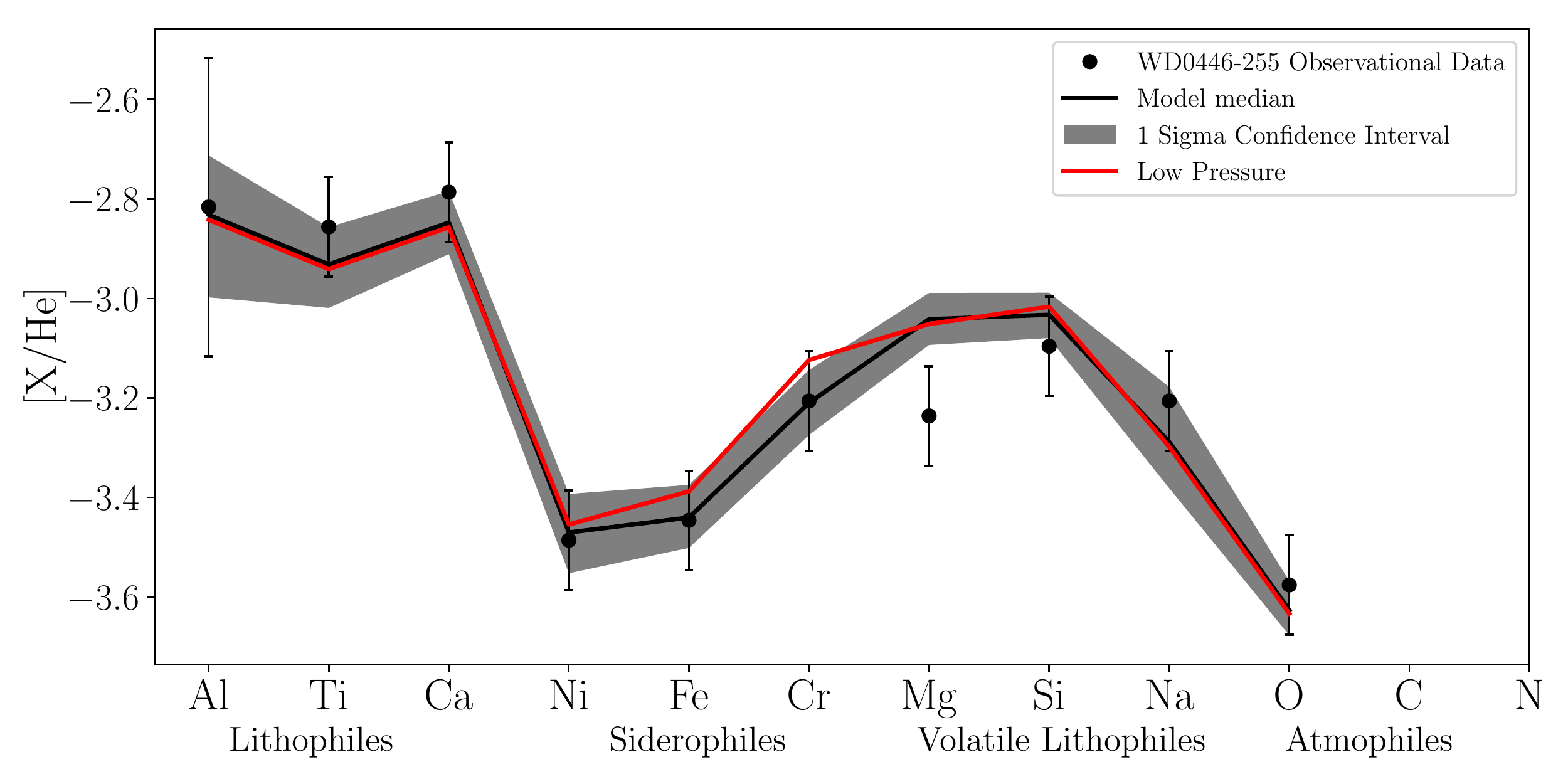}
    \caption{Median fit (with 1 sigma errors) to the observations of WD0446-255 pollution for the model with highest Bayesian evidence. The fit was generated by sampling 10,000 sets of posterior values for each parameter and calculating the median of the resulting abundances. The red line is the median fit when setting pressure to 0\;GPa.}
    \label{fig:WD0446_comp}
\end{figure}

\begin{figure}
    \centering
    \includegraphics[width=8cm]{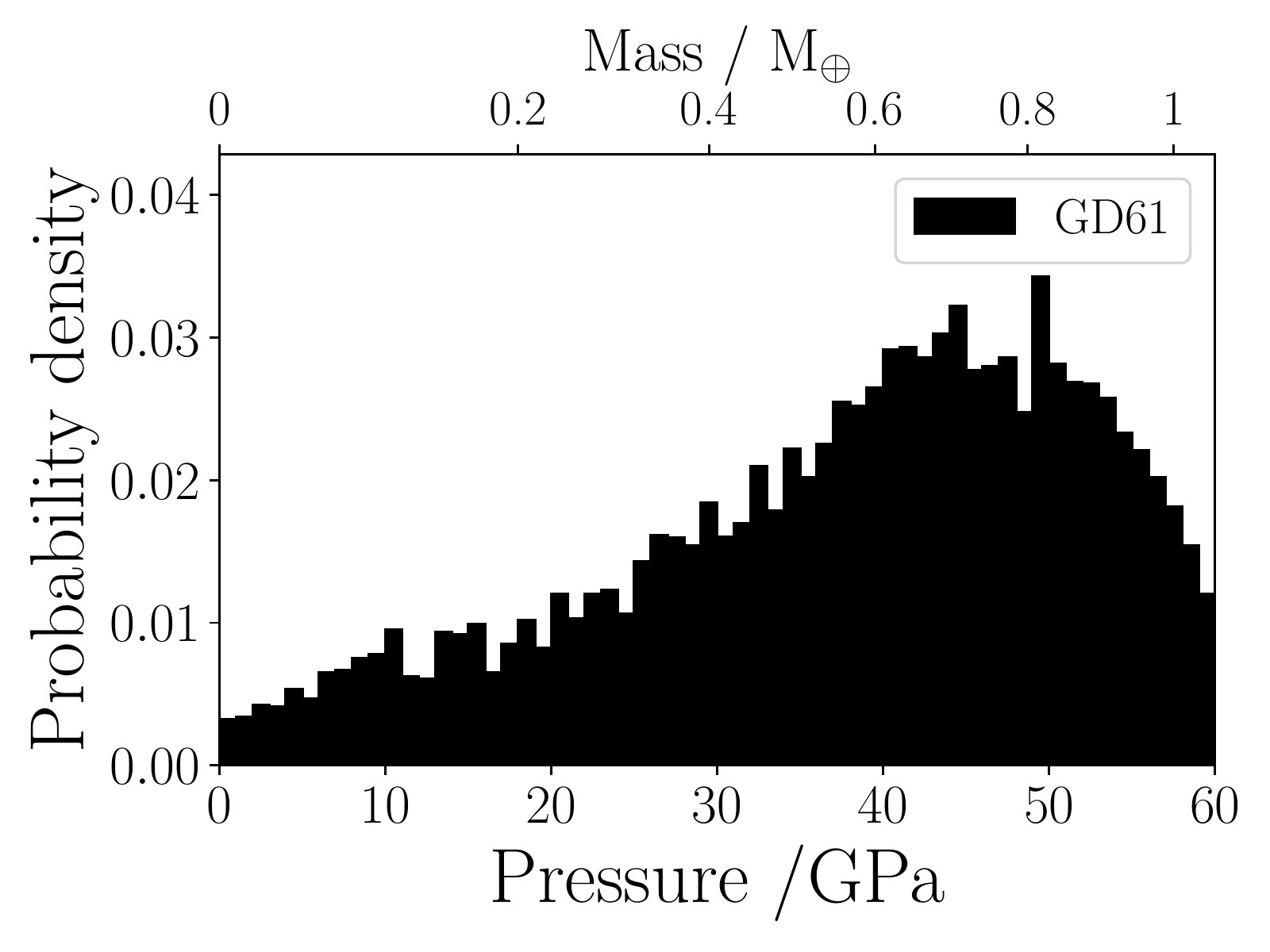}
    \caption{Posterior pressure distribution for GD61. The median pressure is $40^{+12}_{-18}\;\textrm{GPa}$, which corresponds to a planet of mass 0.61 $M_\oplus$.}
    \label{fig:GD61_pressuredist}
\end{figure}

We find 2 systems (GD61 and WD0446-255) for which the model with highest Bayesian evidence invoked both core--mantle differentiation and accretion of a mantle-rich fragment, as well as favouring high pressure (i.e., a large parent body) over low pressure. We retrieve median pressures of $40^{+12}_{-18}\;\textrm{GPa}$ and $37^{+15}_{-22}\;\textrm{GPa}$ for GD61 and WD0446-255 respectively, corresponding to masses of $0.61\;M_\oplus$ and $0.59\;M_\oplus$. This is illustrated for GD61 by Figure \ref{fig:GD61_pressuredist}. We do not rule out a low pressure solution in either case because we find a non-negligible probability that the pressure is below 10\;GPa (5\% and 10\% for GD61 and WD0446-255 respectively). For GD61, the high pressure preference is driven by Fe and Si, while for WD0446-255 the most important elements are Fe and Cr. The median (high pressure) model fits for GD61 and WD0446-255 are shown in Figures \ref{fig:GD61_comp} and \ref{fig:WD0446_comp} respectively, which also illustrate how those fits are improved at high pressure. Despite the pollutants being mantle-rich, the Ni abundance does not vary significantly with pressure - we discuss this effect in Appendix Section \ref{sec:p_sensitivity}.

We also find that the data for NLTT43806 can be fitted with mantle-rich material. However, it can be alternatively explained with crust-rich material \citep{Zuckerman2011,Harrison2018}, a possibility which we don't consider.


These systems are shown in Figure \ref{fig:bowtie_multipanel} with red circles. They have super-solar Mg/Fe and a Ni abundance which is generally not low enough to clearly favour low pressure.

\subsection{Differentiation pressure degenerate with oxygen fugacity}
\label{sec:Pdegen}

We find 4 systems which favour core--mantle differentiation, but for which we are unable to constrain the conditions under which it occurred due to a degeneracy between pressure (i.e., parent body size) and oxygen fugacity: PG0843+516, HE0106-3253, PG1015+161 and SDSSJ0512-0505. These systems are shown in Figure \ref{fig:bowtie_multipanel} with red crosses. In these cases, constraints on pressure could be determined given independent constraints on oxygen fugacity. \citet{Doyle2019} and \citet{Doyle2020}, for example, provide estimates of $\textrm{fO}_2$ based on the inferred FeO and Fe content of white dwarf pollutants. These calculations are in principle equivalent to our treatment of $\textrm{fO}_2$. However, we note that in a more realistic treatment which allows $\textrm{fO}_2$ to vary over the course of accretion, this equivalence may not hold for high valence elements, such as W, that exhibit a significant change in metal-silicate preference with $\textrm{fO}_2$. For further discussion of individual systems, we refer the reader to Appendix Section \ref{sec:individual_systems}.

\begin{figure}
    \centering
    \includegraphics[width=8cm]{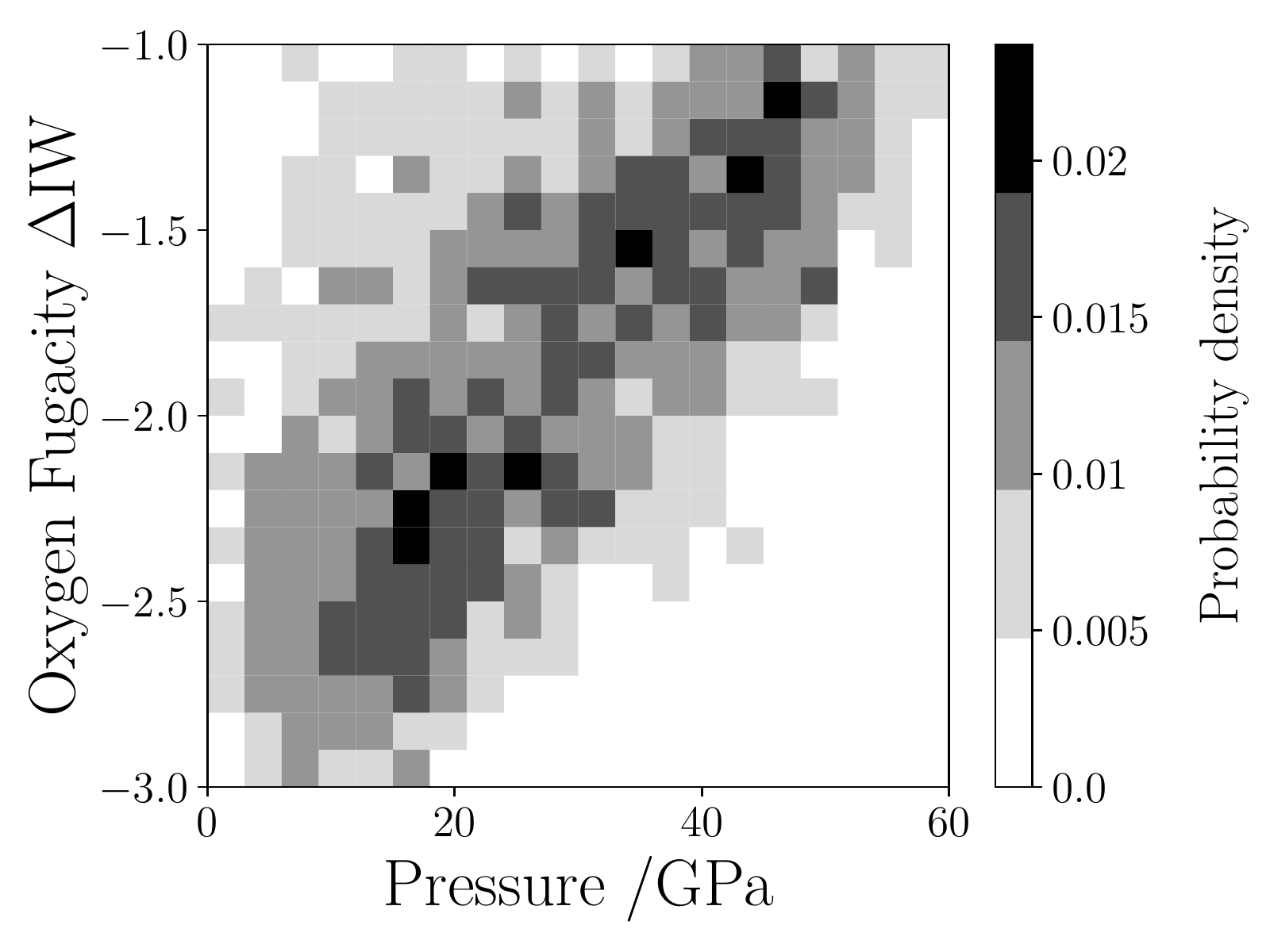}
    \caption{Posterior probability distribution in pressure/oxygen fugacity space for PG0843+516, showing a degeneracy between pressure and oxygen fugacity.}
    \label{fig:PG0843+516_p_v_fO2}
\end{figure}

Degeneracy between pressure and oxygen fugacity is inherent to our partitioning model, since the effect of both variables is to alter the elemental partition coefficients. If only one pressure-sensitive element is observed, degeneracy is therefore inevitable (unless the abundance of that element is so extreme that it requires both pressure and oxygen fugacity to adopt values near the edge of their priors). The systems mentioned above are degenerate because only one pressure-sensitive element, either Si or Cr, is observed. We illustrate an example of the resulting degeneracy in Figure \ref{fig:PG0843+516_p_v_fO2}. The degeneracy can be broken if multiple elements are observed, because the lines in pressure--oxygen fugacity space which can fit each element will (in general) be different. For mantle--rich pollutants, the ideal elements to observe would be Ni and Si, while for core--rich fragments Cr and Si are optimal.

\subsection{Differentiation pressure unconstrained}
\label{sec:Puncon}

We find 5 systems for which, although core-mantle differentiation is favoured, there is insufficient information available to constrain the conditions under which it occurred: SDSSJ0823+0546, SDSSJ0738+1835, SDSSJ0845+2257, WD0122-227 and WD1145+288. These are shown with red plus signs in Figure \ref{fig:bowtie_multipanel}. This occurs when the fragment is only slightly core- or mantle- rich (SDSSJ0845+2257), or when the only pressure sensitive element in the data does not place strong constraints on pressure (SDSSJ0823+0546, WD0122-227 and WD1145+288), or when degeneracy leads to a bimodal pressure distribution (SDSSJ0738+1835). The resulting posterior on pressure is approximately uniform. As an example, the pressure posterior for SDSSJ0845+2257 is shown in Figure \ref{fig:SDSSJ0845_pressuredist}. Further comments on individual systems can be found in the Appendix.

\begin{figure}
    \centering
    \includegraphics[width=8cm]{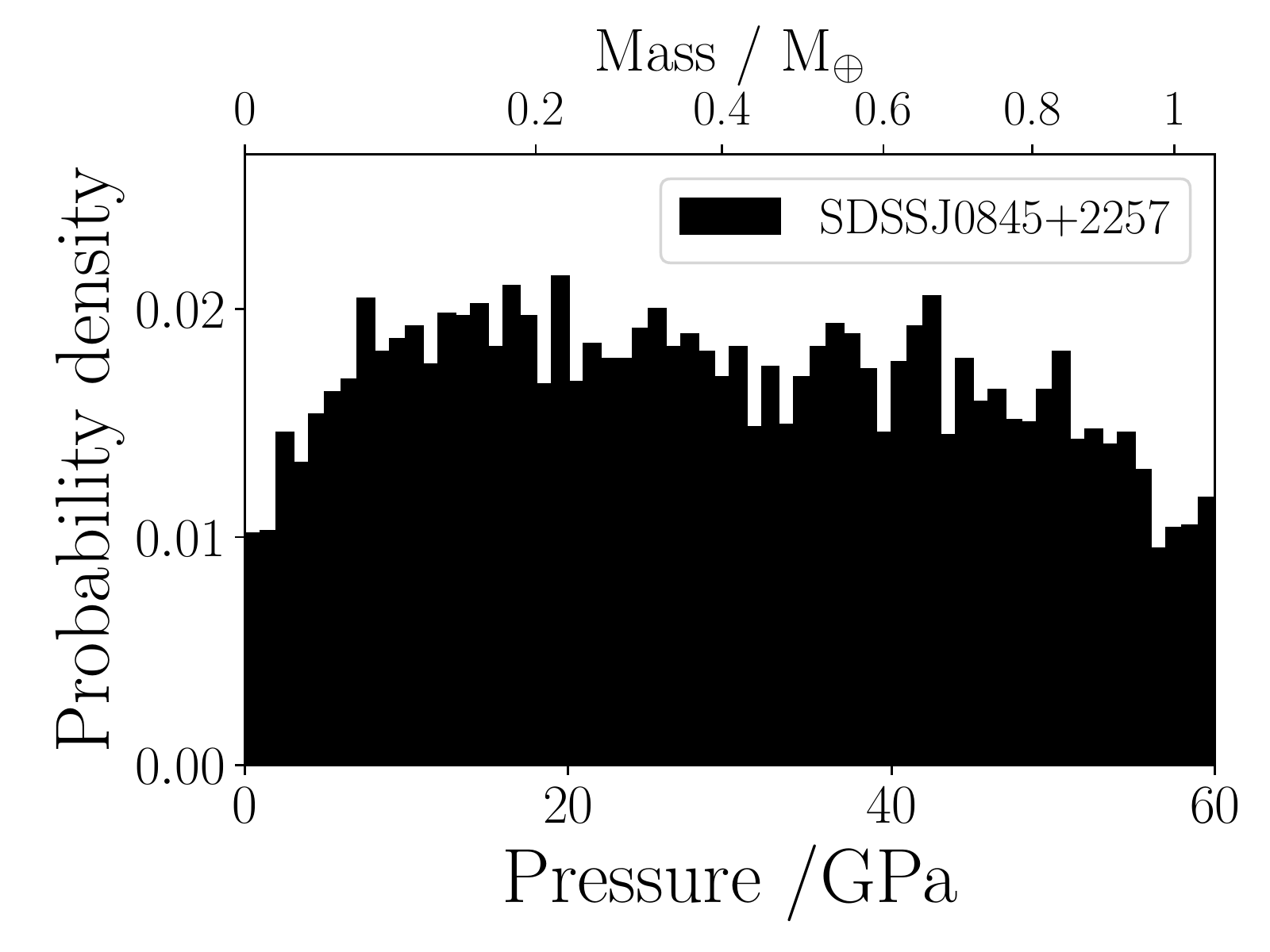}
    \caption{Posterior pressure distribution for SDSSJ0845+2257, showing the model's inability to constrain pressure (i.e., parent body size) for this system}
    \label{fig:SDSSJ0845_pressuredist}
\end{figure}

\subsection{No evidence of differentiation}
\label{sec:NED}

We find that 23 systems show no evidence of accretion from a differentiated body. These are shown with black crosses in Figure \ref{fig:bowtie_multipanel}. Note that these systems tend to be clustered closely around the pinch point of the contours in each panel. Systems in this region of abundance space have an abundance roughly matching bulk Earth, and hence are more likely to be explained by accretion of undifferentiated material.

It is possible that these systems have in fact accreted material from a differentiated body, but if the fragment has not been processed in a way that alters the core:mantle ratio away from that of its parent there is no observational signature of this process. This could occur if the parent body has been accreted directly on to the white dwarf without significant collisional evolution, or if multiple bodies were accreted (see Section \ref{sec:ModelCaveats2}).

Some of the white dwarf pollutants in this category show evidence of post-nebular processes which we don't investigate in this paper as they are not directly relevant to core--mantle differentiation. For example, \citet{Harrison2021b} proposed that post-nebula volatilisation occurred in GD362 to explain its high Mn abundance. Similarly, the discovery of Be in GD378 and GALEXJ2339-0424 \citep{Klein2021} has been interpreted as evidence of spallation products in icy exomoons \citep{Doyle2021}.

We note that our chi--squared per data point for GALEX1931+0117 and WD1232+563 is 1.8 and 1.5 respectively, indicating a noticeable degree of mismatch to the data. This is due to a low Ca/Mg ratio, which cannot be reproduced while also matching the other data points. For GALEX1931+0117, we used the data from \citet{Gaensicke2012} because our solutions when using the other data sets \citep{Vennes2011b,Melis2011} were either unphysical or yielded a poor fit.

\subsection{Unphysical solution}
\label{sec:PF}

We find that the model's fit to LHS2534 is unphysical and does not account for the possibility that the pollution is caused by crustal material \citep{Hollands2021}. This system is shown with a black circle in the top panel of Figure \ref{fig:bowtie_multipanel}, and we discuss it further in Appendix Section \ref{sec:individual_systems}.

\section{Discussion}
\label{sec:Discussion}

Our results highlighted that the abundances observed in five white dwarfs show evidence of the conditions under which core-mantle differentiation occurred. For two systems (GD61 and WD0446-255), the most likely model is the accretion of mantle-rich fragments of large (terrestrial planet) parent bodies. For three systems (WD0449-259, WD1350-162 and WD2105-820), the most likely model is the accretion of core-rich fragment of a smaller (less than Mars/Moon) parent bodies. Here we discuss the strength of our conclusions, noting that in all cases, the evidence in favour of a particular model is not conclusive.

Tighter constraints are possible given serendipitous observation of (almost) pure core or mantle material, and/or reduction in observational error. We discuss this further in Section \ref{sec:future_observations}.

\subsection{Robustness of constraints placed on pollutant size (pressure)}
\label{sec:robustness}

Observational errors on abundances of, for example, Cr are often comparable (typically 0.1-0.4 dex) to changes in the Cr abundance between the core of an asteroid, the Moon and Earth (which we estimate to be $\approx$ 0.5 dex). This can render definitive conclusions difficult to make.

Crucially, for GD61 and WD0446-255, whilst the most likely explanation (highest Bayesian evidence) for the observed abundances is the accretion of mantle-rich fragments of large (terrestrial planet mass) parent bodies, we cannot rule out accretion of undifferentiated material. The composition of this material would be similar to the Sun or other stars, modified by incomplete condensation and (for WD0446-255) a feeding zone. The requirement to invoke differentiation is less than $3\sigma$ in each case ($1.3\sigma$ and $2.6\sigma$ respectively, with corresponding Bayes factors of 1.1 and 9.2 over the best undifferentiated model).

For WD0449-259, WD1350-162 and WD2105-820, the white dwarfs classified as having accreted core-rich fragments of small parent bodies, the need to invoke the accretion of differentiated material is strong (> $3\sigma$). The posterior distribution peaks towards smaller planetary bodies (lower pressure) in each case, with approximately 50 \% of the posterior distribution lying below $0.2\;M_{\oplus}$ (20\;GPa) in each case (e.g., Figure \ref{fig:WD0449_pressuredist}). We note, however, that the model cannot resolve down to asteroidal sizes (barring extreme cases, as in Section \ref{sec:lowpressurecorerich}) because the abundances of Cr, Ni and Si typically change by only ~0.1 dex between the pressures corresponding to an asteroid and a Mars-sized planet (roughly 0 - 10\;GPa), as shown in Figure \ref{fig:bowtie_multipanel}. 

The most informative conclusions we can make pertain to WD0449-259. Although this object relies on a Ni detection and the Ni/Fe abundances is less sensitive to changes in pressure for core-rich objects (see Section \ref{sec:d_behaviour} and Figure \ref{fig:bowtie_multipanel}), Figure \ref{fig:WD0449_comp} shows that there is a detectable difference in Ni between low and high pressure. Our results for WD2105-820 are reliant on an Si upper bound, for which nucleosynthetic anomalies could provide an alternative explanation (see Section \ref{sec:silicon}). For WD1350-162, the conclusions are potentially affected by a degeneracy as described in Section \ref{sec:retrieval}.



\subsection{Implications of white dwarf pollutant masses}

White dwarfs that have accreted planetary material provide a unique means to probe the composition, and potentially the differentiation, of exoplanetary bodies. In order to interpret any inferred planetary compositions, it is important to ascertain whether the observations arise from the sum of many small asteroids or a fragment of a large terrestrial planet. The presence of core/mantle fragments has implications for subsequent crustal composition and thickness \citep{Dyck2021}, the redox conditions of core-mantle segregation, and, importantly, the heat source required for such large-scale melting (Bonsor et al. in prep). The inference that the abundances of some polluted white dwarfs are best explained by the loss of volatiles during the nebula phase, whilst some point to post-nebula volatisation \cite{Harrison2021b} has different implications for planet formation, depending on whether the accreted body is a minor planet or asteroid. 

Whilst the models presented here suggest that white dwarfs show evidence for the accretion of a range of planetary body sizes - from small undifferentiated asteroids to larger planetary fragments that have undergone core formation -  there is insufficient information to discard, nor to prove, the model of \citet{Jura2003} whereby all white dwarfs exhibit pollution by small asteroids. Whilst this may be possible with future observations (see Section \ref{sec:future_observations}), there is currently no single object where the alternative scenario can be ruled out (see discussion in Section \ref{sec:robustness}). It may be the case that multiple mechanisms exist for white dwarf pollution and the conclusion that one object has accreted a fragment of a terrestrial planet does not necessarily mean that this is the case for all white dwarfs. 

Terrestrial planet analogs could plausibly supply all white dwarf pollutants. Accretion rates can be in the region of $10^{10}\;\textrm{g}\;\textrm{s}^{-1}$ for the most heavily polluted WDs \citep{Farihi2009}. If this rate were to be maintained for 5\;Gyr, the total accreted mass would be on the order of $10^{24}\;\textrm{kg}$ (0.25\;$\textrm{M}_{\oplus}$). This is likely to be an overestimate, since accretion rates are lower for cooler (i.e., older) white dwarfs \citep{Farihi2009}. The delivery mechanism would need to be efficient, supplying on the order of 10\% of the total terrestrial planet mass (assuming it is equal to that of the solar system). It would also need to be slow, with accretion events spread out across multiple Gyr. Dynamical delivery processes could in future be investigated by coupling constraints on pollutants' parent body mass to the mass accreted by DBZ white dwarfs over $\sim$Myr timescales (see Figure 6 of \citealt{Veras2016}). This is beyond the scope of our current work. Dynamical mechanisms for the liberation of exomoons have been suggested by \citet{Payne2016a,Payne2016b}, and high Be abundances in GD378 and GALEXJ2339-0424 \citep{Klein2021} led \citet{Doyle2021} to conclude that they have likely accreted an exomoon. Intermediate differentiation pressures would in principle support this theory, although the resolution of our model would make it difficult to distinguish this case (roughly 3.5\;GPa, \citet{Righter1996}) from the very low pressure, asteroidal case.

However, dynamical mechanisms that lead to the accretion of fragments of planets \citet{Veras2013} or moons \cite{Payne2016a, Payne2016b} occur infrequently and struggle to supply the ubiquitously observed pollution (25-50\% of white dwarfs \cite{Zuckerman2003,Koester2014}. Asteroids, comets and other small planetary bodies are much more common in exoplanetary systems (in terms of number). Models that invoke dynamical instabilities following stellar mass loss or due to the presence of a companion have been shown to scatter sufficient material onto star-grazing orbits to explain the observed pollution \cite{Bonsor2011, Debes2012}. Future serendipitous observations that can shed light on the size of the planetary bodies accreted by white dwarfs will play an important role in our understanding of the origin of white dwarf pollution.  

\subsection{Usefulness of Silicon}
\label{sec:silicon}

One of our key results is that Si is potentially extremely useful as an indicator of pressure, because its partitioning behaviour is sensitive to temperature which is in turn linked to pressure via the silicate liquidus (equation \ref{eq:liquidus}). This conclusion may be unexpected, given that partitioning experiments show, at most, a weak dependence on pressure (e.g., \citet{Siebert2013,Fischer2015}). Given that Si is detectable in white dwarf atmospheres, this may give it a prominent role in any future effort to constrain the masses of white dwarf pollutants. This is contingent on other elements also being present so that the fragment core fraction and oxygen fugacity can also be constrained.

However, we also note that there are significant caveats. The pressure dependence of Si is inherited almost entirely from the pressure-temperature relationship imposed onto it. Hence, one's choice of pressure-temperature profile can affect any conclusions drawn from Si. We also encountered problems associated with modelling Si and O together (see Section \ref{sec:ModelCaveats2}). Also, nucleosynthetic isotope anomalies suggest that Earth's Si abundance may not be as simplistically related to the initial Si abundance of the molecular cloud which formed the Solar System as we have assumed here \citep{Tanaka2021}. These confounding factors make it difficult to be confident in our interpretation of Si abundances.

\subsection{Model limitations}
\label{sec:ModelCaveats2}

We note in Section \ref{sec:ModelCaveats} some simplifying assumptions made in our partitioning model. However there are also caveats associated with the experiments on which it is based, as well as assumptions within our broader pollution model. 

\begin{itemize}
\item The empirical data on which our partitioning model is based were obtained by experiments operating within a limited range of pressures, temperatures and oxygen fugacities. Partitioning behaviour at high pressure is based on regression to comparatively low quantities of data and so our model becomes decreasingly valid at increasingly high pressure. \citet{Schaefer2017} demonstrated that different parametrizations can deviate significantly at pressures above 100\;GPa. We only explore pressures up to 60\;GPa, limiting us to roughly Earth-sized objects. In principle, our method could be applied to super-Earths.
\item Sulphur is a chalcophile, so may be expected to affect the partitioning behaviour of Ni. However, including S in our model is non-trivial since stellar abundances of S are difficult to constrain. We implemented the parametrization given by \citet{Boujibar2014} and included it for comparisons to Mars (which is thought to have high S content in its core) but did not include it in our white dwarf modelling.
\item We assume Ti to be highly lithophilic, and set $D_{\textrm{Ti}} = 0$ as in \citet{Rudge2010}. 
\item The peridotite liquidus is assumed to not change with composition in our model. However, in reality the liquidus temperature should decrease with increasing mantle FeO content, as in \citet{Dyck2021}. The magnitude of the temperature change is on the order of 200\;K. We consider temperatures between 2000\;K and 4000\;K (see Figure \ref{fig:d_el_3d_plot}), so neglect this effect.
\item We found that under certain conditions, Si and O could enter a positive feedback loop due to their mutual interactions, and become extremely highly concentrated in the core. We therefore cap $D_{\textrm{Si}}$ at 1 and $D_{\textrm{O}}$ at 0.3 to prevent this unrealistic behaviour.
\item As noted in Section \ref{sec:ModelCaveats}, we ignore the possibility of partial differentiation. If some mantle and/or core material were to remain unequilibrated, as in \citet{Brennan2020} for example, key pressure sensitive elements would end up distributed more evenly between the core and mantle. This would reduce the model's sensitivity to pressure. The resulting fragment abundances would resemble fragments with less extreme core fractions (i.e., closer to that of their parent), which may help explain why some fragments in our sample appear to have significant components of both mantle and core material.
\item We assume that pollution is due to accretion of a single fragment from a single parent body. If multiple bodies are accreted, then in general the observed composition will be a weighted average of the bodies. Extreme pollution signatures of any description therefore become less likely, the implication being that we may be underestimating the number of fragments derived from differentiated parent bodies, as was pointed out by \citet{Turner2019}. Accretion of multiple bodies would greatly reduce our ability to infer pressures of core--mantle differentiation, unless one body dominates the mixture. As will be shown in Section \ref{sec:future_observations}, high material purity is crucial. Additionally, unless the core fractions of accreted fragments are all very similar (as well as the $P_{\textrm{diff}}$ of their parent bodies, if there are multiple), their pressure signatures will be contradictory to some extent.
\item Large parent bodies do not necessarily imply high accretion rates, because accreted fragments may be much smaller than their parent body. We therefore can't use accretion rate to independently estimate the size of parent bodies. However, we do use accretion rates to estimate limits on the total mass accreted (and hence compatibility with pollution sources) towards the beginning of our Discussion section.
\item When all parameters are included, our model has 9 variables. In many cases, we are therefore modelling systems with more variables than data points (typically, we model 6-7 elements). This may account for the large number of degeneracies we found. The Bayesian nature of the model minimises the number of parameters invoked, however. In particular, the parameters describing differentiation (fragment core fraction, core--mantle differentiation pressure and oxygen fugacity) are treated as a set, which means there is a large statistical penalty for its inclusion. However, a large number of data points is not necessary to infer differentiation (extreme Ca/Fe and Mg/Fe ratios would be sufficient), and the additional observation of 2 pressure sensitive elements may then be enough to break the degeneracy between pressure and oxygen fugacity and give a constraint on pressure.
\item As noted in \citet{Koester2014}, H-dominated white dwarfs with $\textrm{T}_{\textrm{eff}} \gtrsim $15,000-18,000\;K have negligible convection zones. The concept of a sinking time-scale (i.e., the time-scale over which material sinks out from a convection zone) is then poorly defined. In these cases, we instead use the diffusion time-scales at a Rosseland mean optical depth of unity. However, observations of elements in a white dwarf atmosphere probe different depths depending on the specific element and the wavelength it is observed at. The time-scales used may therefore differ from the actual diffusion time-scales.
\end{itemize}

\subsection{Potential for future observations}
\label{sec:future_observations}

In our present data set, there were no objects for which we were able to comprehensively rule out a high or low pressure solution, but such constraints may be possible with future data. In this section, we investigate the model's potential performance on synthetic data sets to identify the circumstances under which this might be possible.

To construct the synthetic data, we used our partitioning model to calculate the mantle and core composition of a body (of bulk Earth composition) with $P_{\textrm{diff}}$ of either 0\;GPa or 60\;GPa. The corresponding oxygen fugacity was set to be either IW\;-\;1 (for a synthetic core-rich fragment, in order to make Cr and Si as lithophilic as possible) or IW\;-\;3 (for a synthetic mantle-rich fragment, in order to make Ni as siderophilic as possible). The choice of oxygen fugacity is favourable to the model. The mantle-rich fragments were composed of purely mantle material, while the core-rich fragments were composed of 99\% core material (so that the purely lithophile elements would still be present). We also created a variant of each synthetic observation in which the material was less pure: either increasing the fragment core fraction to 2\% for the mantle-rich fragments, or decreasing the fragment core fraction to 75\% for core-rich fragments, for a total of 8 synthetic observations.

\begin{figure*}
    \centering
    \includegraphics[width=\textwidth,keepaspectratio=true]{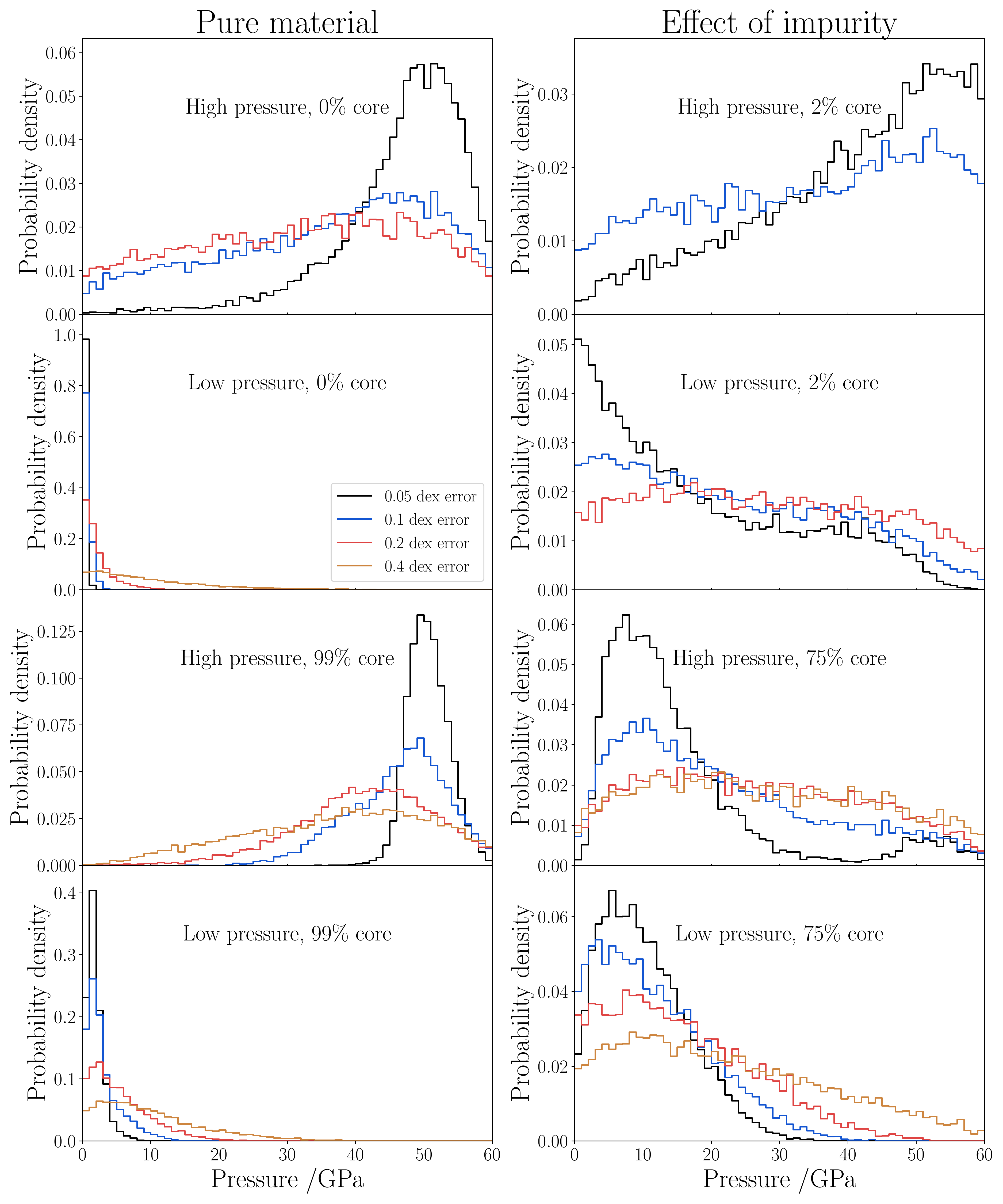}
    \caption{Posterior distributions on pressure for synthetic observations of white dwarf pollution. The synthetic abundances were defined by the pressure within the differentiating parent body and the core fraction of the fragment. The pressure and fragment core fraction are indicated in each panel. The top 4 panels correspond to mantle-rich material, and the bottom 4 panels correspond to core-rich material. The left hand column of panels is for pure (or nearly pure) material, while the right hand column shows the effect of impurity. Within each panel, the different distributions are for the same data set, but with different assumed errors on each of the data points. Errors of 0.05, 0.1, 0.2 and 0.4 dex were used. Missing data sets indicate that for the given parameters, core-mantle differentiation was not invoked by the model.}
    \label{fig:idealposteriors}
\end{figure*}

Our model can generally constrain pressure more tightly when observational errors are smaller, and the material is purer. This is shown in Figure \ref{fig:idealposteriors}. Within each panel, each of the distributions comes from the same synthetic observation - we changed only the size of the assumed error on each data point to produce the different distributions. In each case, a good fit to the data was obtained.

We find that the purity of the core- or mantle- material is typically the most important requirement. The presence of pure material can be inferred from extreme values of log(Mg/Fe) or log(Ca/Fe). These ratios do not necessarily have to be determined by direct detection: upper bounds on Ca, Mg and Fe can be used as well.

While we have focussed on Ni, Cr and Si as tracers of pressure, other elements which are sensitive to pressure could in principle be used, such as V, W and Mo. Of these, W and Mo have yet to be detected in white dwarf atmospheres and are therefore candidates for detection efforts.

\subsubsection{High pressure, mantle-rich material}

Assuming typical observational errors of 0.2 dex, no strong constraints on pressure can be made. A preference for high pressure is suggested in the case of pure mantle (see the red line in the top left panel of Figure \ref{fig:idealposteriors}, which peaks at 46\;GPa). With slight contamination by mantle material (top right panel), the model does not invoke differentiation. Low pressure can be ruled out only if the pollution is pure mantle and errors are 0.05 dex (black line in top left panel). In this case, the probability that pressure is below 10\;GPa is 0.7\%.



\subsubsection{Low pressure, mantle-rich material}

This offers the best prospect for tight constraints on pressure, and hence parent body mass, given serendipitous observation of pure mantle material. The signature of this material is a very significant Ni depletion which the model can only approach by decreasing pressure (and oxygen fugacity).

Assuming standard errors of 0.2 dex, high pressure can be ruled out in the pure mantle case. This gives a 99\% probability that pressure is below 10\;GPa. With 2\% core material, the model can't constrain pressure to low values (see the red lines in the second row of panels in Figure \ref{fig:idealposteriors}). A low pressure preference is still evident when errors are reduced to 0.05 dex, however.


Our ability to infer low pressure mantle material is limited by the low log(Ni/Hx) which must be observed. In the pure mantle case, we predict a log(Ni/Hx) value of -9.2 for a heavily polluted white dwarf (see Figure \ref{fig:observability}). In our sample, only WD1425+540 has log(Ni/Hx) below -9.2. We do not infer low pressure, mantle-rich material for any system. To observe log(Ni/Hx) values below about -9, the white dwarf's effective temperature needs to be below roughly 10,000\;K (based on data extracted from the Montreal White Dwarf Database \citep{MWDD}).

\subsubsection{High pressure, core-rich material}
\label{sec:highpressurecorerich}

A critically important degeneracy occurs when the pollution has a significant mantle impurity of 25\% (corresponding to $\textrm{log}(\textrm{Mg}/\textrm{Fe}) \approx -1$ in our synthetic data). High pressure fragments can masquerade as low pressure fragments - we describe this behaviour in more detail in Section \ref{sec:retrieval}. This can be seen in the right hand panel third from the top in Figure \ref{fig:observability}, in which the model generally does not recover the high-pressure signature but instead favours low pressure. This may explain why we did not infer accretion of high pressure core-rich material in any system. We advise extreme caution when assessing core rich fragments which are not completely (or almost completely) pure. In particular, a full analysis should consider any apparent high pressure solutions, even if those solutions appear to be highly disfavoured. We find that WD1350-162 admits a disfavoured solution at high pressure and oxygen fugacity, which may be the true solution.


By contrast, when assuming (almost) pure core material (99\%), and standard errors of 0.2 dex, we can rule out an asteroidal solution. We find a 99\% probability that the pressure is above 15\;GPa in this case (shown by the red line in the left hand panel third from the top in Figure \ref{fig:observability}). The most significant factor limiting our model's ability to constrain pressure is therefore the material's purity. However, for a 99\% core-rich fragment, the log(Mg/Fe) ratio is roughly -2.6 (and log(Ca/Fe) is roughly -3.7), both of which are far lower than any pollutant in our sample. Serendipitous observation of very low Mg/Fe and/or Ca/Fe is required (and would be a reliable indicator of very core-rich material, since white dwarf atmospheric effects increase these ratios).

\subsubsection{Low pressure, core-rich material}
\label{sec:lowpressurecorerich}

\begin{figure}
    \centering
    \includegraphics[width=8cm]{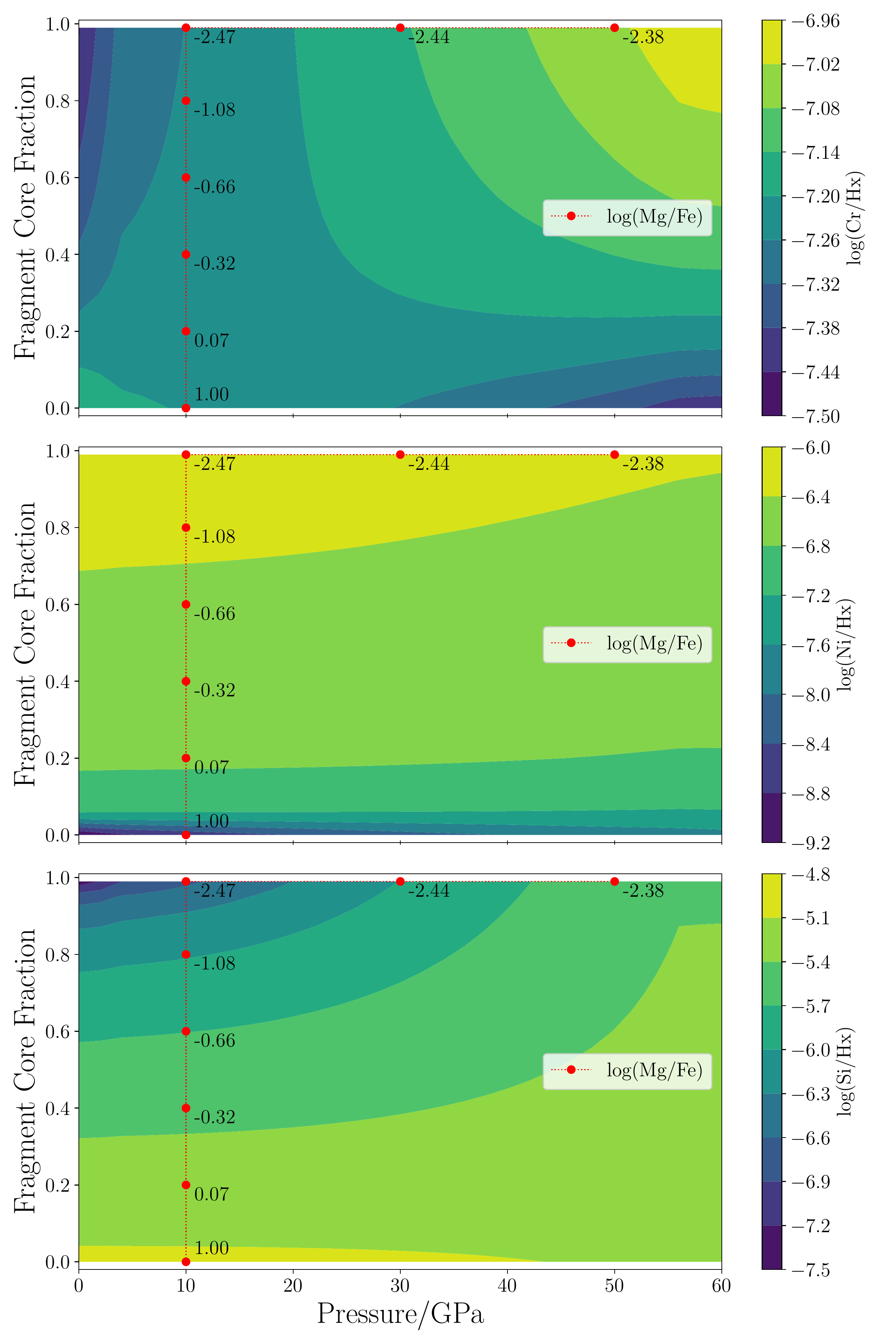}
    \caption{Top panel: Expected abundance of Cr in a white dwarf atmosphere as a function of the core fraction of a pollutant fragment and the core--mantle differentiation pressure in its parent. Logarithms are in base 10. Hx represents the dominant element in the white dwarf atmosphere, either H or He. Mg/Fe ratios are shown at selected locations in pressure/fragment core fraction space - this quantity (as well as Ca/Fe, not shown here) acts as a proxy for the fragment core fraction. When this quantity is known, Cr, Ni and/or Si can be used to infer pressure. Middle panel: Similar to top panel, but for Ni instead of Cr. Bottom panel: Similar to top panel, but for Si instead of Cr. Note that these plots assume a very high level of pollution, with the the total pollution of Al, Ti, Ca, Ni, Fe, Cr, Mg, Si, Na, O, C and N varying between -3 and -4 log units relative to Hx, which is in the range of the most highly polluted white dwarfs known. However, this is partially due to the additional assumption that we observe the white dwarf in the steady state of accretion. Observation in the declining phase, or shortly after accretion begins, would yield lower abundances of all elements relative to Hx. Observation in the declining phase would also increase the Mg/Fe ratio, due to Mg's longer sinking time-scale. An extract of these data, supplemented with Ca/Fe ratios, can be found in Table \ref{tab:observability}.}
    \label{fig:observability}
\end{figure}

Our model was able to recover a low pressure preference in all cases. This is shown in the bottom row of Figure \ref{fig:observability}, in which all distributions peak below 10\;GPa. Given standard errors of 0.2 dex and (almost) pure core material, we find a 90\% probability that pressure is below 12\;GPa, which strongly suggests a low pressure solution but does not rule out high pressure.


In the best case scenario (99\% core, 0.05 dex errors), the constraint is tight enough to rule out a parent object as large as Mercury, but would allow for objects larger than Eris.


This scenario is the most likely to be affected by undetectably low Cr and Si abundances. Figure \ref{fig:observability} shows that, for a highly polluted white dwarf, log(Cr/Hx) and log(Si/Hx) would both be approximately -7.5. Our sample contains multiple systems with log(Cr/Hx) below -7.5, but only one with log(Si/Hx) below -7.5 (WD1425+540). A few systems with $\textrm{log}(\textrm{Si}/\textrm{Hx}) < -7.5$ are recorded in the Montreal White Dwarf Database. All of these systems have He-dominated atmospheres, suggesting that such low Si abundances are not currently detectable in DA white dwarfs.


\section{Conclusions}
\label{sec:Conclusions}

White dwarfs that have accreted planetary material provide unique insights regarding the composition of exoplanetary bodies. In order to interpret these observations, it is crucial to know whether we are observing asteroids, moons or terrestrial planets. In this work we present a novel technique to distinguish between objects of different masses based on the Ni, Cr or Si content of planetary cores and mantles. The abundances of these elements in fragments of differentiated rocky bodies is a function of core--mantle differentiation pressure, which is in turn a function of the body's mass. We apply our model to observations of polluted white dwarfs. For those which have accreted core- or mantle-rich fragments of rocky objects, we investigated the inferred mass of the parent bodies from which the pollutant fragments were derived.

Our model uses Ni, Cr and (notably) Si to constrain the size of planetary bodies via their differentiation pressure. The pressure scale over which the abundances of these elements change appreciably (relative to typical observational error) is roughly 10\;GPa, meaning that our model can distinguish between Mars-sized ($\sim10\;\textrm{GPa}$) and Earth-sized ($\sim50\;\textrm{GPa}$) bodies, but is unable to resolve sub-Mars objects. In 2/42 systems analysed here (GD61 and WD0446-255), we find evidence that the parent bodies of the pollutants were Earth-sized and differentiated at correspondingly high pressure. However, we are unable to rule out accretion of undifferentiated material in either case. 3 systems (WD0449-259, WD1350-162 and WD2105-820) show evidence of differentiation at lower pressure, and hence smaller parent bodies. For WD0449-259, the most informative system, the key element suggesting low pressure is Ni. The results from WD2105-820 are reliant on a Si upper bound, while those from WD1350-162 may be misleading due to a degeneracy in our model. While we are unable to rule out large (Earth-sized) parent bodies, all 3 of these systems show a clear preference for having accreted a fragment of a comparatively low mass parent.

Our model is subject to inevitable degeneracies and bias, which can be partially mitigated by serendipitous observation of pure (or nearly pure) core- or mantle-like material. We find that large bodies ($\sim1M_\oplus$) can masquerade as small bodies ($\lesssim1M_{\textrm{Mars}}$) if the resulting pollutant is only moderately core--rich. There is also an inherent degeneracy between pressure and oxygen fugacity which must be broken in order to constrain pressure. This degeneracy is still present when the pollutant is highly pure, but it can be broken by inclusion of multiple elements, ideally Ni and Si (for mantle--rich material) or Cr and Si (for core--rich material).

Observation of high purity core- or mantle-like material also provides the best opportunity for tight constraints on pollutant mass. Given observations of pure mantle-like material (including a low Ni abundance or upper bound, with $\log(\textrm{Ni}/\textrm{Hx})\lesssim-9.2$) derived from a small parent body, our model could rule out high pressure (i.e., $P\gtrsim10\;\textrm{GPa}$). Similarly, we could rule out low pressure (i.e., $P\lesssim10\;\textrm{GPa}$) given observations of (almost) pure core-like material derived from a large parent body. This assumes standard observational errors of 0.2 dex; smaller errors would reduce the need for high purity. We emphasise that, while well-constrained abundance estimates of Ni, Cr and Si are ideal for constraining pressure, upper bounds on these abundances are also useful.

\section{Data Availability}
\label{sec:data}

The data and code used in this work are publicly available at \url{https://github.com/andrewmbuchan4/PyllutedWD_Public}. Modelling results are presented in the Supplementary Information.

\section*{Acknowledgements}

We would like to thank the anonymous referee for their useful comments, which have helped us to make several improvements to this manuscript. AMB thanks Laura Rogers, Matthew Auger, Marc Brouwers and Elliot Lynch for helpful discussions. AMB acknowledges a Travel Grant from Wolfson College, Cambridge. AMB is grateful for the support of a PhD studentship funded by a Royal Society Enhancement Award, RGF\textbackslash EA\textbackslash 180174. AB acknowledges support of a Royal Society Dorothy Hodgkin Fellowship, DH150130.




\bibliographystyle{mnras}
\bibliography{references.bib}

\appendix

\onecolumn
\begin{landscape}

\section{TABLES}

\begin{centering}
\renewcommand{\arraystretch}{1.5}
\begin{longtable}{|p{2.5cm}|p{8cm}|p{0.6cm}|p{2cm}|p{1cm}|p{1.8cm}|p{1.5cm}|}
\caption{\label{tab:sample} Summary of white dwarf sample} \\
\hline
         
         System & Source (WD properties) & Type & Mass /$\textrm{M}_{\odot}$ & $\textrm{T}_{\textrm{eff}}$ /K & log(g /$\textrm{m} \textrm{s}^{-2}$) & IR Excess\\

\hline
\endfirsthead
\hline

\hline
\endhead
\hline
\endfoot
G166-58 & \citet{Gianninas2011}, IR excess from \citet{Farihi2008} & H & 0.58 & 7390 & 7.99 & Y \\
G241-6 & \citet{Jura2012} (T, logg), \citet{Bergeron2011} (mass), \citet{Zuckerman2010} (IR excess) & He & 0.71 & 15300 & 8 & N \\
G29-38 & \citet{Xu2014}, IR excess from \citet{Zuckerman1987} & H & 0.85 & 11800 & 8.4 & Y \\
GALEX1931+0117 & \citet{Koester2014}, \citet{Vennes2010a} (IR excess) & H & 0.573 & 21457 & 7.9 & Y \\
GALEXJ2339 & \citet{Klein2021} & He & 0.548 & 13735 & 7.93 & N \\
GD362 & \citet{Leggett2018}, \citet{Becklin2005} and \citet{Kilic2005} (IR excess) & He & 0.551 & 10057 & 7.95 & Y \\
GD378 & \citet{Klein2021} & He & 0.551 & 15620 & 7.93 & N \\
GD40 & \citet{Coutu2019}, \citet{Jura2007} (IR excess) & He & 0.6 & 13594 & 8.02 & Y \\
GD424 & \citet{Izquierdo2020} & He & 0.77 & 16560 & 8.25 & N \\
GD56 & \citet{Gianninas2011}, IR excess from \citet{Jura2007} & H & 0.67 & 15270 & 8.09 & Y \\
GD61 & \citet{Farihi2011} & He & 0.71 & 17280 & 8.2 & Y \\
HE0106-3253 & \citet{Xu2019} (T, logg), \citet{Farihi2010} (Mass, IR excess) & H & 0.62 & 17350 & 8.12 & Y \\
HS2253+8023 & \citet{Klein2011}, IR excess absence from \citet{Farihi2009} & He & 0.84 & 14400 & 8.4 & N \\
LHS2534 & \citet{Hollands2021} & He & 0.55 & 4780 & 7.97 & N \\
NLTT43806 & \citet{Kilic2020}, IR excess from \citet{Farihi2009} & H & 0.704 & 5838 & 8.186 & N \\
PG0843+516 & \citet{Koester2014}, \citet{Xu/Jura2012} (excess) & H & 0.577 & 22412 & 7.902 & Y \\
PG1015+161 & \citet{Kilic2020}, \citet{Jura2007} (IR excess) & H & 0.642 & 19226 & 8.04 & Y \\
PG1225-079 & \citet{Klein2011}, IR excess from \citet{Farihi2010b} & He & 0.58 & 10800 & 8 & Y \\
SDSSJ0512-0505 & \citet{Harrison2021}, mass from MWDD \citep{MWDD}. Unable to find mention of IR excess, so assume absent & He & 0.803 & 5560 & 8.05 & N \\
SDSSJ0738+1835 & \citet{Dufour2012}, IR excess from \citet{Dufour2010} & He & 0.841 & 13950 & 8.4 & Y \\
SDSSJ0823+0546 & \citet{Harrison2021}. Unable to find mention of IR excess, so assume absent. Unable to find value for Mass, so assuming a typical value & He & 0.6 & 5920 & 7.945 & N \\
SDSSJ0845+2257 & \citet{Wilson2015} & He & 0.679 & 19780 & 8.18 & Y \\
SDSSJ1043+0855 & \citet{Tremblay2011}. IR excess from \citet{Farihi2010} & H & 0.65 & 18320 & 8.05 & Y \\
SDSSJ1228+1040 & \citet{Tremblay2011} (mass), IR excess from \citet{Brinkworth2009}, others from \citet{Gaensicke2012} & H & 0.73 & 20900 & 8.15 & Y \\
SDSSJ1242+5226 & \citet{Raddi2015} & He & 0.59 & 13000 & 8 & N \\
SDSSJ2047-1259 & \citet{Hoskin2020} & He & 0.617 & 17970 & 8.04 & N \\
WD0122-227 & \citet{Swan2019} & He & 0.61 & 8380 & 8.06 & N \\
WD0446-255 & \citet{Swan2019} & He & 0.58 & 10120 & 8 & N \\
WD0449-259 & \citet{Swan2019} & He & 0.61 & 9850 & 8.04 & N \\
WD1145+017 & \citet{Fortin-Archambault2020}, IR excess from \citet{Vanderburg2015} & He & 0.656 & 14500 & 8.11 & Y \\
WD1145+288 & \citet{Xu2019} (T, logg), \citet{Kleinman2012} (Mass), \citet{Barber2014} (IR excess) & H & 0.685 & 12140 & 8.14 & Y \\
WD1232+563 & \citet{Coutu2019}, IR excess from \citet{Xu2019} & He & 0.77 & 11787 & 8.3 & Y \\
WD1350-162 & \citet{Swan2019} & He & 0.6 & 11640 & 8.02 & N \\
WD1425+540 & \citet{Bergeron2011}, IR excess absence from \citet{Xu2017} & He & 0.56 & 14490 & 7.95 & N \\
WD1536+520 & \citet{Farihi2016}, IR excess - identified as candidate by \citet{Debes2011b}, confirmed by \citet{Barber2014} & He & 0.58 & 20800 & 7.96 & Y \\
WD1551+175 & \citet{Xu2019}, \citet{Bergeron2011} (Mass), \citet{Bergfors2014} (IR excess) & He & 0.57 & 14756 & 8.02 & Y \\
WD2105-820 & \citet{Swan2019} & H & 0.86 & 10890 & 8.41 & N \\
WD2115-560 & \citet{Swan2019} & H & 0.58 & 9600 & 7.97 & Y \\
WD2157-574 & \citet{Swan2019} & H & 0.63 & 7010 & 8.06 & N \\
WD2207+121 & \citet{Coutu2019}, IR excess from \citet{Xu/Jura2012} & He & 0.57 & 14752 & 7.97 & Y \\
WD2216-657 & \citet{Swan2019} & He & 0.61 & 9120 & 8.05 & N \\
WDJ1814-7354 & \citet{GonzalezEgea2020} & H & 0.59 & 10090 & 8 & Y \\
\end{longtable}
\small
\begin{longtable}{|p{2.6cm}|p{1.4cm}|p{1.6cm}|p{1.5cm}|p{1.4cm}|p{1.6cm}|p{1.6cm}|p{1.7cm}|p{1.7cm}|p{1.4cm}|p{1.7cm}|p{1.6cm}|p{1cm}|}

\caption{\label{tab:abundances} Elemental abundances used in our modelling. Hx refers to the dominant component in the white dwarf atmosphere, either H or He as appropriate. Superscripted indices in the System column indicate the source and any additional notes; see table notes for details.} \\
\hline

         
         System & log(Al/Hx) & log(Ti/Hx) & log(Ca/Hx) & log(Ni/Hx) & log(Fe/Hx) & log(Cr/Hx) & log(Mg/Hx) & log(Si/Hx) & log(Na/Hx) & log(O/Hx) & log(C/Hx) & log(N/Hx) \\

\hline
\endfirsthead
\hline

\hline
\endhead
\hline
\endfoot
G166-58$^{a}$ & - & - & -9.33 $\pm$ 0.08 & -9.5 $\pm$ 0.2 & -8.22 $\pm$ 0.13 & - & -8.06 $\pm$ 0.05 & <-8.2 & - & - & - & - \\
G241-6$^{b, 1}$ & <-7.7 & -8.97 $\pm$ 0.1 & -7.3 $\pm$ 0.2 & -8.15 $\pm$ 0.4 & -6.82 $\pm$ 0.14 & -8.46 $\pm$ 0.1 & -6.26 $\pm$ 0.1 & -6.62 $\pm$ 0.2 & - & -5.64 $\pm$ 0.11 & - & - \\
G29-38$^{c, 2}$ & <-6.1 & -7.9 $\pm$ 0.16 & -6.58 $\pm$ 0.12 & <-7.3 & -5.9 $\pm$ 0.1 & -7.51 $\pm$ 0.12 & -5.77 $\pm$ 0.13 & -5.6 $\pm$ 0.17 & <-6.7 & -5 $\pm$ 0.12 & -6.9 $\pm$ 0.12 & <-5.7 \\
GALEX1931+0117$^{d, 3}$ & -6.2 $\pm$ 0.2 & - & - & -6.7 $\pm$ 0.3 & -4.5 $\pm$ 0.3 & -6.1 $\pm$ 0.3 & - & -4.75 $\pm$ 0.2 & - & -4.1 $\pm$ 0.3 & - & - \\
GALEX1931+0117$^{e}$ & - & - & -6.11 $\pm$ 0.05 & - & -4.43 $\pm$ 0.09 & - & -4.42 $\pm$ 0.06 & -4.24 $\pm$ 0.07 & - & -3.62 $\pm$ 0.05 & - & - \\
GALEX1931+0117$^{f, 3}$ & <-5.85 & <-7 & -5.83 $\pm$ 0.1 & <-5.6 & -4.1 $\pm$ 0.1 & -5.92 $\pm$ 0.14 & -4.1 $\pm$ 0.1 & -4.35 $\pm$ 0.11 & - & -3.68 $\pm$ 0.1 & - & - \\
GALEXJ2339$^{g, 4}$ & <-7.7 & -9.58 $\pm$ 0.4 & -8.03 $\pm$ 0.75 & <-8 & -6.99 $\pm$ 0.3 & -8.73 $\pm$ 0.26 & -6.58 $\pm$ 0.14 & -6.59 $\pm$ 0.08 & <-8 & -5.52 $\pm$ 0.05 & - & - \\
GD362$^{h, 5}$ & -6.4 $\pm$ 0.2 & -7.95 $\pm$ 0.1 & -6.24 $\pm$ 0.1 & -7.07 $\pm$ 0.15 & -5.65 $\pm$ 0.1 & -7.41 $\pm$ 0.1 & -5.98 $\pm$ 0.25 & -5.84 $\pm$ 0.3 & -7.79 $\pm$ 0.2 & <-5.14 & - & <-4.14 \\
GD378$^{g, 6}$ & <-7.7 & -10.13 $\pm$ 0.46 & -8.7 $\pm$ 0.76 & <-8.3 & -7.51 $\pm$ 0.36 & -9.72 $\pm$ 0.68 & -7.44 $\pm$ 0.2 & -7.49 $\pm$ 0.12 & <-7.2 & -6.04 $\pm$ 0.31 & - & - \\
GD40$^{b, 7}$ & -7.35 $\pm$ 0.12 & -8.61 $\pm$ 0.2 & -6.9 $\pm$ 0.2 & -7.84 $\pm$ 0.26 & -6.47 $\pm$ 0.12 & -8.31 $\pm$ 0.16 & -6.2 $\pm$ 0.16 & -6.44 $\pm$ 0.3 & - & -5.62 $\pm$ 0.1 & - & - \\
GD424$^{i, 8}$ & -6.3 $\pm$ 0.1 & -7.78 $\pm$ 0.09 & -6.15 $\pm$ 0.05 & -6.93 $\pm$ 0.1 & -5.53 $\pm$ 0.12 & -7.19 $\pm$ 0.07 & -5.15 $\pm$ 0.04 & -5.29 $\pm$ 0.04 & <-6.5 & -4.59 $\pm$ 0.12 & - & - \\
GD56$^{a}$ & - & - & -6.86 $\pm$ 0.2 & - & -5.44 $\pm$ 0.2 & - & -5.55 $\pm$ 0.2 & -5.69 $\pm$ 0.2 & - & - & - & - \\
GD61$^{j, 9}$ & <-7.8 & <-8.6 & -7.9 $\pm$ 0.0634 & <-8.8 & -7.6 $\pm$ 0.0667 & <-8.0 & -6.69 $\pm$ 0.0467 & -6.82 $\pm$ 0.0367 & <-6.8 & -5.95 $\pm$ 0.0434 & - & - \\
HE0106-3253$^{a}$ & - & - & -5.93 $\pm$ 0.11 & - & -4.7 $\pm$ 0.06 & - & -5.57 $\pm$ 0.2 & -5.48 $\pm$ 0.05 & - & - & - & - \\
HS2253+8023$^{k, 10}$ & <-6.7 & -8.74 $\pm$ 0.04 & -7 $\pm$ 0.1 & -7.32 $\pm$ 0.2 & -6.17 $\pm$ 0.05 & -8 $\pm$ 0.06 & -6.12 $\pm$ 0.08 & -6.28 $\pm$ 0.06 & <-6.8 & -5.38 $\pm$ 0.12 & - & - \\
LHS2534$^{l, 11}$ & - & - & -10.08 $\pm$ 0.11 & - & -9.06 $\pm$ 0.08 & -10.28 $\pm$ 0.06 & -8.62 $\pm$ 0.06 & - & -9.53 $\pm$ 0.06 & - & - & - \\
NLTT43806$^{m}$ & -7.6 $\pm$ 0.17 & -9.55 $\pm$ 0.14 & -7.9 $\pm$ 0.19 & -9.1 $\pm$ 0.17 & -7.8 $\pm$ 0.17 & -9.55 $\pm$ 0.22 & -7.1 $\pm$ 0.13 & -7.2 $\pm$ 0.14 & -8.1 $\pm$ 0.14 & - & - & - \\
PG0843+516$^{d, 12}$ & -6.5 $\pm$ 0.2 & - & - & -6.3 $\pm$ 0.3 & -4.6 $\pm$ 0.2 & -5.8 $\pm$ 0.3 & -5 $\pm$ 0.2 & -5.2 $\pm$ 0.2 & - & -5 $\pm$ 0.3 & - & - \\
PG0843+516$^{a}$ & - & - & -6.26 $\pm$ 0.2 & - & -3.84 $\pm$ 0.18 & - & -4.82 $\pm$ 0.2 & -4.59 $\pm$ 0.12 & - & - & - & - \\
PG1015+161$^{d, 13}$ & - & - & -6.45 $\pm$ 0.2 & - & -5.5 $\pm$ 0.3 & <-5.8 & -5.3 $\pm$ 0.2 & -6.4 $\pm$ 0.2 & - & -5.5 $\pm$ 0.2 & - & - \\
PG1015+161$^{a}$ & - & - & -6.4 $\pm$ 0.2 & - & -4.92 $\pm$ 0.2 & - & -5.6 $\pm$ 0.2 & -5.42 $\pm$ 0.21 & - & - & - & - \\
PG1225-079$^{n, 14}$ & <-7.84 & -9.45 $\pm$ 0.02 & -8.06 $\pm$ 0.03 & -8.76 $\pm$ 0.14 & -7.42 $\pm$ 0.07 & -9.27 $\pm$ 0.06 & -7.5 $\pm$ 0.2 & -7.45 $\pm$ 0.1 & <-8.26 & <-5.54 & -7.8 $\pm$ 0.1 & - \\
SDSSJ0512-0505$^{o, 15}$ & - & - & -8.9 $\pm$ 0.1 & - & -7.75 $\pm$ 0.1 & -10 $\pm$ 0.2 & -7.65 $\pm$ 0.1 & - & -9.65 $\pm$ 0.1 & - & - & - \\
SDSSJ0738+1835$^{p, 16}$ & -6.39 $\pm$ 0.11 & -7.95 $\pm$ 0.11 & -6.23 $\pm$ 0.15 & -6.31 $\pm$ 0.1 & -4.98 $\pm$ 0.09 & -6.76 $\pm$ 0.12 & -4.68 $\pm$ 0.07 & -4.9 $\pm$ 0.16 & -6.36 $\pm$ 0.16 & -3.81 $\pm$ 0.19 & - & - \\
SDSSJ0823+0546$^{o, 17}$ & - & <-10 & -9.8 $\pm$ 0.1 & -8.6 $\pm$ 0.1 & -7.35 $\pm$ 0.1 & - & -7.85 $\pm$ 0.1 & - & - & - & - & - \\
SDSSJ0845+2257$^{q, 18}$ & -5.7 $\pm$ 0.15 & <-7.15 & -5.95 $\pm$ 0.1 & -5.65 $\pm$ 0.3 & -4.6 $\pm$ 0.2 & -6.4 $\pm$ 0.3 & -4.7 $\pm$ 0.15 & -4.8 $\pm$ 0.3 & - & -4.25 $\pm$ 0.2 & - & - \\
SDSSJ1043+0855$^{r, 19}$ & -7.06 $\pm$ 0.3 & <-7.00 & -5.96 $\pm$ 0.2 & -7.38 $\pm$ 0.3 & -6.15 $\pm$ 0.3 & <-6.5 & -5.11 $\pm$ 0.2 & -5.33 $\pm$ 0.5 & - & -4.9 $\pm$ 0.2 & - & - \\
SDSSJ1228+1040$^{d, 20}$ & -5.75 $\pm$ 0.2 & - & -5.94 $\pm$ 0.2 & <-6.5 & -5.2 $\pm$ 0.3 & <-6.00 & -5.2 $\pm$ 0.2 & -4.7 $\pm$ 0.2 & - & -4.55 $\pm$ 0.2 & - & - \\
SDSSJ1228+1040$^{d, 21}$ & -5.75 $\pm$ 0.2 & - & -5.94 $\pm$ 0.2 & <-6.5 & -5.2 $\pm$ 0.3 & <-6.00 & -5.2 $\pm$ 0.2 & -5.2 $\pm$ 0.2 & - & -4.55 $\pm$ 0.2 & - & - \\
SDSSJ1242+5226$^{s, 22}$ & <-6.5 & -8.2 $\pm$ 0.2 & -6.53 $\pm$ 0.1 & <-7.3 & -5.9 $\pm$ 0.15 & -7.5 $\pm$ 0.2 & -5.26 $\pm$ 0.15 & -5.3 $\pm$ 0.06 & -7.2 $\pm$ 0.2 & -4.3 $\pm$ 0.1 & - & <-5.00 \\
SDSSJ2047-1259$^{t, 23}$ & <-6.5 & - & -6.9 $\pm$ 0.1 & -7.4 $\pm$ 0.1 & -6.4 $\pm$ 0.2 & - & -5.6 $\pm$ 0.1 & -5.6 $\pm$ 0.1 & - & -4.8 $\pm$ 0.1 & - & - \\
WD0122-227$^{u}$ & - & - & -10.1 $\pm$ 0.1 & - & -8.5 $\pm$ 0.2 & - & -8.5 $\pm$ 0.4 & <-7.6 & - & <-5.2 & - & - \\
WD0446-255$^{u, 24}$ & -7.3 $\pm$ 0.3 & -8.8 $\pm$ 0.1 & -7.4 $\pm$ 0.1 & -8.2 $\pm$ 0.1 & -6.9 $\pm$ 0.1 & -8.5 $\pm$ 0.1 & -6.6 $\pm$ 0.1 & -6.5 $\pm$ 0.1 & -7.9 $\pm$ 0.1 & -5.8 $\pm$ 0.1 & - & - \\
WD0449-259$^{u, 25}$ & - & -10.7 $\pm$ 0.2 & -9.1 $\pm$ 0.1 & -8.4 $\pm$ 0.2 & -7.9 $\pm$ 0.2 & - & -8.3 $\pm$ 0.4 & <-7.3 & - & <-6.6 & - & - \\
WD1145+017$^{v, 3}$ & -6.89 $\pm$ 0.2 & -8.57 $\pm$ 0.2 & -7 $\pm$ 0.2 & -7.02 $\pm$ 0.3 & -5.61 $\pm$ 0.2 & -7.92 $\pm$ 0.4 & -5.91 $\pm$ 0.2 & -5.89 $\pm$ 0.2 & - & -5.12 $\pm$ 0.35 & - & <-7.00 \\
WD1145+288$^{a}$ & - & - & -6.88 $\pm$ 0.08 & - & -5.43 $\pm$ 0.2 & - & -6 $\pm$ 0.2 & <-4.7 & - & - & - & - \\
WD1232+563$^{a, 26}$ & <-7.50 & -8.96 $\pm$ 0.11 & -7.69 $\pm$ 0.05 & <-7.3 & -6.45 $\pm$ 0.11 & -8.16 $\pm$ 0.07 & -6.09 $\pm$ 0.05 & -6.36 $\pm$ 0.13 & - & -5.14 $\pm$ 0.15 & - & - \\
WD1350-162$^{u, 25}$ & - & -10 $\pm$ 0.1 & -8.7 $\pm$ 0.1 & - & -7.1 $\pm$ 0.1 & -9 $\pm$ 0.2 & -6.8 $\pm$ 0.1 & -7.3 $\pm$ 0.2 & - & -6.2 $\pm$ 0.1 & - & - \\
WD1425+540$^{w, 27}$ & - & - & -9.26 $\pm$ 0.1 & -9.67 $\pm$ 0.2 & -8.15 $\pm$ 0.14 & - & -8.16 $\pm$ 0.2 & -8.03 $\pm$ 0.31 & - & -6.62 $\pm$ 0.23 & -7.29 $\pm$ 0.17 & - \\
WD1536+520$^{x, 28}$ & -5.38 $\pm$ 0.15 & -6.84 $\pm$ 0.15 & -5.28 $\pm$ 0.15 & - & -4.5 $\pm$ 0.15 & -5.93 $\pm$ 0.15 & -4.06 $\pm$ 0.15 & -4.32 $\pm$ 0.15 & - & -3.4 $\pm$ 0.15 & <-4.2 & - \\
WD1551+175$^{a, 26}$ & -6.99 $\pm$ 0.15 & -8.68 $\pm$ 0.11 & -6.93 $\pm$ 0.07 & <-7.5 & -6.6 $\pm$ 0.1 & -8.25 $\pm$ 0.07 & -6.29 $\pm$ 0.05 & -6.33 $\pm$ 0.1 & - & -5.48 $\pm$ 0.15 & - & - \\
WD2105-820$^{u}$ & - & - & -8.2 $\pm$ 0.1 & - & -6 $\pm$ 0.2 & - & -6 $\pm$ 0.2 & <-5.5 & - & - & - & - \\
WD2115-560$^{u}$ & -7.6 $\pm$ 0.1 & - & -7.4 $\pm$ 0.1 & - & -6.4 $\pm$ 0.1 & - & -6.4 $\pm$ 0.1 & -6.2 $\pm$ 0.1 & - & <-5.0 & <-4.3 & <-4.0 \\
WD2157-574$^{u}$ & -8.1 $\pm$ 0.1 & - & -8.1 $\pm$ 0.1 & -8.8 $\pm$ 0.1 & -7.3 $\pm$ 0.1 & - & -7 $\pm$ 0.1 & -7 $\pm$ 0.1 & - & <-3.8 & <-3.6 & <-3.0 \\
WD2207+121$^{a, 26}$ & -7.08 $\pm$ 0.15 & -8.84 $\pm$ 0.14 & -7.4 $\pm$ 0.08 & -7.55 $\pm$ 0.2 & -6.46 $\pm$ 0.13 & -8.16 $\pm$ 0.19 & -6.15 $\pm$ 0.1 & -6.17 $\pm$ 0.11 & - & -5.32 $\pm$ 0.15 & - & - \\
WD2216-657$^{u}$ & - & -10.6 $\pm$ 0.1 & -9 $\pm$ 0.1 & - & -8 $\pm$ 0.2 & - & -7.1 $\pm$ 0.1 & <-7 & <-8.5 & <-6.5 & - & - \\
WDJ1814-7354$^{y, 29}$ & <-7.3 & <-8 & -7.22 $\pm$ 0.15 & <-6.3 & -6.06 $\pm$ 0.19 & - & -6.14 $\pm$ 0.08 & <-6 & <-7.4 & - & - & - \\
\end{longtable}

\begin{itemize}
      \small
      \item Sources for the abundances quoted here are as follows: (a) \citet{Xu2019}, (b) \citet{Jura2012}, (c) \citet{Xu2014}, (d) \citet{Gaensicke2012}, (e) \citet{Vennes2011b}, as given in \citet{Gaensicke2012}, (f) \citet{Melis2011}, (g) \citet{Klein2021}, (h) \citet{Zuckerman2007}, (i) \citet{Izquierdo2020}, (j) \citet{Farihi2013}, (k) \citet{Klein2011}, (l) \citet{Hollands2021}, (m) \citet{Zuckerman2011}, (n) \citet{Klein2011} and \citet{Xu2013}, (o) \citet{Harrison2021}, (p) \citet{Dufour2012}, (q) \citet{Wilson2015}, (r) \citet{Melis2016}, (s) \citet{Raddi2015}, (t) \citet{Hoskin2020}, (u) \citet{Swan2019}, (v) \citet{Fortin-Archambault2020}, (w) \citet{Xu2017}, (x) \citet{Farihi2016}, (y) \citet{GonzalezEgea2020}
      \item Notes on specific systems: (1) Excluding C (subsolar), N (subsolar), P, S, Cl, Mn, Cu, Ga, Ge; (2) Excluding S, Mn; (3) Excluding C (subsolar); (4) Using larger error in case of asymmetry. Excluding Be, Mn, Li, V; (5) Excluding C (subsolar), Sc, V, Mn, Co, Cu, Sr; (6) Using larger error in case of asymmetry. Excluding Be, P, S, Mn, Li, V, C (subsolar), N (subsolar); (7) Using larger error in case of asymmetry. Excluding C (subsolar), N (subsolar), P, S, Cl, Mn, Cu, Ga, Ge; (8) Using larger error in case of asymmetry, using Keck data where available (and WHT for O); (9) Excluding S, P, Sc, C (subsolar), N (subsolar), using lower Al upper bound, 1 sigma errors calculated to 3sf from 3 sigma errors given and rounded up where necessary; (10) Excluding Mn, Sc, V, Sr; (11) Excluding Li, K; (12) Using favoured stratified abundances for Ca and Mg, excluding P and S. Excluding C (subsolar); (13) Excluding P, S, C (subsolar); (14) Excluding Mn, Sc, V, Sr, S, Zn; (15) Abundances were refitted by \citet{Harrison2021}, Si not previously included; (16) Excluding Sc, V, Mn, Co; (17) Abundances were refitted by \citet{Harrison2021}, Ti not previously included; (18) Excluding C (subsolar), N (subsolar), S, Sc, Mn; (19) Excluding P, S, Sc, V, Mn, C (subsolar); (20) Excluding P, S, C (subsolar), using favoured strat values, using optical Si abundance; (21) Excluding P, S, C (subsolar), using favoured strat values, using UV Si abundance; (22) Excluding P, S, Sc, V, Mn, C (subsolar); (23) Excluding P, S, C (subsolar), N (subsolar); (24) Excluding Sc, V, Mn, Sr; (25) Excluding very high Na point; (26) Excluding Mn; (27) Excluding N (subsolar). Using model I; (28) Excluding P, S. Errors reported to be typically 0.1 - 0.2 dex; (29) Excluding P, S
\end{itemize}

\renewcommand{\arraystretch}{2}
\normalsize
\begin{longtable}{|p{2.8cm}|p{1.7cm}|p{1.7cm}|p{2.2cm}|p{2cm}|p{1.8cm}|p{1.8cm}|p{2cm}|p{2cm}|p{2cm}|}
\caption{\label{tab:results1} Results from Bayesian model. Errors given are 1 sigma. N/A indicates that a parameter was not invoked. A description of the parameters is given in Section \ref{sec:Methods} in the main text. Superscripted indices have the same meaning as in Table \ref{tab:abundances}}\\
\hline
         System & $[\textrm{Fe}/\textrm{H}]_{\textrm{index}}$ & $t$/Myrs & log($d_{\textrm{formation}}$/AU) & $z_{\textrm{formation}}$/AU & $f_c$ & log($f_{\textrm{pol}}$) & log($t_{\textrm{event}}$/Yrs) & $P$/GPa & $f_{\textrm{O}_{2}}$ ($\Delta$IW) \\

\hline
\endfirsthead
\hline

\hline
\endhead
\hline
\endfoot
G166-58$^{a}$ & $274_{-69}^{+462}$ & $0.69_{-0.68}^{+8.48}$ & N/A & N/A & N/A & $-7.92_{-0.04}^{+0.04}$ & $6.20_{-2.02}^{+1.08}$ & N/A & N/A \\
G241-6$^{b, 1}$ & $320_{-72}^{+164}$ & $6.45_{-4.41}^{+9.19}$ & $0.24_{-0.22}^{+0.16}$ & N/A & N/A & $-5.52_{-0.07}^{+0.06}$ & $7.24_{-0.56}^{+0.41}$ & N/A & N/A \\
G29-38$^{c, 2}$ & $338_{-124}^{+126}$ & $0.02_{-0.02}^{+2.29}$ & $-1.27_{-0.39}^{+0.31}$ & $0.14_{-0.01}^{+0.01}$ & N/A & $-4.94_{-0.02}^{+0.02}$ & $4.59_{-2.56}^{+2.20}$ & N/A & N/A \\
GALEX1931+0117$^{d, 3}$ & $137_{-98}^{+462}$ & $0.00_{-0.00}^{+1.15}$ & $0.55_{-0.03}^{+0.03}$ & N/A & $0.16_{-0.03}^{+0.04}$ & $-3.46_{-0.03}^{+0.03}$ & $3.94_{-2.49}^{+2.54}$ & $45.0_{-13.3}^{+9.8}$ & $-2.6_{-0.2}^{+0.3}$ \\
GALEX1931+0117$^{e}$ & $459_{-167}^{+250}$ & $0.00_{-0.00}^{+1.77}$ & N/A & N/A & N/A & $-3.74_{-0.10}^{+0.10}$ & $3.86_{-2.56}^{+2.82}$ & N/A & N/A \\
GALEX1931+0117$^{f, 3}$ & $561_{-0}^{+0}$ & $0.00_{-0.00}^{+0.02}$ & $0.43_{-0.06}^{+0.04}$ & $0.07_{-0.04}^{+0.05}$ & $0.31_{-0.03}^{+0.03}$ & $-3.58_{-0.04}^{+0.03}$ & $3.07_{-1.36}^{+1.52}$ & $51.6_{-7.2}^{+5.0}$ & $-2.7_{-0.1}^{+0.2}$ \\
GALEXJ2339$^{g, 4}$ & $413_{-159}^{+178}$ & $1.94_{-1.02}^{+2.84}$ & N/A & $0.07_{-0.05}^{+0.05}$ & N/A & $-5.42_{-0.03}^{+0.03}$ & $5.12_{-3.44}^{+2.19}$ & N/A & N/A \\
GD362$^{h, 5}$ & $568_{-177}^{+180}$ & $0.23_{-0.15}^{+0.27}$ & $-0.81_{-0.23}^{+0.13}$ & $0.12_{-0.03}^{+0.02}$ & N/A & $-5.18_{-0.03}^{+0.02}$ & $4.49_{-2.93}^{+2.24}$ & N/A & N/A \\
GD378$^{g, 6}$ & $481_{-311}^{+311}$ & $2.29_{-1.54}^{+14.25}$ & N/A & N/A & N/A & $-6.08_{-0.11}^{+0.12}$ & $6.11_{-4.07}^{+1.46}$ & N/A & N/A \\
GD40$^{b, 7}$ & $369_{-187}^{+262}$ & $1.90_{-1.08}^{+13.58}$ & $-0.57_{-0.07}^{+0.36}$ & $0.09_{-0.05}^{+0.04}$ & N/A & $-5.38_{-0.06}^{+0.06}$ & $6.50_{-4.39}^{+1.10}$ & N/A & N/A \\
GD424$^{i, 8}$ & $381_{-95}^{+141}$ & $5.21_{-4.61}^{+19.86}$ & $-0.60_{-0.01}^{+0.02}$ & N/A & N/A & $-4.34_{-0.02}^{+0.02}$ & $7.13_{-0.96}^{+0.57}$ & N/A & N/A \\
GD56$^{a}$ & $563_{-305}^{+263}$ & $0.01_{-0.01}^{+3.17}$ & N/A & N/A & N/A & $-5.06_{-0.09}^{+0.09}$ & $4.39_{-2.84}^{+2.51}$ & N/A & N/A \\
GD61$^{j, 9}$ & $450_{-197}^{+260}$ & $3.09_{-2.76}^{+10.41}$ & $0.30_{-0.20}^{+0.12}$ & N/A & $0.03_{-0.01}^{+0.01}$ & $-5.83_{-0.03}^{+0.02}$ & $6.91_{-3.55}^{+0.63}$ & $40.5_{-18.5}^{+11.7}$ & $-2.5_{-0.3}^{+0.4}$ \\
HE0106-3253$^{a}$ & $515_{-302}^{+282}$ & $0.00_{-0.00}^{+1.41}$ & $-1.47_{-0.34}^{+0.41}$ & $0.09_{-0.01}^{+0.01}$ & $0.60_{-0.06}^{+0.06}$ & $-4.56_{-0.04}^{+0.04}$ & $4.09_{-2.54}^{+2.45}$ & $25.6_{-15.5}^{+18.7}$ & $-1.9_{-0.6}^{+0.6}$ \\
HS2253+8023$^{k, 10}$ & $597_{-287}^{+216}$ & $0.26_{-0.20}^{+9.81}$ & $-0.44_{-0.16}^{+0.64}$ & N/A & N/A & $-5.19_{-0.06}^{+0.07}$ & $6.04_{-3.83}^{+1.35}$ & N/A & N/A \\
LHS2534$^{l, 11}$ & $926_{-15}^{+22}$ & $4.71_{-0.85}^{+8.56}$ & $0.55_{-0.02}^{+0.02}$ & N/A & $0.52_{-0.12}^{+0.14}$ & $-8.38_{-0.03}^{+0.03}$ & $3.62_{-2.30}^{+3.66}$ & $35.7_{-16.5}^{+14.8}$ & $-1.3_{-0.2}^{+0.2}$ \\
NLTT43806$^{m}$ & $817_{-272}^{+110}$ & $0.78_{-0.75}^{+10.53}$ & $-0.16_{-0.30}^{+0.35}$ & $0.10_{-0.06}^{+0.04}$ & $0.03_{-0.01}^{+0.01}$ & $-6.56_{-0.05}^{+0.05}$ & $6.34_{-2.52}^{+1.07}$ & $45.6_{-15.0}^{+9.7}$ & $-2.7_{-0.2}^{+0.4}$ \\
PG0843+516$^{d, 12}$ & $417_{-210}^{+280}$ & $0.01_{-0.01}^{+1.81}$ & $-0.07_{-0.29}^{+0.35}$ & N/A & $0.64_{-0.09}^{+0.08}$ & $-4.25_{-0.08}^{+0.09}$ & $4.13_{-2.55}^{+2.50}$ & $29.3_{-17.1}^{+21.1}$ & $-2.5_{-0.3}^{+0.6}$ \\
PG0843+516$^{a}$ & $463_{-295}^{+313}$ & $0.00_{-0.00}^{+1.50}$ & $-0.01_{-0.38}^{+0.36}$ & N/A & $0.75_{-0.10}^{+0.08}$ & $-3.77_{-0.12}^{+0.12}$ & $3.97_{-2.40}^{+2.64}$ & $25.8_{-15.0}^{+18.7}$ & $-1.9_{-0.7}^{+0.6}$ \\
PG1015+161$^{d, 13}$ & $377_{-232}^{+350}$ & $0.00_{-0.00}^{+1.12}$ & $-1.30_{-0.44}^{+0.42}$ & $0.11_{-0.02}^{+0.02}$ & $0.34_{-0.12}^{+0.13}$ & $-5.00_{-0.09}^{+0.10}$ & $3.79_{-2.31}^{+2.68}$ & $17.8_{-11.5}^{+19.0}$ & $-1.7_{-0.7}^{+0.4}$ \\
PG1015+161$^{a}$ & $499_{-316}^{+293}$ & $0.00_{-0.00}^{+1.27}$ & $-1.09_{-0.55}^{+0.56}$ & $0.11_{-0.03}^{+0.02}$ & $0.59_{-0.13}^{+0.11}$ & $-4.72_{-0.12}^{+0.12}$ & $3.98_{-2.51}^{+2.57}$ & $31.9_{-18.8}^{+17.4}$ & $-2.0_{-0.6}^{+0.6}$ \\
PG1225-079$^{n, 14}$ & $494_{-250}^{+260}$ & $1.49_{-1.00}^{+2.71}$ & $-0.87_{-0.32}^{+0.21}$ & $0.12_{-0.04}^{+0.02}$ & N/A & $-6.93_{-0.07}^{+0.07}$ & $5.74_{-3.51}^{+1.56}$ & N/A & N/A \\
SDSSJ0512-0505$^{o, 15}$ & $499_{-319}^{+305}$ & $3.65_{-1.35}^{+1.60}$ & $-1.04_{-0.55}^{+0.33}$ & $0.13_{-0.02}^{+0.01}$ & $0.54_{-0.21}^{+0.17}$ & $-7.38_{-0.06}^{+0.06}$ & $3.20_{-2.04}^{+2.53}$ & $25.5_{-16.6}^{+20.0}$ & $-1.8_{-0.7}^{+0.5}$ \\
SDSSJ0738+1835$^{p, 16}$ & $375_{-219}^{+185}$ & $0.49_{-0.07}^{+0.07}$ & $0.10_{-0.41}^{+0.27}$ & N/A & $0.44_{-0.07}^{+0.07}$ & $-3.79_{-0.07}^{+0.07}$ & $2.39_{-1.51}^{+1.78}$ & $32.3_{-19.6}^{+19.2}$ & $-2.5_{-0.3}^{+0.6}$ \\
SDSSJ0823+0546$^{o, 17}$ & $438_{-297}^{+341}$ & $17.66_{-3.54}^{+3.41}$ & $0.03_{-0.38}^{+0.34}$ & N/A & $0.96_{-0.04}^{+0.02}$ & $-7.23_{-0.06}^{+0.06}$ & $3.25_{-2.05}^{+2.34}$ & $29.0_{-18.9}^{+19.8}$ & $-2.1_{-0.6}^{+0.6}$ \\
SDSSJ0845+2257$^{q, 18}$ & $586_{-350}^{+250}$ & $0.47_{-0.43}^{+12.05}$ & $-0.21_{-0.29}^{+0.36}$ & N/A & $0.31_{-0.09}^{+0.10}$ & $-3.76_{-0.09}^{+0.08}$ & $6.23_{-3.45}^{+1.17}$ & $28.4_{-18.2}^{+19.8}$ & $-2.1_{-0.6}^{+0.7}$ \\
SDSSJ1043+0855$^{r, 19}$ & $342_{-178}^{+245}$ & $0.00_{-0.00}^{+1.92}$ & $-0.22_{-0.38}^{+0.51}$ & N/A & N/A & $-4.60_{-0.09}^{+0.10}$ & $4.03_{-2.59}^{+2.64}$ & N/A & N/A \\
SDSSJ1228+1040$^{d, 20}$ & $421_{-245}^{+284}$ & $0.01_{-0.01}^{+1.82}$ & $-0.69_{-0.30}^{+0.42}$ & $0.12_{-0.05}^{+0.02}$ & N/A & $-4.28_{-0.07}^{+0.06}$ & $4.11_{-2.59}^{+2.55}$ & N/A & N/A \\
SDSSJ1228+1040$^{d, 21}$ & $486_{-270}^{+291}$ & $0.00_{-0.00}^{+1.77}$ & $-0.90_{-0.52}^{+0.25}$ & $0.12_{-0.04}^{+0.02}$ & N/A & $-4.38_{-0.09}^{+0.08}$ & $3.97_{-2.68}^{+2.66}$ & N/A & N/A \\
SDSSJ1242+5226$^{s, 22}$ & $374_{-81}^{+111}$ & $4.93_{-0.39}^{+0.86}$ & $0.05_{-0.55}^{+0.34}$ & N/A & N/A & $-4.32_{-0.04}^{+0.05}$ & $4.51_{-3.09}^{+1.60}$ & N/A & N/A \\
SDSSJ2047-1259$^{t, 23}$ & $523_{-367}^{+297}$ & $0.62_{-0.12}^{+0.15}$ & $-0.01_{-0.41}^{+0.34}$ & N/A & N/A & $-4.67_{-0.06}^{+0.06}$ & $2.83_{-1.95}^{+2.18}$ & N/A & N/A \\
WD0122-227$^{u}$ & $467_{-299}^{+311}$ & $7.84_{-4.78}^{+6.14}$ & $0.03_{-0.40}^{+0.37}$ & N/A & $0.57_{-0.20}^{+0.19}$ & $-8.23_{-0.15}^{+0.15}$ & $3.96_{-2.58}^{+2.92}$ & $29.6_{-18.7}^{+19.1}$ & $-2.0_{-0.6}^{+0.6}$ \\
WD0446-255$^{u, 24}$ & $814_{-339}^{+113}$ & $1.09_{-0.75}^{+1.42}$ & $-0.58_{-0.11}^{+0.13}$ & $0.13_{-0.02}^{+0.01}$ & $0.05_{-0.02}^{+0.02}$ & $-5.62_{-0.04}^{+0.05}$ & $4.54_{-3.05}^{+2.42}$ & $36.9_{-22.5}^{+15.5}$ & $-2.2_{-0.5}^{+0.6}$ \\
WD0449-259$^{u, 25}$ & $581_{-387}^{+267}$ & $7.11_{-4.28}^{+5.58}$ & $-1.03_{-0.61}^{+0.94}$ & $0.10_{-0.04}^{+0.03}$ & $0.65_{-0.23}^{+0.18}$ & $-7.47_{-0.15}^{+0.14}$ & $3.99_{-2.53}^{+2.70}$ & $19.9_{-13.1}^{+19.3}$ & $-1.4_{-0.7}^{+0.3}$ \\
WD1145+017$^{v, 3}$ & $640_{-327}^{+214}$ & $0.54_{-0.39}^{+8.70}$ & $0.13_{-0.44}^{+1.31}$ & N/A & N/A & $-4.87_{-0.12}^{+0.32}$ & $5.90_{-3.78}^{+1.50}$ & N/A & N/A \\
WD1145+288$^{a}$ & $502_{-297}^{+285}$ & $0.01_{-0.01}^{+1.47}$ & $-0.96_{-0.58}^{+1.21}$ & $0.12_{-0.04}^{+0.02}$ & $0.48_{-0.15}^{+0.14}$ & $-5.30_{-0.12}^{+0.14}$ & $4.16_{-2.57}^{+2.41}$ & $31.7_{-19.9}^{+17.5}$ & $-2.0_{-0.6}^{+0.6}$ \\
WD1232+563$^{a, 26}$ & $559_{-0}^{+0}$ & $13.26_{-9.00}^{+24.86}$ & N/A & N/A & N/A & $-5.10_{-0.03}^{+0.02}$ & $7.49_{-0.44}^{+0.34}$ & N/A & N/A \\
WD1350-162$^{u, 25}$ & $161_{-114}^{+440}$ & $6.84_{-1.13}^{+1.25}$ & $0.00_{-0.33}^{+0.31}$ & N/A & $0.42_{-0.08}^{+0.08}$ & $-6.04_{-0.06}^{+0.06}$ & $3.11_{-2.02}^{+2.28}$ & $22.7_{-14.9}^{+22.7}$ & $-2.1_{-0.6}^{+0.7}$ \\
WD1425+540$^{w, 27}$ & $470_{-275}^{+295}$ & $1.34_{-0.82}^{+4.38}$ & N/A & N/A & N/A & $-6.59_{-0.08}^{+0.10}$ & $5.48_{-3.73}^{+1.90}$ & N/A & N/A \\
WD1536+520$^{x, 28}$ & $445_{-232}^{+294}$ & $0.19_{-0.14}^{+8.80}$ & $0.10_{-0.40}^{+0.29}$ & N/A & N/A & $-3.26_{-0.07}^{+0.08}$ & $5.79_{-3.78}^{+1.54}$ & N/A & N/A \\
WD1551+175$^{a, 26}$ & $589_{-221}^{+171}$ & $4.19_{-3.44}^{+20.66}$ & $-0.66_{-0.01}^{+0.01}$ & N/A & N/A & $-5.38_{-0.04}^{+0.03}$ & $7.08_{-4.24}^{+0.64}$ & N/A & N/A \\
WD2105-820$^{u}$ & $210_{-159}^{+486}$ & $0.00_{-0.00}^{+0.08}$ & $0.07_{-0.35}^{+0.29}$ & N/A & $0.84_{-0.26}^{+0.11}$ & $-5.96_{-0.13}^{+0.14}$ & $1.53_{-0.98}^{+3.79}$ & $18.8_{-12.6}^{+21.4}$ & $-1.7_{-0.7}^{+0.4}$ \\
WD2115-560$^{u}$ & $472_{-181}^{+217}$ & $0.00_{-0.00}^{+0.30}$ & $0.20_{-0.47}^{+1.10}$ & N/A & N/A & $-5.84_{-0.04}^{+0.04}$ & $3.20_{-1.74}^{+2.73}$ & N/A & N/A \\
WD2157-574$^{u}$ & $364_{-199}^{+275}$ & $0.01_{-0.00}^{+2.44}$ & N/A & N/A & N/A & $-6.55_{-0.04}^{+0.04}$ & $3.09_{-2.18}^{+3.70}$ & N/A & N/A \\
WD2207+121$^{a, 26}$ & $673_{-438}^{+214}$ & $5.42_{-4.17}^{+22.74}$ & $0.12_{-0.43}^{+0.28}$ & N/A & N/A & $-5.22_{-0.06}^{+0.06}$ & $7.13_{-4.58}^{+0.62}$ & N/A & N/A \\
WD2216-657$^{u}$ & $388_{-272}^{+345}$ & $7.18_{-0.97}^{+0.97}$ & $-0.18_{-0.21}^{+0.28}$ & $0.07_{-0.05}^{+0.06}$ & N/A & $-7.25_{-0.05}^{+0.04}$ & $3.18_{-2.09}^{+2.34}$ & N/A & N/A \\
WDJ1814-7354$^{y, 29}$ & $382_{-147}^{+199}$ & $0.02_{-0.02}^{+3.22}$ & N/A & N/A & N/A & $-5.94_{-0.05}^{+0.04}$ & $4.76_{-2.43}^{+2.15}$ & N/A & N/A \\

\end{longtable}
\renewcommand{\arraystretch}{1.5}
\begin{longtable}{|p{2.8cm}|p{1.3cm}|p{1.3cm}|p{1cm}|p{1cm}|p{1cm}|p{1.5cm}|p{1.5cm}|p{1.5cm}|p{2cm}|p{2cm}|p{1.5cm}|}
\caption{\label{tab:results2} Results from Bayesian model. N/A indicates that a parameter was not invoked, or (for the category column) that another data set was used for that system. Category abbreviations are as follows: HPM = High Pressure Mantle-rich, LPC = Low Pressure Core-rich, PD = Pressure degenerate with oxygen fugacity, PU = Pressure unconstrained, NED = No Evidence of Differentiation, U = Unphysical. * Previously explained as crust-rich. $^\dagger$ May be subject to a degeneracy which causes high pressure fragments to appear to be low pressure; see section \ref{sec:highpressurecorerich}. Superscripted indices have the same meaning as in Table \ref{tab:abundances}} \\
\hline
         System & Good Fit? & Primitive? & Core Rich? & Mantle Rich? & Volatile Rich? & Volatile Depleted? & Moderate Volatile Depleted? & Temperature & Inferred Parent Core Number Fraction & Differentiation Sigma & Category\\

\hline
\endfirsthead
\hline

\hline
\endhead
\hline
\endfoot
G166-58$^{a}$ & Y & Y & N & N & Y & N & N & N/A & N/A & N/A & NED \\
G241-6$^{b, 1}$ & Y & Y & N & N & Y & N & N & 223 & N/A & N/A & NED \\
G29-38$^{c, 2}$ & Y & Y & N & N & N & N & Y & 1952 & N/A & N/A & NED \\
GALEX1931+0117$^{d, 3}$ & N & N & Y & N & Y & N & N & 130 & 0.01 & 2.7 & N/A \\
GALEX1931+0117$^{e}$ & N & N & Y & N & Y & N & N & 150 & 0.14 & 4.9 & NED \\
GALEX1931+0117$^{f, 3}$ & N & Y & N & N & Y & N & N & N/A & N/A & N/A & N/A \\
GALEXJ2339$^{g, 4}$ & Y & Y & N & N & Y & N & N & N/A & N/A & N/A & NED \\
GD362$^{h, 5}$ & Y & Y & N & N & N & N & Y & 1516 & N/A & N/A & NED \\
GD378$^{g, 6}$ & Y & Y & N & N & Y & N & N & N/A & N/A & N/A & NED \\
GD40$^{b, 7}$ & Y & Y & N & N & N & Y & N & 1197 & N/A & N/A & NED \\
GD424$^{i, 8}$ & N & Y & N & N & N & Y & N & 1268 & N/A & N/A & NED \\
GD56$^{a}$ & Y & Y & N & N & Y & N & N & N/A & N/A & N/A & NED \\
GD61$^{j, 9}$ & Y & N & N & Y & Y & N & N & 198 & 0.16 & 1.3 & HPM \\
HE0106-3253$^{a}$ & Y & N & Y & N & N & N & Y & 2170 & 0.04 & 9.2 & PD \\
HS2253+8023$^{k, 10}$ & Y & Y & N & N & Y & N & N & 912 & N/A & N/A & NED \\
LHS2534$^{l, 11}$ & Y & N & Y & N & Y & N & N & 130 & 0.0 & 4.3 & U \\
NLTT43806$^{m}$ & Y & N & N & Y & Y & N & N & 514 & 0.17 & 2.8 & HPM* \\
PG0843+516$^{d, 12}$ & N & N & Y & N & Y & N & N & 429 & 0.17 & 4.6 & N/A \\
PG0843+516$^{a}$ & Y & N & Y & N & Y & N & N & 373 & 0.13 & 5.6 & PD \\
PG1015+161$^{d, 13}$ & N & N & Y & N & N & N & Y & 1976 & 0.07 & 1.4 & N/A \\
PG1015+161$^{a}$ & Y & N & Y & N & N & N & Y & 1761 & 0.1 & 3.3 & PD \\
PG1225-079$^{n, 14}$ & Y & Y & N & N & N & N & Y & 1570 & N/A & N/A & NED \\
SDSSJ0512-0505$^{o, 15}$ & Y & N & Y & N & N & N & Y & 1714 & 0.1 & 2.1 & PD \\
SDSSJ0738+1835$^{p, 16}$ & Y & N & Y & N & Y & N & N & 299 & 0.15 & 2.9 & PU \\
SDSSJ0823+0546$^{o, 17}$ & Y & N & Y & N & Y & N & N & 349 & 0.13 & 10.1 & PU \\
SDSSJ0845+2257$^{q, 18}$ & Y & N & Y & N & Y & N & N & 565 & 0.14 & 1.6 & PU \\
SDSSJ1043+0855$^{r, 19}$ & N & Y & N & N & Y & N & N & 587 & N/A & N/A & NED \\
SDSSJ1228+1040$^{d, 20}$ & Y & Y & N & N & N & N & Y & 1419 & N/A & N/A & N/A \\
SDSSJ1228+1040$^{d, 21}$ & Y & Y & N & N & N & N & Y & 1588 & N/A & N/A & NED \\
SDSSJ1242+5226$^{s, 22}$ & Y & Y & N & N & Y & N & N & 330 & N/A & N/A & NED \\
SDSSJ2047-1259$^{t, 23}$ & Y & Y & N & N & Y & N & N & 376 & N/A & N/A & NED \\
WD0122-227$^{u}$ & Y & N & Y & N & Y & N & N & 349 & 0.1 & 2.2 & PU \\
WD0446-255$^{u, 24}$ & Y & N & N & Y & N & Y & N & 1217 & 0.15 & N/A & HPM \\
WD0449-259$^{u, 25}$ & Y & N & Y & N & N & N & Y & 1711 & 0.1 & N/A & LPC \\
WD1145+017$^{v, 3}$ & Y & Y & N & N & Y & N & N & 279 & N/A & N/A & NED \\
WD1145+288$^{a}$ & Y & N & Y & N & N & N & Y & 1642 & 0.16 & 1.5 & PU \\
WD1232+563$^{a, 26}$ & N & Y & N & N & Y & N & N & N/A & N/A & N/A & NED \\
WD1350-162$^{u, 25}$ & Y & N & Y & N & Y & N & N & 364 & 0.09 & N/A & LPC$^\dagger$ \\
WD1425+540$^{w, 27}$ & Y & Y & N & N & Y & N & N & N/A & N/A & N/A & NED \\
WD1536+520$^{x, 28}$ & Y & Y & N & N & Y & N & N & 300 & N/A & N/A & NED \\
WD1551+175$^{a, 26}$ & Y & Y & N & N & N & Y & N & 1395 & N/A & N/A & NED \\
WD2105-820$^{u}$ & Y & N & Y & N & Y & N & N & 318 & 0.09 & N/A & LPC \\
WD2115-560$^{u}$ & Y & Y & N & N & Y & N & N & 242 & N/A & N/A & NED \\
WD2157-574$^{u}$ & Y & Y & N & N & Y & N & N & N/A & N/A & N/A & NED \\
WD2207+121$^{a, 26}$ & Y & Y & N & N & Y & N & N & 287 & N/A & N/A & NED \\
WD2216-657$^{u}$ & N & Y & N & N & Y & N & N & 540 & N/A & N/A & NED \\
WDJ1814-7354$^{y, 29}$ & Y & Y & N & N & Y & N & N & N/A & N/A & N/A & NED \\

\end{longtable}
\end{centering}
\end{landscape}

\begin{centering}
\begin{longtable}{|p{1.5cm}|p{1.5cm}|p{1.5cm}|p{1.5cm}|p{2cm}|p{1.5cm}|p{3cm}|p{2.5cm}|}
\caption{\label{tab:partition} Summary of coefficients used in parametrisations of partitioning behaviour. For coefficients sourced from \citet{Fischer2015}, those marked with $^\dagger$ are results quoted from epsilon modelling. *S partitioning was implemented following the procedure described by \citet{Boujibar2014}, but was not included when running the model on white dwarf systems.} \\
\hline
         
         Element & Valence & a & b $/\textrm{K}$ & c $/\textrm{K}\;\textrm{GPa}^{-1}$ & d & Source & Equation Used\\

\hline
\endfirsthead
\hline

\hline
\endhead
\hline
\endfoot
Fe & 2 & 0 & 0 & 0 & 0 & N/A & (\ref{eq:logdm_dfe}) \\
Mn & 2 & -0.02 & -5600 & 38 & 0.036 & \citet{Corgne2008} & (\ref{eq:logdm_rudge}) \\
Ni & 2 & 0.46 & 2700 & -61 & 0 & \citet{Fischer2015} & (\ref{eq:logdm_fischer2}) \\
Cr & 2 & -0.3 & -2200 & -5 & 0 & \citet{Fischer2015} $^\dagger$ & (\ref{eq:logdm_fischer2}) \\
Ga & 3 & 3.5 & -4800 & -126 & -0.97 & \citet{Corgne2008} & (\ref{eq:logdm_rudge}) \\
Si & 4 & 0.549 & -12324 & 0 & 0 & \citet{Siebert2013} & (\ref{eq:logdm_rudge}) \\
Nb & 5 & 4.09 & -15500 & -166 & -0.75 & \citet{Corgne2008} & (\ref{eq:logdm_rudge}) \\
Ta & 5 & 7.74 & -20000 & -264 & -1.69 & \citet{Corgne2008} & (\ref{eq:logdm_rudge}) \\
Cu & 1 & 0.3 & 2300 & -37 & 0.14 & \citet{Corgne2008} & (\ref{eq:logdm_rudge}) \\
Zn & 2 & -1.11 & 600 & -23 & -0.21 & \citet{Corgne2008} & (\ref{eq:logdm_rudge}) \\
V & 3 & -1.5 & -2300 & 9 & 0 & \citet{Fischer2015} $^\dagger$ & (\ref{eq:logdm_fischer2}) \\
Co & 2 & 0.36 & 1500 & -33 & 0 & \citet{Fischer2015} & (\ref{eq:logdm_fischer2}) \\
P & 5 & 0.64 & -1593 & -74.95 & 0 & \citet{Wade2005} & (\ref{eq:logdm_rudge}) \\
Tl & 1 & -0.118 & -783 & 0 & 0 & \citet{Wood2008Pb} & (\ref{eq:logdm_rudge}) \\
W & 4.52 & 3.2 & -1605 & -115 & -0.85 & \citet{Cottrell2009} & (\ref{eq:logdm_rudge}) \\
O & -2 & 0.986 & -3275 & 0 & 0 & \citet{Siebert2013} & (\ref{eq:logdm_rudge}) \\
C & 4 & -1 & 4842 & 31 & 0 & \citet{Blanchard2019} & (\ref{eq:logdm_fischer2}) \\
S & -2 & 0 & 405 & 136 & 0 & \citet{Boujibar2014} & N/A * \\
\end{longtable}
\end{centering}

\begin{centering}
\begin{longtable}{|p{2.5cm}|p{2.5cm}|p{1.5cm}|p{1.5cm}|p{1.5cm}|p{1.5cm}|p{1.5cm}|}
\caption{\label{tab:observability} Tabulated extract of Figure \ref{fig:observability}, showing predicted elemental abundances in a white dwarf atmosphere as a function of the core fraction of a pollutant fragment and the core--mantle differentiation pressure in its parent. Hx represents the dominant element in the white dwarf atmosphere, either H or He. Mg/Fe and Ca/Fe act as proxies for the fragment core fraction. When this quantity is known, Cr, Ni and/or Si can be used to infer pressure. Note that this table assumes steady-state accretion and heavy pollution (meaning that the total pollution of the elements we model varies between -3 and -4 log units relative to Hx) - see caption of Figure \ref{fig:observability} and the main text for further details.} \\
\hline
         Fragment Core Fraction & Pressure /GPa & log(Mg/Fe) & log(Ca/Fe) & log(Cr/Hx) & log(Ni/Hx) & log(Si/Hx)\\

\hline
\endfirsthead
\hline

\hline
\endhead
\hline
\endfoot
0 & 0 & 0.99 & -0.06 & -7.15 & -9.19 & -5.05 \\
0 & 10 & 1.0 & -0.05 & -7.2 & -8.67 & -5.06 \\
0 & 20 & 1.01 & -0.04 & -7.23 & -8.39 & -5.06 \\
0 & 30 & 1.03 & -0.03 & -7.26 & -8.16 & -5.07 \\
0 & 40 & 1.06 & 0.0 & -7.3 & -7.99 & -5.09 \\
0 & 50 & 1.1 & 0.04 & -7.36 & -7.85 & -5.12 \\
0 & 60 & 1.13 & 0.07 & -7.42 & -7.73 & -5.14 \\
0.2 & 0 & 0.06 & -0.99 & -7.24 & -6.74 & -5.27 \\
0.2 & 10 & 0.07 & -0.98 & -7.22 & -6.74 & -5.26 \\
0.2 & 20 & 0.08 & -0.97 & -7.22 & -6.75 & -5.26 \\
0.2 & 30 & 0.1 & -0.96 & -7.21 & -6.77 & -5.25 \\
0.2 & 40 & 0.13 & -0.93 & -7.21 & -6.79 & -5.24 \\
0.2 & 50 & 0.17 & -0.89 & -7.22 & -6.81 & -5.22 \\
0.2 & 60 & 0.2 & -0.85 & -7.23 & -6.84 & -5.22 \\
0.4 & 0 & -0.33 & -1.39 & -7.31 & -6.53 & -5.49 \\
0.4 & 10 & -0.32 & -1.38 & -7.24 & -6.54 & -5.47 \\
0.4 & 20 & -0.31 & -1.37 & -7.21 & -6.54 & -5.45 \\
0.4 & 30 & -0.29 & -1.35 & -7.19 & -6.56 & -5.41 \\
0.4 & 40 & -0.27 & -1.32 & -7.16 & -6.58 & -5.37 \\
0.4 & 50 & -0.23 & -1.28 & -7.14 & -6.6 & -5.31 \\
0.4 & 60 & -0.19 & -1.25 & -7.12 & -6.62 & -5.28 \\
0.6 & 0 & -0.67 & -1.73 & -7.37 & -6.43 & -5.74 \\
0.6 & 10 & -0.66 & -1.72 & -7.25 & -6.44 & -5.7 \\
0.6 & 20 & -0.65 & -1.71 & -7.21 & -6.44 & -5.66 \\
0.6 & 30 & -0.63 & -1.69 & -7.17 & -6.45 & -5.58 \\
0.6 & 40 & -0.61 & -1.66 & -7.13 & -6.47 & -5.49 \\
0.6 & 50 & -0.57 & -1.62 & -7.09 & -6.49 & -5.4 \\
0.6 & 60 & -0.53 & -1.59 & -7.06 & -6.51 & -5.33 \\
0.8 & 0 & -1.09 & -2.15 & -7.41 & -6.37 & -6.1 \\
0.8 & 10 & -1.08 & -2.14 & -7.25 & -6.37 & -6.02 \\
0.8 & 20 & -1.07 & -2.13 & -7.2 & -6.38 & -5.91 \\
0.8 & 30 & -1.05 & -2.11 & -7.16 & -6.39 & -5.77 \\
0.8 & 40 & -1.03 & -2.08 & -7.11 & -6.4 & -5.62 \\
0.8 & 50 & -0.99 & -2.04 & -7.05 & -6.42 & -5.47 \\
0.8 & 60 & -0.95 & -2.01 & -7.01 & -6.44 & -5.38 \\
0.99 & 0 & -2.48 & -3.54 & -7.45 & -6.33 & -7.37 \\
0.99 & 10 & -2.47 & -3.53 & -7.26 & -6.34 & -6.67 \\
0.99 & 20 & -2.46 & -3.52 & -7.2 & -6.34 & -6.3 \\
0.99 & 30 & -2.44 & -3.5 & -7.15 & -6.35 & -5.99 \\
0.99 & 40 & -2.42 & -3.47 & -7.09 & -6.36 & -5.75 \\
0.99 & 50 & -2.38 & -3.43 & -7.03 & -6.38 & -5.54 \\
0.99 & 60 & -2.34 & -3.4 & -6.98 & -6.39 & -5.42 \\

\end{longtable}
\end{centering}

\twocolumn

\section{Comments on individual systems}
\label{sec:individual_systems}

\subsection{GD61}

The depletion in Ni and Fe illustrated in Figure \ref{fig:GD61_comp} suggests that the pollution is rich in mantle-like material. The red line shows how the median fit would change if the pressure was fixed to 0\;GPa while using the same posterior distribution for the other parameters. While the difference due to pressure is very much a secondary effect, the fit is noticeably improved at higher pressure. This is driven by Si partitioning and the indirect change in mantle Fe content. Our median model hits the artificial $D_{\textrm{Si}}$ cap, which implies that our pressure constraint could be stronger if we allowed Si to partition more strongly into the core.

The median pressure is 40\;GPa, which (given the composition inferred by our model) corresponds to a planet of mass 0.6 $\textrm{M}_{\oplus}$. We illustrate the posterior distribution of pressure in Figure \ref{fig:GD61_pressuredist}, which exhibits a peak between 40 and 60 GPa - i.e., roughly Earth-like pressure, and hence Earth-like mass.

Our results are generally in agreement with previous work by \citet{Farihi2013}. We find that GD61 is in a steady-state phase of accretion, which is consistent with \citet{Farihi2013} and the presence of an infra-red excess. \citet{Farihi2013} additionally found that the pollution contained excess O, implying the accretion of water. We follow a similar analysis, computing the amount of O in the parent body's mantle which would be left over if all other components were fully oxidised. Assuming that all remaining O is contained in $\textrm{H}_{2}\textrm{O}$, and that the parent body's mass is indeed 0.61M$_\oplus$ as implied by our model, we find that there is enough excess O (34\% of the mantle's O budget, by number) to account for significant amounts of water. The implied total mass of water in the parent body is $5.1^{+2.1}_{-2.9} \times 10^{23}\;\textrm{kg}$, where the error comes from propagating our error on pressure. \citet{Farihi2013} found that GD61 has accreted (at most) $5 \times 10^{19}\;\textrm{kg}$ of water (based on the trace H content in the atmosphere of GD61). This discrepancy implies that either our inferred parent body differentiation pressure (and hence mass) is an overestimate, or that GD61 has so far accreted only a very small amount of the parent body's water budget (roughly 0.01\%).

\subsection{WD0446-255}

High pressure and low oxygen fugacity is favoured. Compared to the median fit (with pressure = $37^{+15}_{-22}\;\textrm{GPa}$), the log-likelihood is decreased by 0.66 when fixing pressure to 0. All elements, apart from Mg, fit better at high pressure. Cr and Fe show the greatest change with pressure, but as with GD61, the Ni abundance is not strongly affected despite the fragment being mantle-rich. We discuss this behaviour in the Appendix.

The depletion in siderophile elements pushes the model to a mantle rich fragment with 5.5\% core. Compared to GD61 it has a low Mg/Fe ratio, only slightly above solar, as shown in Figure \ref{fig:bowtie_multipanel}. However, our model finds that the Mg abundance is anomalously low: the model is able to match the abundance of all 9 other elements modelled to within 1 sigma, but predicts an Mg abundance roughly 2 sigma above that observed. This may be down to random chance: the probability that at least one of the 10 data points is observed at least 2 sigma away from its true value is 37\%. The sensitivity to pressure is therefore greater than suggested by Figure \ref{fig:bowtie_multipanel}. To best match the low Mg, the model favours a build-up phase solution. This is consistent with the results of \citet{Swan2019}. The high refractory element abundances (and slight O depletion) suggest incomplete condensation.




\subsection{NLTT43806}

Our model finds a good fit to the data by invoking the accretion of mantle-rich material. Again, high pressure is preferred. However, the pollutant has previously been explained as being composed (at least partially) of crust-rich material due to the high Al and Ca abundances \citep{Zuckerman2011}. Our model does not reproduce these Al or Ca observations because we don't include crustal material. We infer that our fit could be further improved by including crustal material, as was found by \citet{Harrison2018}.

\subsection{WD0449-259}

The model finds 2 degenerate solutions for this system. The first solution is a highly core-rich fragment, with extreme heating, accreting in build up/steady state. This roughly follows the heating effects arrow from the ~80\% core contour to the data point in the middle panel of Figure \ref{fig:bowtie_multipanel}. The second solution invokes a less core-rich fragment, with decreased heating, accreting in the declining phase. This is similar to the first solution, but starting at a lower core fraction, and following part of both the heating and sinking arrows to the data point.

Importantly, both solutions favour low pressure (and high oxygen fugacity) in order to make Ni as siderophilic as possible. Given that this is a core-rich fragment (as suggested by the super-solar Fe/Mg and Ni/Mg ratios), moving as much Ni into the core as possible helps to match the high observed Ni/Fe.

\citet{Swan2019} reported an anomalously high Na abundance for this system which our model cannot reproduce. The cause is unknown. We therefore leave open the question of how to explain these anomalous lines and consider only the other observed (or constrained) elements.


\subsection{WD1350-162}

The median model is moderately core-rich (41\%), with low pressure favoured by the low Si abundance. The high observed Mg value relative to other elements is explained by accretion in the declining phase. This is consistent with the lack of observed infra-red excess. However, the model seems to also admit a (less favoured) high pressure solution (with additional heating and a slightly higher fragment core fraction). We discuss this important degeneracy in section \ref{sec:highpressurecorerich}.

As with WD0449-259, an anomalously high Na abundance was reported for this system by \citet{Swan2019} which we exclude from our modelling.

\subsection{WD2105-820}

This system has a strong requirement for differentiation (>4$\sigma$) due to its low Ca/Fe ratio, which suggests a core-rich fragment (84\%). Our model explains the comparatively high Mg abundances as being caused by accretion in the declining phase. The preference for low pressure is driven by an upper bound on Si: core-like fragments become more Si rich as pressure increases.

\subsection{PG0843+516}

The model is able to fit this data by the accretion of a core rich fragment (73\%), driven by the high Fe relative to Ca, Mg and Si. We find a degeneracy between pressure and oxygen fugacity, which is caused by Si. Increasing pressure makes Si more siderophilic, but increasing oxygen fugacity makes Si less siderophilic. It's possible to increase (or decrease) both in such a way as to keep the final Si abundance within the errors. Including a Cr or Ni data point may help break the degeneracy.

As noted by \citet{Xu2019}, there is a known discrepancy between metal abundances based on UV and optical data. We ran our model on two data sets for this system, taken from \citet{Gaensicke2012} and \citet{Xu2019}. \citet{Gaensicke2012} used UV data from HST's COS instrument, and we use their abundances reported for Al, Ni, Fe, Cr, Mg, Si and O. The abundances of \citet{Xu2019} are based on optical data from HIRES at Keck, from which we use Ca, Fe, Mg and Si. The Mg abundances are consistent to within the reported errors, but the Fe and Si abundances are both significantly higher in \citet{Xu2019}. Our results are based on the version presented in \citet{Xu2019}, because we were unable to simultaneously fit the extreme Cr and Ni abundances in \citet{Gaensicke2012}. 

It is possible that the material in the WD's photosphere may have been completely recycled in between the UV and optical observations. The sinking timescales for this WD are very short - of the observed elements, the longest sinking timescale belongs to Si (25 days). This implies that the material in the WD's photosphere may have been completely recycled in between the UV and optical observations. However, as a DA white dwarf with $\textrm{T}_{\textrm{eff}} \gtrsim $15,000-18,000\;K, our interpolated sinking timescales for this white dwarf should be treated with caution (see section \ref{sec:ModelCaveats2}).

\subsection{HE0106-3253 and PG1015+161}

These systems are similar to PG0843+516: the model infers a core-rich fragment with the only pressure-sensitive element being Si. This leads to a pressure/oxygen fugacity degeneracy. Along with PG0843+516, they appear towards the bottom left of the bottom panel in Figure \ref{fig:bowtie_multipanel}, in the region where Si is reasonably sensitive to pressure. This suggests that these objects would potentially yield tight pressure constraints, if another element were present to break the pressure/oxygen fugacity degeneracy.

PG1015+161 has multiple data sets, from \citet{Gaensicke2012} and \citet{Xu2019}. As with PG0843+516, we use the set from \citet{Xu2019} since the model is unable to fit the data from \citet{Gaensicke2012} (in particular, the low Si/Mg ratio is difficult to reconcile with the other detected abundances).

\subsection{SDSSJ0512-0505}

This system has roughly solar Fe/Mg and Ca/Mg ratios, but a depletion in Cr, which the model can explain as the accretion of a primarily core-rich fragment. To match the low Cr/Fe ratio, the model reduces the partition coefficient of Cr. It can do this by either lowering pressure or increasing oxygen fugacity, leading to a degeneracy.

\subsection{SDSSJ0823+0546}

Our model finds that SDSSJ0823+0546 has accreted a core-rich (96\%) fragment. The only pressure-sensitive element detected is Ni. As discussed in section \ref{sec:d_behaviour}, Ni is typically poor for constraining the pressure of core-rich fragments, and we correspondingly find that the model is unable to constrain pressure. A Si detection would strongly constrain pressure: our model predicts that the Si abundance would vary by more than an order of magnitude across the range of pressures we consider.

\subsection{SDSSJ0738+1835}

Cr, Ni and Si are all present, but the model can't constrain pressure because it finds multiple high-likelihood regions of parameter space, leading to a bimodal pressure distribution. All of these solutions invoke accretion of a moderately core-rich fragment (44\%) in the declining phase. However, there are 3 possible combinations of pressure, oxygen fugacity and stellar metallicity. Two of these solutions invoke high pressure, while the other invokes low pressure.

\subsection{SDSSJ0845+2257}

Pressure could not be constrained for this system due to a combination of both the large error bars and the inference that the fragment is only slightly core-rich (33\%), reducing the effect of changing pressure.

\subsection{WD0122-227 and WD1145+288}

WD0122-227 and WD1145+288 are polluted by core-rich fragments, but in each case the only potential for information about pressure comes from an upper bound on Si. Since these upper bounds are not strong, pressure is unconstrained.

\subsection{LHS2534}

The model was able to fit the data, but the result was deemed unphysical. The median model corresponds to a parent body composed of mostly O (79\%), with very low Fe content (6\%) - consequently it has an unrealistically small core. This behaviour is a consequence of moving to an unusually high metallicity/high formation distance part of parameter space. \citet{Hollands2021} concluded that this system was polluted by crustal material. We don't consider crustal material in our model, which may explain why we were unable to find a satisfactory fit to the data. We also note that this object is magnetic, which complicates spectral analysis \citep{Hollands2021}.

\subsection{GALEX1931+0117}

We adopted data from \citet{Gaensicke2012} for this system, from which we found no evidence of differentiation. However, data were also available from \citet{Vennes2011b} and \citet{Melis2011}. The fit to the data from \citet{Vennes2011b} was deemed to be unphysical, with a parent body composed of mostly O (89\%) and low Fe (3\%), similar to LHS2534. This was likely driven by the relative enrichment in Si and O compared to the other data sets. The data set from \citet{Melis2011} has a low Ca/Mg ratio which the model was unable to fit given that the short sinking timescales effectively restrict it to steady state accretion, and that the O abundance constrains the effect of incomplete condensation. \citet{Melis2011} suggested that the abundances may be explained by accretion of a parent body which has been stripped of its crust (and most of its mantle) by the AGB stellar wind. Similarly, \citet{Harrison2018} found that their fit to the data was improved if the parent body had a large crust which was removed by collisions. Our inability to fit this dataset may be because we don't include selective depletion of crustal material. We also note that our sinking timescales for hot DAs such as this may not be accurate (see \ref{sec:ModelCaveats2}).

\section{Predicted core/mantle composition}

\begin{figure*}
    \centering
    \includegraphics[width=13cm,keepaspectratio=true]{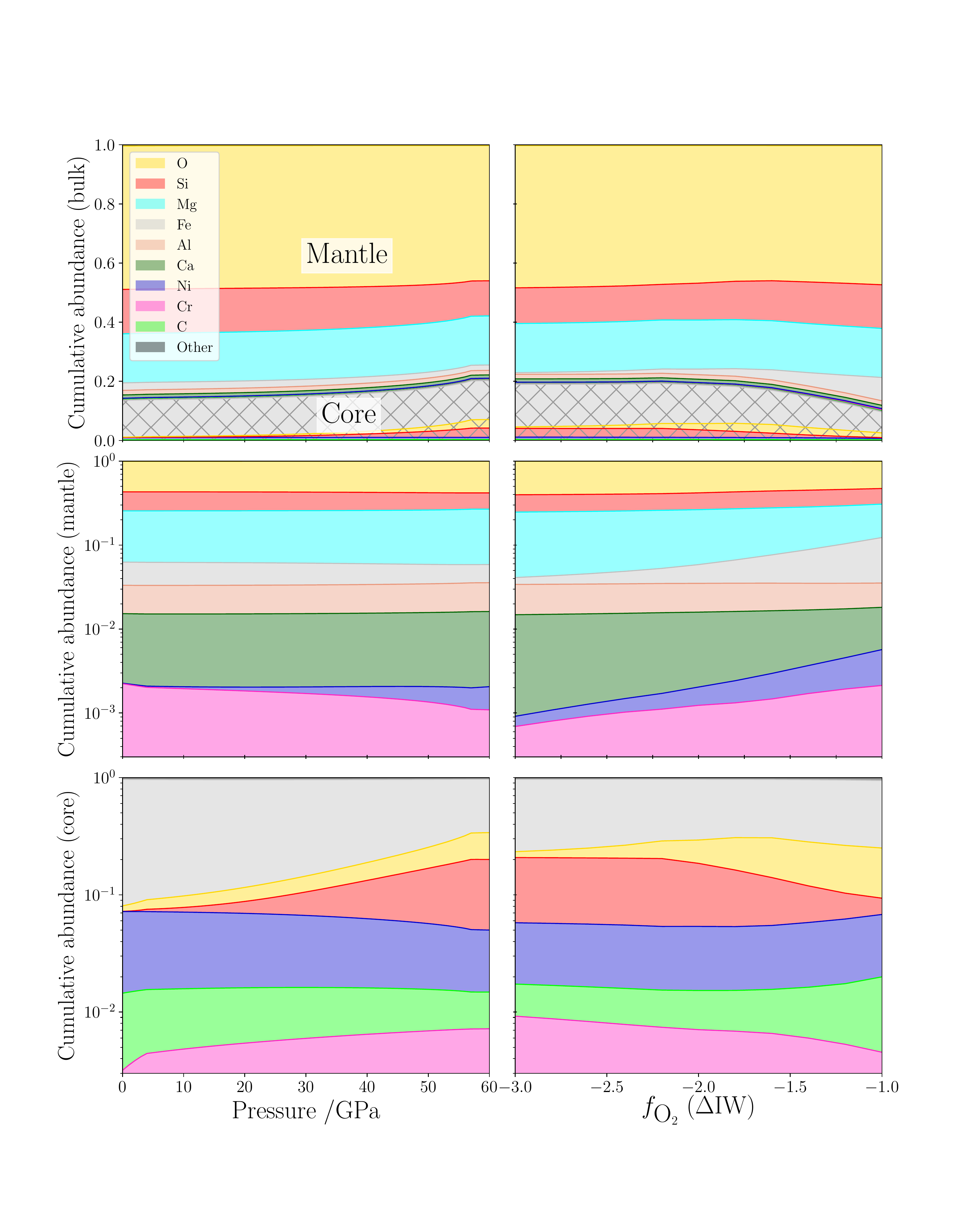}
    \caption{Illustration of the effect of model's predicted elemental partitioning behaviours on the resulting core and mantle composition in the parent body of a White Dwarf pollutant. In each panel, the vertical thickness of each shaded segment indicates the abundance (by number) of the corresponding element. The left hand column of panels shows how the elemental abundances vary with pressure, while the right hand panels show variation with oxygen fugacity (measured relative to the Iron W\"ustite buffer). The top panels show both the core (hatched area) and mantle (unhatched area) compositions of a body with bulk Earth composition. The cumulative abundance of elements in the core is equal to the predicted core number fraction, which for reference is approximately 0.17 for Earth. The middle and lower plots show just the mantle and core compositions respectively, with abundances plotted on a log scale (base 10) to better the behaviour of specific elements of interest. See the main text for a discussion of the salient features. When varying pressure, oxygen fugacity was held at IW\;-\;2, and when varying oxygen fugacity the pressure was held at 54\;GPa. These figures were chosen to roughly correspond to Earth-like values \citep{Fischer2015}.}
    \label{fig:stacked_composition_plot}
\end{figure*}

In this section, we show predictions for the core and mantle composition of an Earth-like body which forms under different pressure/oxygen fugacity conditions. Figure \ref{fig:stacked_composition_plot} illustrates the change in number abundance of various elements in the core and mantle as a function of pressure and oxygen fugacity. In each panel, the vertical thickness of each shaded segment indicates the abundance (by number) of the corresponding element. The top panels show both the core (hatched area) and mantle (unhatched area) compositions. Our model predicts that the core number fraction increases with increasing pressure (as Si and O partition more strongly into the core) and decreases with increasing oxygen fugacity (as metallic Fe is oxidised and incorporated into the mantle).

The behaviour discussed in section \ref{sec:d_behaviour} is visible in Figure \ref{fig:stacked_composition_plot}. In the left-hand middle panel, the dark blue strip shows the mantle abundance of Ni, which increases significantly with increasing pressure (as Ni becomes less siderophilic). In contrast, the core abundance of Ni (shown in the lower left-hand panel) changes very little. The lower left-hand panel also shows a dramatic increase in core Si (red) as pressure increases, while the mantle abundance remains roughly constant. As pressure increases, the increase in core Cr (pink), as well as the noticeable decrease in mantle Cr, are both visible.

\section{Pressure Sensitivity of Different Elements}
\label{sec:p_sensitivity}

Pressure sensitivity varies by element, and also depends on the overall fragment composition. We wish to calculate how the log bulk number abundance of an element, $M$, in the fragment, $X_b$, varies with pressure.

We firstly calculate $X_b$. For an element differentiating with partition coefficient $D$, the normalised core number abundance in the parent, $N_c$ is

\begin{equation}
 N_{c} = \frac{N_{b}D}{Dw + (1-w)},
 \label{eq:N_c}
\end{equation} where $N_b$ is the normalised bulk number abundance of element $M$ in the parent and w is the parent core number fraction. Similarly, the normalised mantle number abundance in the parent, $N_m$ is

\begin{equation}
 N_{m} = \frac{N_{b}}{Dw + (1-w)}
 \label{eq:N_m}
\end{equation}

The fragment's composition is determined by combining the parent core and mantle material in arbitrary proportions. Taking $f$ to be the fraction of core-like material in the fragment, we find that

\begin{equation}
 X_{b} = \frac{fDN_{b}+(1-f)N_{b}}{Dw + (1-w)}
 \label{eq:X_b}
\end{equation}

As the pressure changes, $D$ and $w$ also change. Differentiating equation \ref{eq:X_b} with respect to $P$ gives

\begin{equation}
 \frac{dX_{b}}{dP} = \frac{N_{b}(f-w)\frac{dD}{dP}-N_{b}(f(D-1)+1)(D-1)\frac{dw}{dP}}{(1+(D-1)w)^{2}},
 \label{eq:dX_b_dP}
\end{equation} in which we note that $P$ could be replaced by any variable which affects $D$ and/or $w$. Note that if the parent core fraction does not vary with $P$, the right hand side vanishes when $f=w$. In other words, we recover the intuitive result that if the fragment samples the parent's core and mantle in the same proportion as the parent itself, the abundance of $M$ is unaffected. In our model, the parent core fraction generally increases with pressure (i.e., $\frac{dw}{dP} > 0$, see Figure \ref{fig:stacked_composition_plot}), which means that even when $f=w$ the abundance of $M$ varies with pressure. 

Elemental observations are typically expressed as log number abundances (base 10), so we now find $\frac{d \log X_{b}}{dP}$. Using the identity

\begin{equation}
 \frac{d \log(X_{b})}{dP} = \frac{1}{X_{b}\ln(10)}\frac{dX_{b}}{dP}
 \label{eq:dlogX_identity}
\end{equation} we find

\begin{equation}
 \frac{d \log(X_{b})}{dP} = \frac{(f-w)\frac{dD}{dP}-(f(D-1)+1)(D-1)\frac{dw}{dP}}{\ln(10)(f(D-1)+1)(w(D-1)+1}
 \label{eq:dlogX_b_dP}
\end{equation}

The greater the magnitude of this quantity, the easier it is to resolve the core--mantle differentiation pressure in the parent body. Observational errors are typically on the order of 0.1 log units, so as a rough order of magnitude estimate we require $|\frac{d \log(X_{b})}{dP}| > 0.01 \textrm{GPa}^{-1}$ in order to resolve the pressure to within 10 GPa. This threshold could be lower if multiple elements are detected.

\begin{figure}
    \centering
    \includegraphics[width=8cm]{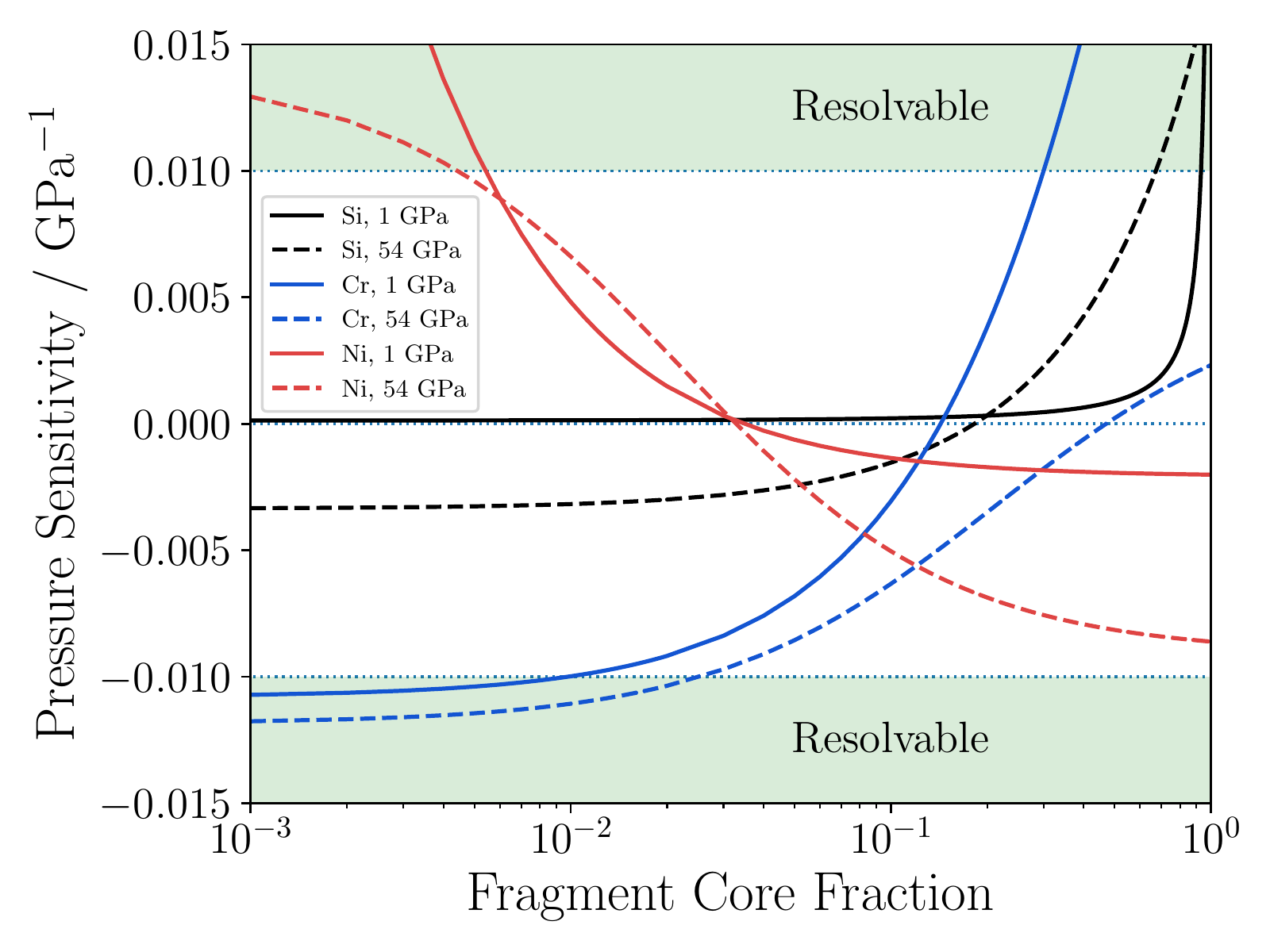}
    \caption{Sensitivity of selected elements to pressure with varying fragment core fraction, calculated using Equation \ref{eq:dlogX_b_dP}. The shaded green regions give an approximate indication of the magnitude of pressure sensitivity which our model can resolve (this threshold could be more generous if multiple elements are present or errors are small). Pure (or nearly pure) core- or mantle-like material is needed in order for pressure induced abundance changes to be resolvable.}
    \label{fig:dlogX_dP_v_fcf}
\end{figure}

Figure \ref{fig:dlogX_dP_v_fcf} shows how pressure sensitivity changes with fragment core fraction for different elements, assuming a bulk Earth parent composition.

Despite Ni having a partition coefficient which spans many orders of magnitude over the pressures we investigate, it only becomes sensitive to pressure at very low fragment core fractions ($\approx$0.01). This is because the value of $f$ at which $\frac{d \log(X_{b})}{dP} = 0$ is reduced by virtue of $D_{\textrm{Ni}}$ being large and decreasing with pressure. Because Ni is highly siderophilic, its abundance in the core is not sensitive to pressure and even a small amount of core-like material overwhelms the mantle's Ni contribution. This is why the Ni abundance of certain systems shows little sensitivity to pressure, despite being polluted by apparently mantle-rich material.

In the context of Figure \ref{fig:bowtie_multipanel}, $\frac{d (\log X_{b})}{dP}$ is the quantity which sets how spread apart the pressure contours are at a given fragment core fraction. Since we plot ratios of 2 elements, the contours will additionally be spread apart further (along the relevant axis direction) if the elements have sensitivities with opposite signs (and vice versa). Pressure sensitivity is a function of pressure, which means that along a line of constant fragment core fraction, the pressure contours are not necessarily evenly spaced.

\bsp	
\label{lastpage}
\end{document}